\documentclass[aps,pra,twocolumn,reprint,10pt]{revtex4-1}
\pdfoutput=1
\usepackage{amsmath,amssymb,braket}
\usepackage{blindtext}
\usepackage{natbib}
\usepackage{hyperref}
\usepackage{graphicx}
\usepackage{color,soul}
\usepackage[caption=false]{subfig}
\renewcommand{\vec}[1]{\mathbf{#1}}
\newcommand{\RN}[1]{%
  \textup{\uppercase\expandafter{\romannumeral#1}}%
}
\begin{document}
\title{ORIGIN AND IMPLICATIONS OF $A^2$-LIKE CONTRIBUTION \\ IN THE QUANTIZATION OF CIRCUIT-QED SYSTEMS}
\author{Moein Malekakhlagh}
\author{Hakan E. T\"ureci}
\affiliation{Department of Electrical Engineering, Princeton University, Princeton, New Jersey, 08540}

\begin{abstract}
By placing an atom into a cavity, the electromagnetic mode structure of the cavity is modified. In Cavity QED, one manifestation of this phenomenon is the appearance of a gauge-dependent diamagnetic term, known as the $A^2$ contribution. Although in atomic Cavity QED, the resulting modification in the eigenmodes is negligible, in recent superconducting circuit realizations, such corrections can be observable and may have qualitative implications. We revisit the canonical quantization procedure of a circuit QED system consisting of a single superconducting transmon qubit coupled to a multimode superconducting microwave resonator. A complete derivation of the quantum Hamiltonian of an open circuit QED system consisting of a transmon qubit coupled to a leaky transmission line cavity is presented. We introduce a complete set of modes that properly conserves the current in the entire structure and present a sum rule for the dipole transition matrix elements of a multi-level transmon qubit coupled to a multi-mode cavity. Finally, an effective multi-mode Rabi  model is derived with coefficients that are given in terms of circuit parameters.   
\end{abstract}
\maketitle
\section{Introduction}

In single-mode realization of Cavity QED (CQED), a single atom coupled to a small high-Q electromagnetic resonator can be well-described by a model wherein the matter is described by a single atomic transition, and its coupling to one of the modes of the resonator can saturate this transition before other modes are populated \cite{haroche_exploring_2013}. A plethora of fundamental physical phenomena and their recent applications in quantum information science has been explored and vigorously pursued with a superconducting circuit-based realization of this setup \cite{blais_cavity_2004, wallraff_strong_2004, chiorescu_coherent_2004, devoret_implementing_2004, girvin_circuit_2009, devoret_superconducting_2013}. In such systems, the existence of the atom leads to a modification in the cavity modal structure due to Rayleigh-like scattering. Such corrections are unobservably small in atomic CQED unless a special cavity structure is chosen. In recent realizations of circuit quantum electrodynamics (cQED) however, such corrections may have observable consequences which we discuss in this article. 

A better known manifestation of the aforementioned scattering corrections is the so-called $A^2$ term in single-mode CQED. There has been a lively debate in recent years \cite{nataf_no-go_2010, viehmann_superradiant_2011, hayn_superradiant_2012, vukics_adequacy_2012, baksic_superradiant_2013, zhang_quantum_2014} about the impact of this term on synthetic realizations of the single-mode superradiant phase transition \cite{hepp_superradiant_1973, wang_phase_1973, carmichael_higher_1973} when instead of one, $N$ identical non-interacting quantum dipoles are coupled with an identical strength to a single cavity mode. This particular instability of the electromagnetic vacuum has originally been discussed \cite{hepp_superradiant_1973, wang_phase_1973, carmichael_higher_1973} within the context of the single-mode version of the Dicke model \cite{dicke_coherence_1954} where the $A^2$ term was not included. Subsequent work shortly thereafter \cite{rzaznewski_phase_1975, knight_are_1978, bialynicki-birula_no-go_1979} pointed out that the $A^2$ term rules out such a transition. Recent theoretical work on superconducting realizations of the Dicke Model \cite{nataf_no-go_2010} has challenged the validity of such "no-go" theorems \cite{bialynicki-birula_no-go_1979}. Leaving this contentious matter aside~\cite{viehmann_superradiant_2011, ciuti_comment_2012}, we note here that the $A^2$ term is a gauge-dependent object, and specifically appears in the Coulomb gauge description of the single-mode atomic CQED. However, the scattering corrections due to the existence of an atom in a cavity are physical and measurable, and hence not dependent on the choice of gauge. In fact, recent realizations \cite{sundaresan_beyond_2015} of the multimode strong coupling regime in a very long coplanar waveguide cavity, as well as cQED systems in the ultra-strong coupling regime \cite{niemczyk_circuit_2010, peropadre_switchable_2010} provide settings where such corrections may be observable, as we discuss below.  

By fabricating a charge qubit close to a transmission line resonator, the resonator's capacitance per unit length is locally altered. This impurity scattering term is typically neglected \cite{koch_time-reversal-symmetry_2010, nunnenkamp_synthetic_2011, schmidt_circuit_2013, devoret_quantum_2014} in the derivation of the quantized Hamiltonian for the multi-mode regime of cQED \cite{krimer_route_2014}, we therefore revisit the quantization procedure in section $\RN{2}$. We discuss how the qubit changes the propagation properties of the resonator and how as a result this modifies its eigenmodes and eigenfrequencies. We show in particular that this new basis is the one which properly fulfills current conservation law at the point of connection to the qubit. Finally, in section $\RN{3}$ we show how our results can be generalized to a leaky cavity, one that is connected capacitively to external waveguides. In section $\RN{4}$, we briefly discuss the comparison to the case of atomic CQED. We point out that including the $A^2$ term in the Hamiltonian will lead to the same type of modification in the modes of a cavity.
\section{Model}

\begin{figure}
\subfloat[\label{subfig:circuit-QED-Closed-Schematic}]{%
\includegraphics[scale=0.26]{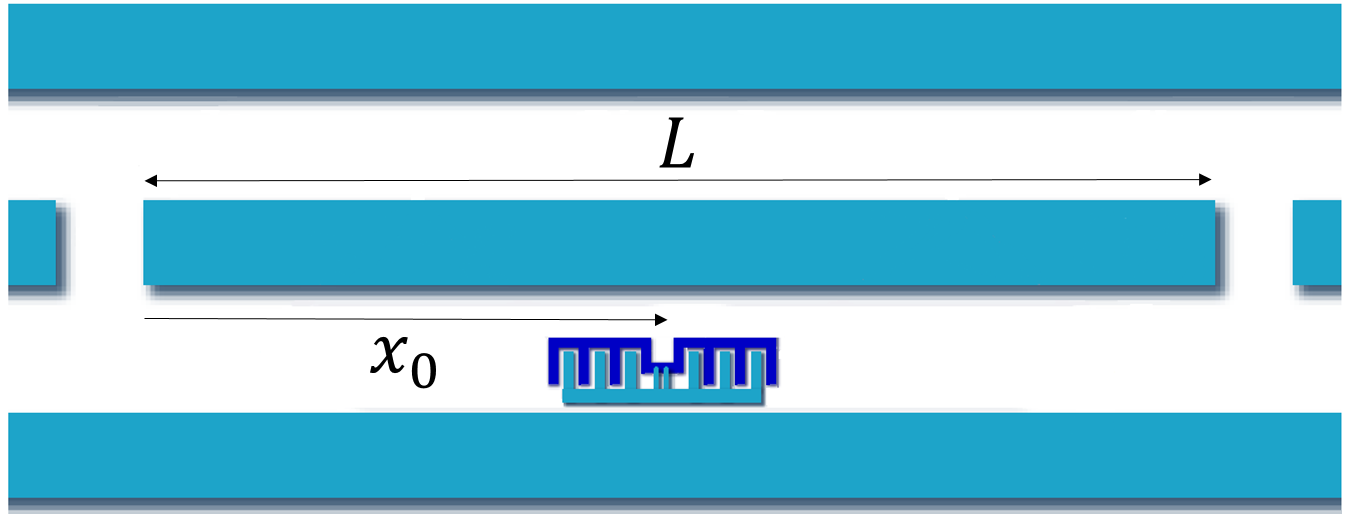}%
}\hfill
\subfloat[\label{subfig:circuit-QED-Closed-EquivalentCircuit}]{%
\includegraphics[scale=0.20]{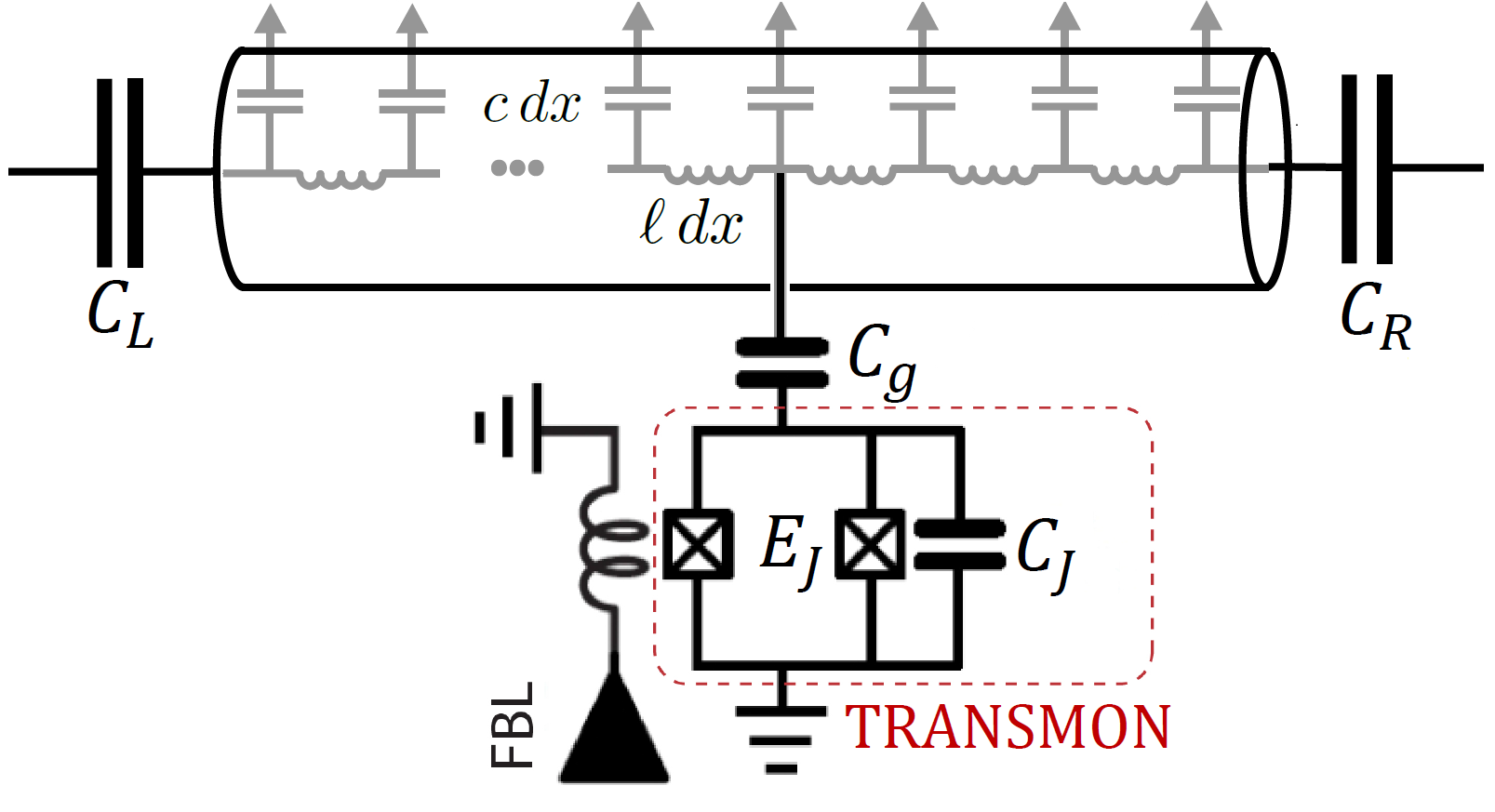}%
}
\caption{A superconducting transmon qubit coupled capacitively to a superconducting coplanar transmission line. (a). Device view (b). Equivalent circuit}
\label{Circuit-QED-Closed}
\end{figure}

As illustrated in Fig.~\ref{subfig:circuit-QED-Closed-Schematic}, we consider a common cQED design \cite{koch_charge-insensitive_2007} consisting of a single transmon qubit that is fabricated at point $x_0$ (in the dipole approximation, as discussed below) inside a superconducting transmission resonator of length $L$. In this section, we assume that $C_{L,R} = 0$, which corresponds to closed (perfectly reflecting) boundary conditions at $x = 0, L$. In section $\RN{3}$ we discuss the open case, where a finite transmission line of length $L$ is coupled through nonzero capacitors $C_{R,L}$ to the rest of the circuit. 

\subsection{Classical Hamiltonian and CC Basis}
As discussed in detail in Appendix B, the Hamiltonian for this circuit can be written as 
\begin{align}
\begin{split}
\mathcal{H}&=\underbrace{\frac{Q_J^2}{2C_J}-E_J\cos\left(2\pi\frac{\Phi_J}{\phi_0}\right)}_{\mathcal{H}_A}\\
&+\underbrace{\int_{0}^{L}dx \left[\frac{\rho^2(x,t)}{2c(x,x_0)}+\frac{1}{2l}\left(\frac{\partial \Phi(x,t)}{\partial x}\right)^2\right]}_{\mathcal{H}_C^{mod}} \\
&+\underbrace{\gamma Q_J \int_{0}^{L}dx \frac{\rho(x,t)}{c(x,x_0)} \delta(x-x_0)}_{\mathcal{H}_{int}}
\end{split}
\end{align}
The notation used here follows the canonical approach to quantization of superconducting electrical circuits \cite{bishop_circuit_2010, devoret_quantum_2014}, briefly reviewed for completeness in Appendix A. In the above Hamiltonian, the canonical variables $\Phi_J$ and $Q_J$ represent the flux and charge of the transmon qubit respectively. In a similar manner, the canonical fields $\Phi(x,t)$ and $\rho(x,t)$ represent the flux field and charge density field of the transmission line. Furthermore, $\phi_0\equiv\frac{h}{2e}$ is the flux quantum and $\gamma\equiv \frac{C_g}{C_g+C_J}$. 

There is a crucial difference between the Hamiltonian we have found here, with respect to earlier treatments \cite{koch_charge-insensitive_2007, devoret_quantum_2014}. We do include the modification in the resonator's capacitance per length at the qubit connection point $x_0$, $c(x,x0) = c + C_s \delta(x-x_0)$ where $C_s$ is the series capacitance of $C_J$ and $C_g$ given as $\frac{C_JC_g}{C_J+C_g}$. The Dirac delta function is the result of treating the qubit as a point object with respect to the resonator, whereas a more realistic model would replace that with a smooth function as we have discussed in Appendix F in great detail. As we see shortly, the delta function appearing in the denominator will not cause any issues in the quantization procedure, since the charge density $\rho(x,t)$ also contains the appropriate information regarding this point object so that $\frac{\rho(x,t)}{c(x,x_0)}$ turns out to be a continuous function in $x$. Once we understand how this correction influences the photonic mode structure of the resonator, we will move on to use that information in constructing the quantization procedure. 

The Hamiltonian equations of motion derived from the Hamiltonian $\mathcal{H}_{C}^{mod}$ of the modified resonator including the impurity scattering term is given by (see Appendix B):
\begin{align}
\frac{\partial \Phi(x,t)}{\partial t}&=\frac{\rho(x,t)}{c(x,x_0)} \\
\frac{\partial \rho(x,t)}{\partial t}&=\frac{1}{l}\frac{\partial^2 \Phi(x,t)}{\partial x^2}
\end{align}
The solution to these linear equations can be written in terms of the Fourier transform $\Phi(x,t) = \frac{1}{2\pi} \int_{-\infty}^{+\infty} dt \, e^{-i\omega t} \tilde{\Phi}(x,\omega)$ where $\tilde{\Phi}(x,\omega)$ is the solution of the 1D Helmholtz equation
\begin{align}
\frac{\partial^2 \tilde{\Phi}(x,\omega)}{\partial x^2}+lc(x,x_0)\omega^2\tilde{\Phi}(x,\omega)&=0
\label{Eq:Modified Wave Equation-Closed Case_body}
\end{align}

We look for solutions that carry zero current across the boundaries, implemented by Neumann-type boundary conditions $\partial_x \tilde{\Phi} (x)|_{x=0,L} = 0$. A solution then exists only at discrete and real values $\omega=\omega_n$. The Dirac delta function hidden in $c(x,x_0)$ can be translated into discontinuity in $\partial_x \tilde{\Phi}(x)$ which is proportional to the current $\tilde{I}(x)=-\frac{1}{l}\frac{\partial \tilde{\Phi}(x)}{\partial x}$ that enters and exits the point of connection to the transmon
\begin{align}
-\frac{1}{l}\left.\frac{\partial \tilde{\Phi}(x,\omega)}{\partial x}\right|_{x_0^+}+\frac{1}{l}\left.\frac{\partial \tilde{\Phi}(x,\omega)}{\partial x}\right|_{x_0^-} =C_s\omega^2 \tilde{\Phi}(x_0,\omega)
\end{align}
where the R.H.S is the current that enters $C_J$ through $C_g$ , therefore the series capacitance $C_s$. This condition amounts to the conservation of current at the point of connection to the qubit and thus it is appropriate to call the set of eigenmodes satisfying this condition the {\it current-conserving (CC) basis}. 
\begin{figure}
\centering
\subfloat[\label{subfig:Eigfreqsx001}]{%
\centerline{\includegraphics[scale=0.24]{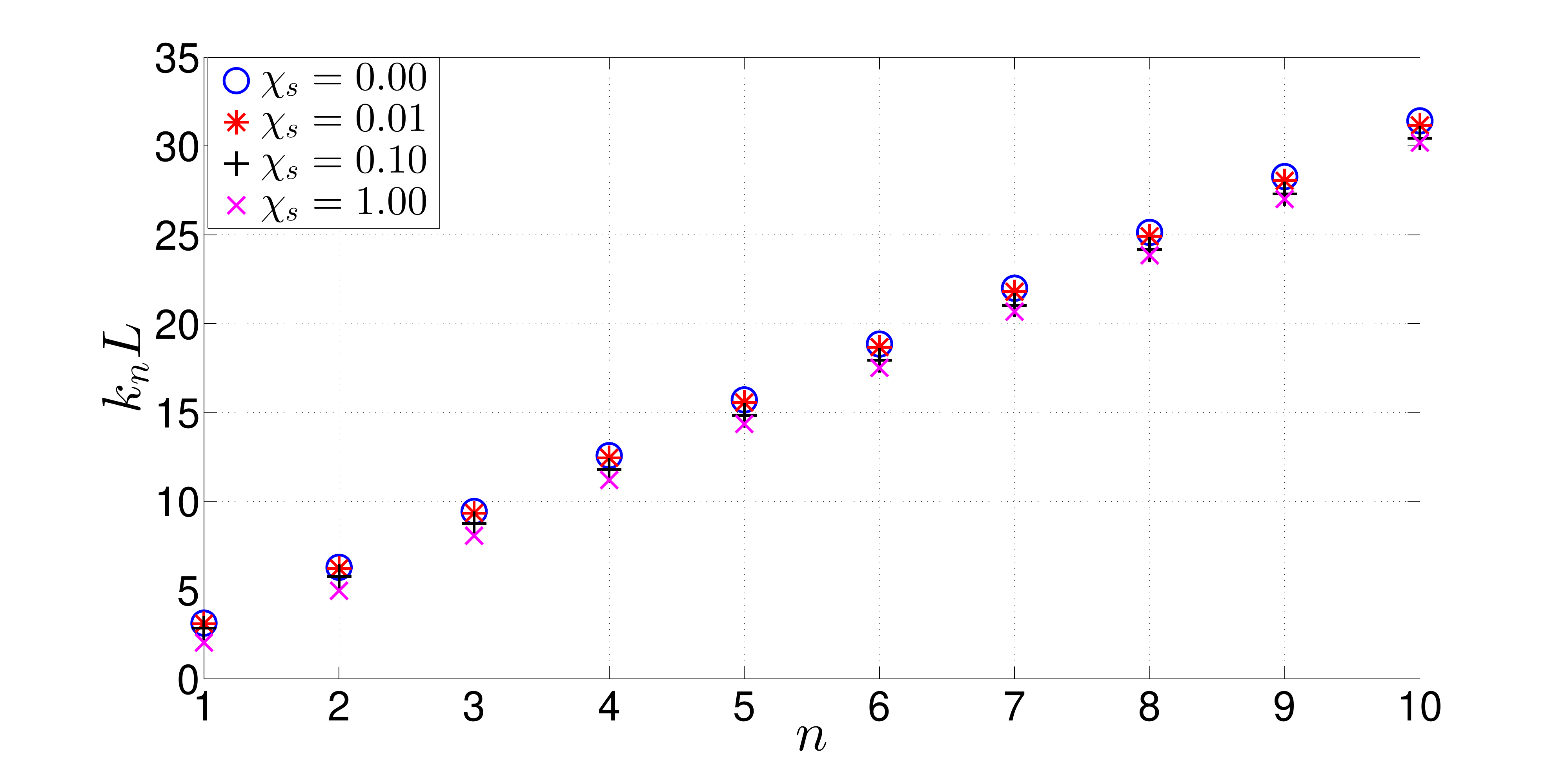}%
}}\hfill
\subfloat[\label{subfig:LevelSpacingchi0001}]{%
\includegraphics[scale=0.107]{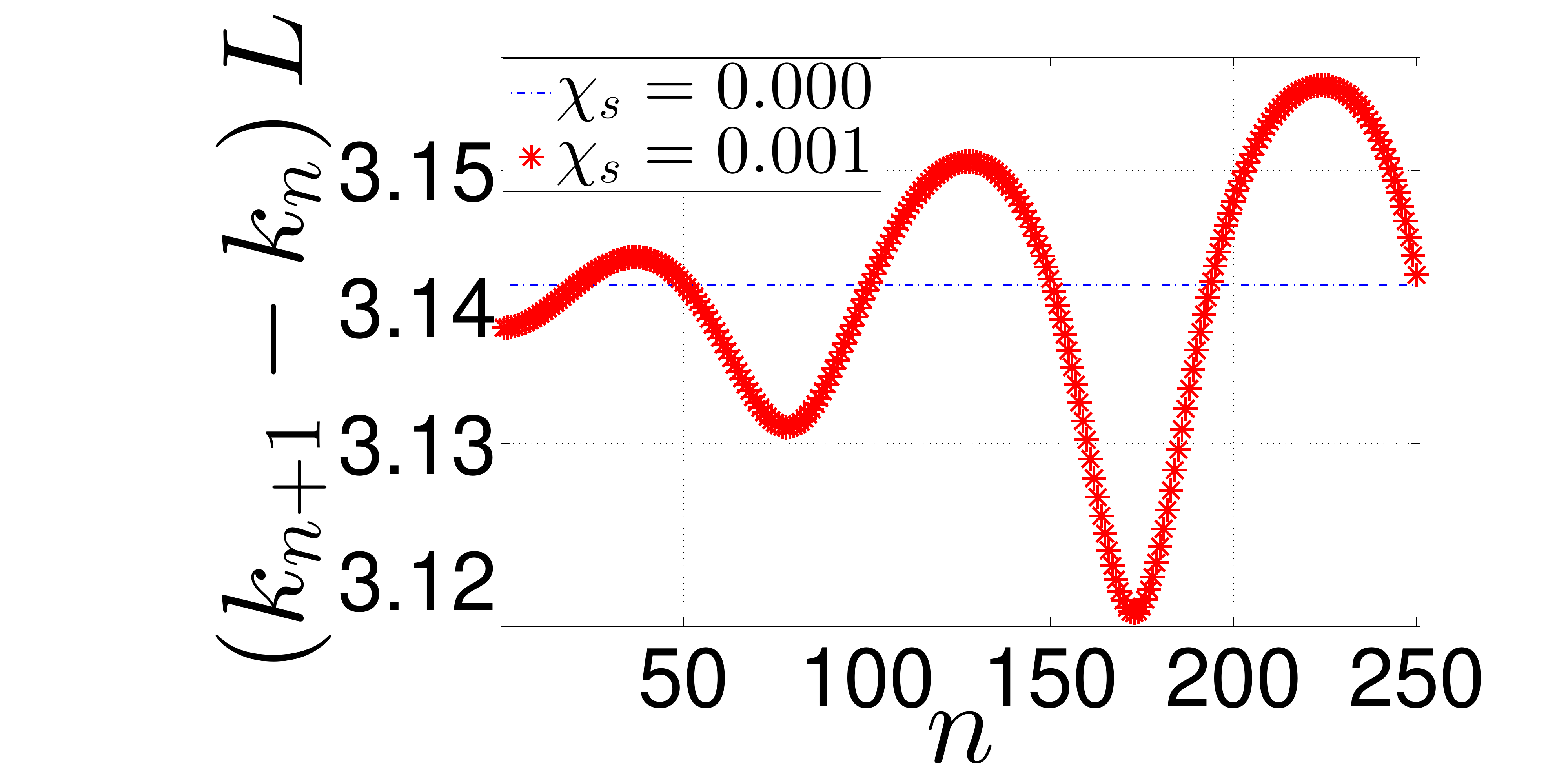}%
}
\subfloat[\label{subfig:LevelSpacingchi001}]{%
\includegraphics[scale=0.107]{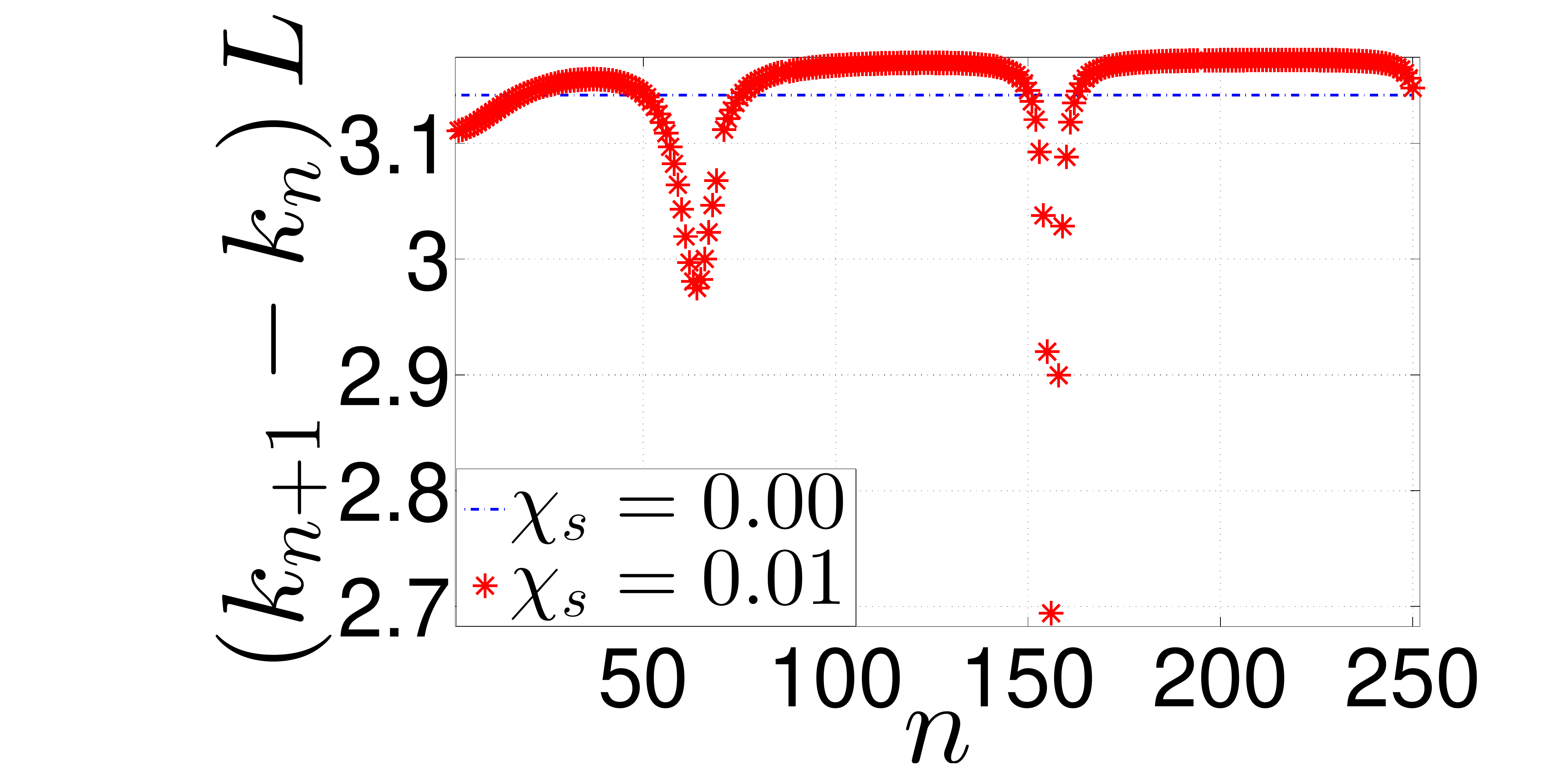}%
}
\caption{a) The first 10 modified resonances for different values of $\chi_s$ and $x_0=0.01L$. Higher modes with larger $\chi_s$ experience a larger shift in frequency. b,c) Normalized level-spacing for $\chi_s=0.001$, and $0.01$ respectively. The blue dashed line shows the constant level spacing for $\chi_s=0$.}
\label{fig: eigenfrequencies for different values of chi_s}
\end{figure}
The solution of the above-stated Neumann problem gives the CC eigenfrequencies $\omega_n$ through the transcendental equation
\begin{align}
\sin(k_nL)+\chi_s k_nL\cos(k_n x_0)\cos(k_n(L-x_0))=0 
\label{Eq: Closed CC Eigenfrequencies_body}
\end{align}
In the equation above, $k_n L= \frac{\omega_n}{v_p}L=\sqrt{lc}\omega_n L$ is the normalized eigenfrequency and $\chi_s=\frac{C_s}{cL}$ is a unitless measure of the transmon-induced modification in eigenfrequencies/eigenstates compared to the conventional cosine basis. The CC eigenfunctions are given by
\begin{align}
\tilde{\Phi}_n(x)\propto
\begin{cases}
\cos{\left(k_n(L-x_0)\right)}\cos{(k_n x)}&0<x<x_0\\
\cos{(k_n x_0)}\cos{(k_n(L-x))} &x_0<x<L
\end{cases}
\label{Eq: Closed CC Eigenmodes_ body}
\end{align}
The proportionality constant has to be set by the orthogonality relations that can be found directly from the modified wave equation \ref{Eq:Modified Wave Equation-Closed Case_body} as
\begin{align}
&\int_0^L dx \frac{c(x,x_0)}{c} \tilde{\Phi}_n(x) \tilde{\Phi}_m(x) = L\delta_{mn}
\label{Eq:Orthogonality of Phi _body}\\
&\int _{0}^{L}dx \frac{\partial\tilde{\Phi}_m(x)}{\partial x}\frac{\partial\tilde{\Phi}_n(x)}{\partial x}=k_m k_n L \delta_{mn}
\label{Eq:Orthogonality of d/dx Phi _body}
\end{align}

Based on these results, eigenfrequencies are not only sensitive to $\chi_s$, but also to the point of connection $x_0$. In order to understand this modification better, first we have plotted the normalized eigenfrequencies in Fig. \ref{subfig:Eigfreqsx001} for different values of $\chi_s$ and the case where qubit is connected very closely to one of the ends. This is a standard location for fabricating a qubit \cite{sundaresan_beyond_2015} to attain a strong coupling strength between the resonator modes and the qubit, since the electromagnetic energy concentration is generally highest near the ends. In this figure, the blue circles representing the eigenfrequencies for $\chi_s=0$ are located at $n\pi$. For $\chi_s \neq 0$, all lower CC eigenfrequencies are red-shifted with respect to the $\chi_s=0$ solutions and by going to higher mode number and higher $\chi_s$, the deviation becomes more visible. In a larger scale however, the behavior of $\chi_s \neq 0$ eigenvalues are non-monotonic and most notably, display a dispersion in frequency. For better visibility of this periodic behavior, in Figs \ref{subfig:LevelSpacingchi0001}-\ref{subfig:LevelSpacingchi001} we have compared the level spacing of CC modes for different values of $\chi_s$ to the constant level spacing of unmodified cosine modes. This behavior is determined by the position of the qubit connection point $x_0$ and is easy to understand. Since $x_0=0.01L$ sits at the local minima of modes 50, 150, 250 and so on, we expect a periodic behavior in the values of CC eigenfrequencies where within some portion of that period set by $\chi_s$, CC solutions are less than the $\chi_s=0$ solutions and vice versa in the remaining portion. 
\begin{figure}
\subfloat[\label{subfig:mode1x001}]{%
\includegraphics[scale=0.28]{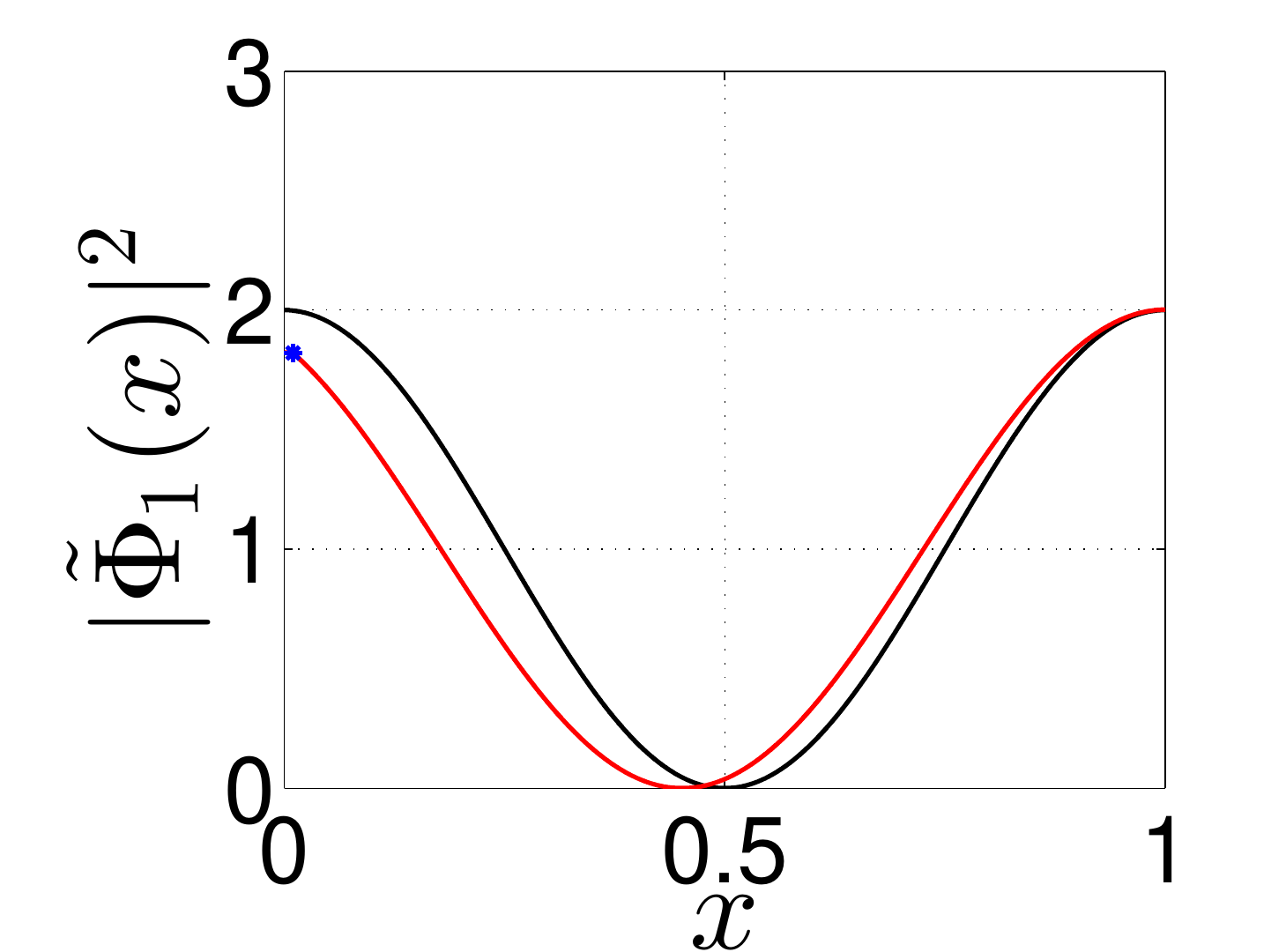}%
}
\subfloat[\label{subfig:mode2x001}]{%
\includegraphics[scale=0.28]{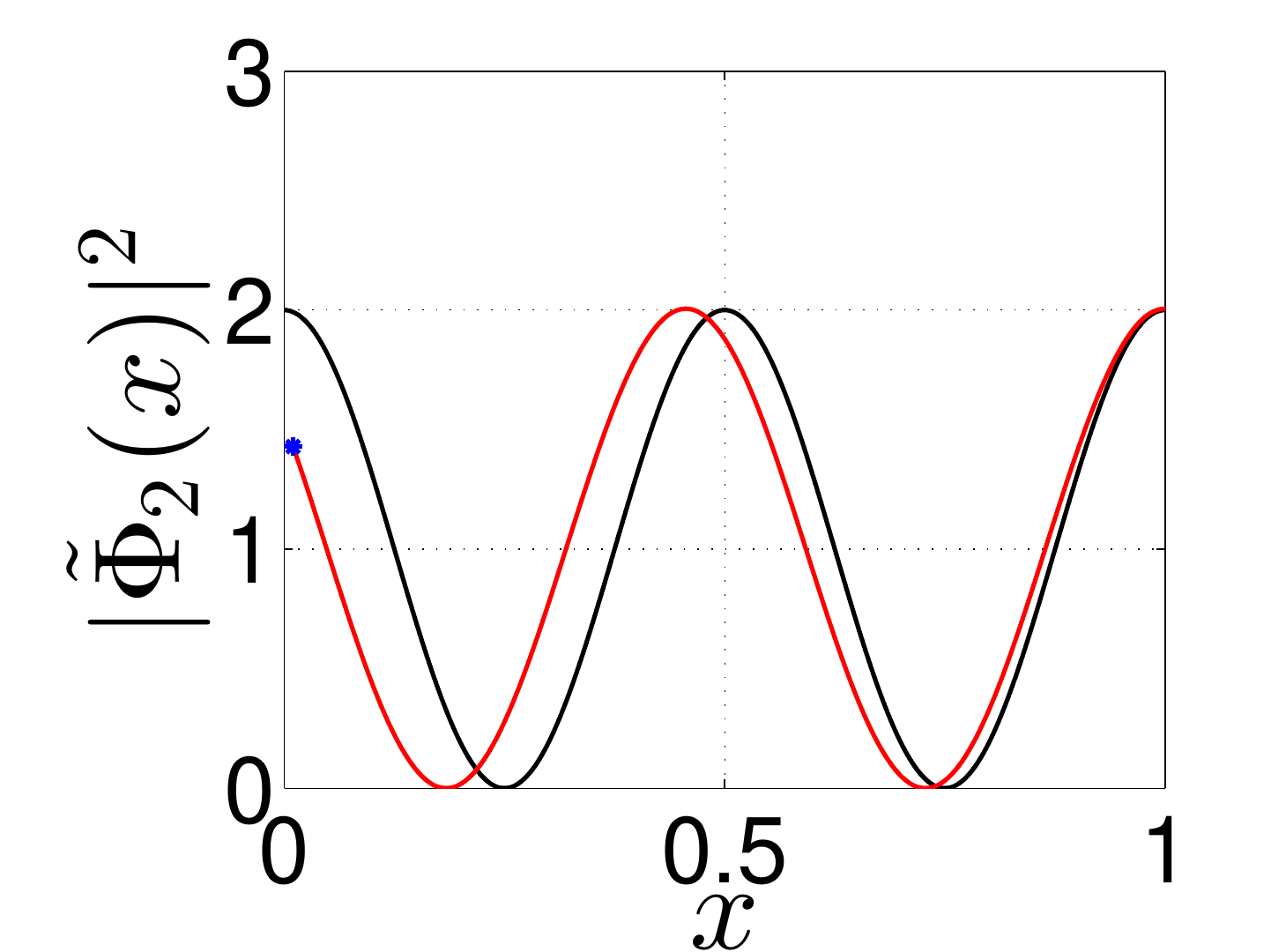}%
}\hfill
\subfloat[\label{subfig:mode3x001}]{%
\includegraphics[scale=0.28]{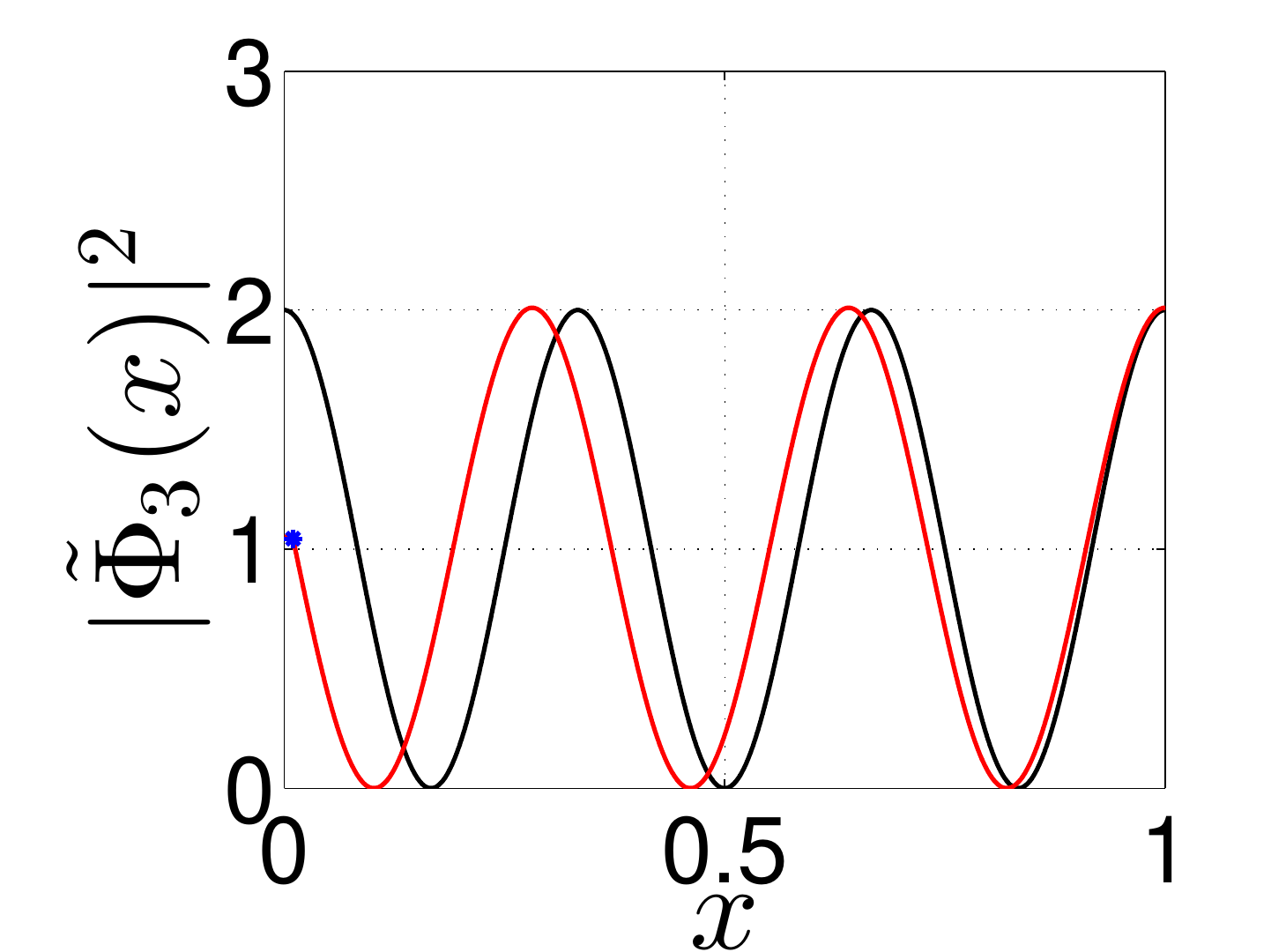}%
}
\subfloat[\label{subfig:mode4x001}]{%
\includegraphics[scale=0.28]{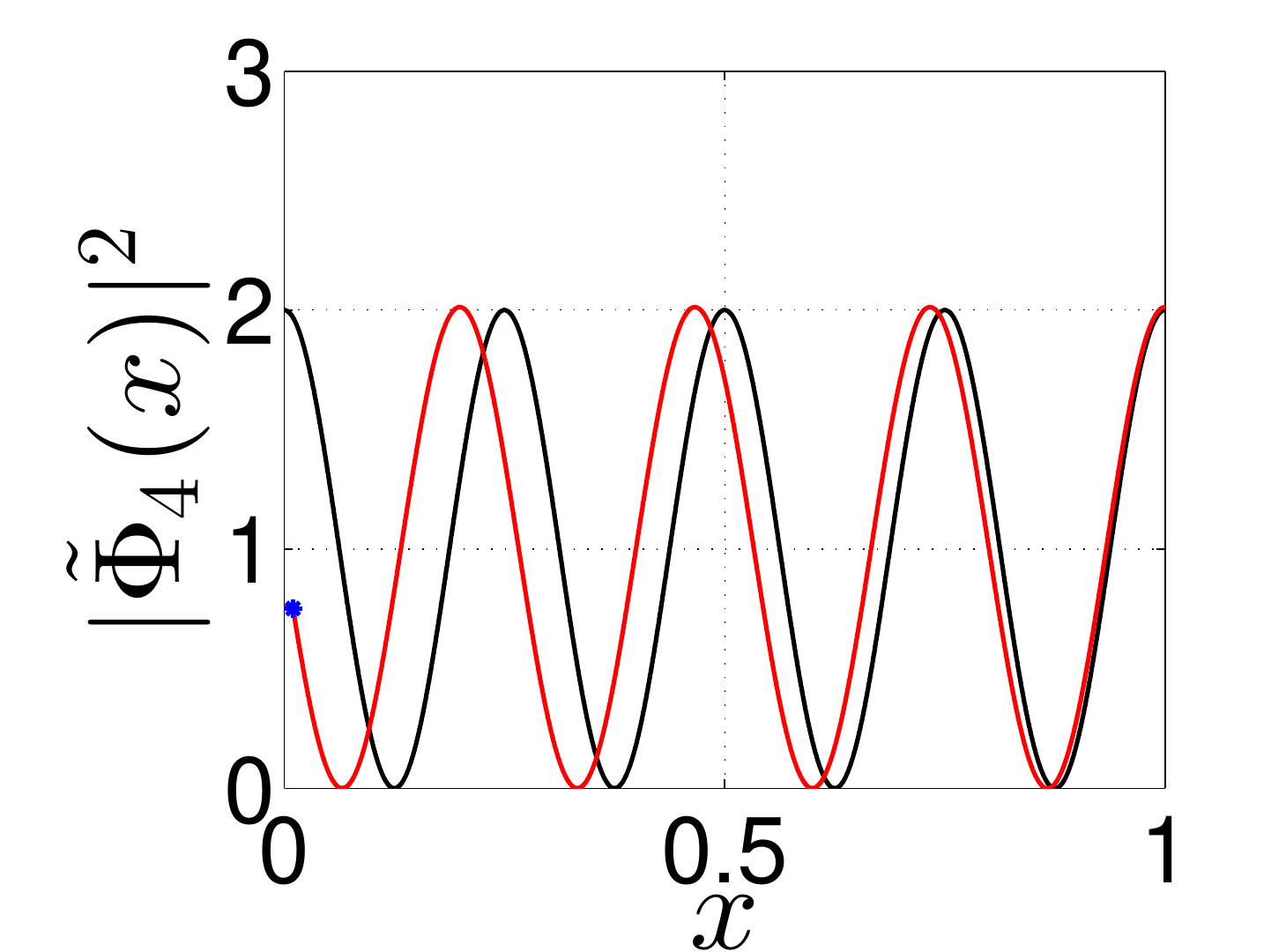}%
}
\caption{ Normalized energy density of the first 4 modes for $x_0=0.01L$ and $\chi_s=0.1$. The black curve shows cosine modes while the red curve represents CC modes. The blue star shows where the qubit is connected.}
\label{fig:Eigx001}
\end{figure}

In Fig. \ref{fig:Eigx001} we show the spatial dependence of the first four modes. The amplitude of the lower CC eigenmodes at the qubit location are consistently less than that for the unmodified cosine eigenmodes, suggesting that the actual coupling strengths of the qubit to these modes are below the ones predicted by the $\chi_s=0$ modes.

We note that similar modification in the modal structure of resonators have also been noted for transmission-line resonators containing inline transmons \cite{bourassa_josephson-junction-embedded_2012} as well as 3D cQED architectures \cite{nigg_black-box_2012}.  In these studies, the modifications result from solving the spectral problem of the quadratic Hamiltonian that in addition to the resonator part also includes the linear part of the JJ's $\cos (2\pi\Phi_J/\phi_0)$-type non-linearity, unlike the situation described here.

In order to see the dependence of these CC modes on $x_0$, we have also investigated two different cases of connecting the qubit to $x_0=\frac{L}{2}$ and $x_0=\frac{L}{4}$ in Appendix C in figures \ref{fig:Eigx050} and \ref{fig:Eigx025} respectively.

\subsection{Canonical Quantization}

As discussed in Appendix C, the conjugate quantum fields of the resonator can be expanded in terms of CC basis as
\begin{align}
&\hat{\Phi}(x,t)=\sum\limits_n \left(\frac{\hbar}{2\omega_n cL}\right)^{\frac{1}{2}}\left(\hat{a}_n(t) + \hat{a}_n^{\dagger}(t)\right)\tilde{\Phi}_n(x)\\
&\hat{\rho}(x,t)=-i\sum\limits_n \left(\frac{\hbar \omega_n}{2cL}\right)^{\frac{1}{2}}\left(\hat{a}_n(t) - \hat{a}_n^{\dagger}(t)\right)c(x,x_0)\tilde{\Phi}_n(x) 
\end{align}

By substituting these expressions into $\hat{\mathcal{H}}_C^{mod}$ and employing orthogonality conditions \ref{Eq:Orthogonality of Phi _body} and \ref{Eq:Orthogonality of d/dx Phi _body}, one finds its diagonal representation
\begin{align}
\hat{\mathcal{H}}_C^{mod}=\sum\limits_n \hbar \omega_n \left( \hat{a}_n^{\dagger} \hat{a}_n + \frac{1}{2} \right)
\end{align}
In a similar manner, the qubit flux operator $\hat{\Phi}_J$ and charge operator $\hat{Q}_J$ can be represented in terms of eigenmodes of transmon Hamiltonian as
\begin{align}
&\hat{\Phi}_J(t)=\sum\limits_{m,n}\bra{m}\hat{\Phi}_J(0)\ket{n}\hat{P}_{mn}(t)\\
&\hat{Q}_J(t)=\sum\limits_{m,n}\bra{m}\hat{Q}_J(0)\ket{n}\hat{P}_{mn}(t)
\end{align}
where $\hat{P}_{mn}(t)$ represent a set of projection operators acting between states $\ket{m}$ and $\ket{n}$. Working in the flux basis, the eigenmodes are found through solving a Schr\"odinger equation as
\begin{align}
\left[-\frac{\hbar^2}{2C_J}\frac{d^2}{d\Phi_J^2}-E_J\cos{\left(2\pi\frac{\Phi_J}{\phi_0}\right)}\right]\Psi_n(\Phi_J)=\hbar\Omega_n\Psi_n(\Phi_J)
\end{align}
whose solution can be characterized in terms of Mathieu functions~\cite{koch_charge-insensitive_2007}. Due to the invariance of the Hamiltonian under flux parity transformation, the eigenmodes are either even or odd functions of $\Phi_J$ and as explained in Appendix C, only off-diagonal elements between states having different parities are non-zero. Consequently, flux and charge matrix elements are purely real and imaginary. Therefore, we can rewrite 
\begin{align}
\hat{\Phi}_J(t)&=\sum\limits_{m<n}\bra{m}\hat{\Phi}_J(0)\ket{n}\left(\hat{P}_{mn}(t)+\hat{P}_{nm}(t)\right)\\
\hat{Q}_J(t)&=\sum\limits_{m<n}\bra{m}\hat{Q}_J(0)\ket{n}\left(\hat{P}_{mn}(t)-\hat{P}_{nm}(t)\right)
\end{align}
Finally, applying the unitary transformation $\hat{a}_l\rightarrow i\hat{a}_l$ and $\hat{P}_{mn}\rightarrow i\hat{P}_{mn}$ for $m<n$, the second quantized representation of the Hamiltonian in its most general form can be expressed as
\begin{align}
\begin{split}
\hat{\mathcal{H}}&=\underbrace{\sum\limits_n \hbar\Omega_n \hat{P}_{nn}}_{\hat{\mathcal{H}}_A}+\underbrace{\sum\limits_n \hbar \omega_n \hat{a}_n^{\dagger} \hat{a}_n}_{\hat{\mathcal{H}}_C^{mod}}\\
&+\underbrace{\sum\limits_{m<n,l} \hbar g_{mnl} \left(\hat{P}_{mn}+\hat{P}_{nm}\right)\left(\hat{a}_l+\hat{a}^{\dagger}_l\right)}_{\hat{\mathcal{H}}_{int}}
\end{split}
\end{align}
where the $g_{mnl}$ stands for the coupling strength between mode $l$ of the resonator and the transition dipole $\hat{P}_{mn}$ and is obtained as
\begin{align}
\hbar g_{mnl} \equiv \gamma \left(\frac{\hbar \omega_l}{2cL}\right)^{\frac{1}{2}}(iQ_{J,mn})\tilde{\Phi}_l(x_0) 
\end{align}

Various TRK sum rules \cite{wang_generalization_1999} can be developed for a transmon qubit, as discussed in detail in Appendix D. For instance, the sum of transition matrix elements of $\hat{Q}_J$ between the ground state and all the excited states obey 
\begin{align}
\begin{split}
\sum\limits_{n>0} 2\left(E_n-E_0 \right)\left|\bra{0}\hat{Q}_J\ket{n}\right|^2&=(2e)^2E_J \bra{0}\cos{\left(\frac{2\pi}{\phi_0}\hat{\Phi}_J\right)}\ket{0}\\
&<(2e)^2E_J
\end{split}
\label{Eq:Sum Rule for Q_J _body}
\end{align}  
Since all terms on the R.H.S are positive, this imposes an upper bound to the strength of $Q_{J,0n}$. A multi-mode Rabi Hamiltonian can be recovered by truncating the transition matrix elements to only one relevant quasi-resonant transition term (assumed here to be the $0 \rightarrow 1$ transition):
\begin{align}
\begin{split}
\hat{\mathcal{H}}&=\frac{1}{2} \hbar \omega_{01} \hat{\sigma}^z + \sum\limits_n \hbar \omega_n \hat{a}_n^{\dagger} \hat{a}_n \\
&+ \sum\limits_n \hbar g_n (\hat{\sigma}^- + \hat{\sigma}^+)(\hat{a}_n+\hat{a}^{\dagger}_n)
\end{split}
\end{align}
The coupling strength $g_n$ is now reduced to
\begin{align}
\hbar g_n \equiv \gamma \left(\frac{\hbar \omega_n}{2cL}\right)^{\frac{1}{2}}(iQ_{J,01})\tilde{\Phi}_n(x_0) 
\label{Eq: coupling strength gn-Body}
\end{align}
and based on Eq. \ref{Eq:Sum Rule for Q_J _body}, $Q_{J,01}$ has to satisfy
\begin{align}
|Q_{J,01}|^2<\frac{2e^2 E_J}{E_1-E_0}\approx \frac{2E_J}{\sqrt{8E_JE_C}-E_C}e^2 
\end{align}
where we have defined the charging energy $E_C\equiv\frac{e^2}{2C_J}$.   
\begin{figure}
\centering
\subfloat[\label{subfig:couplingsx001}]{%
\centerline{\includegraphics[scale=0.24]{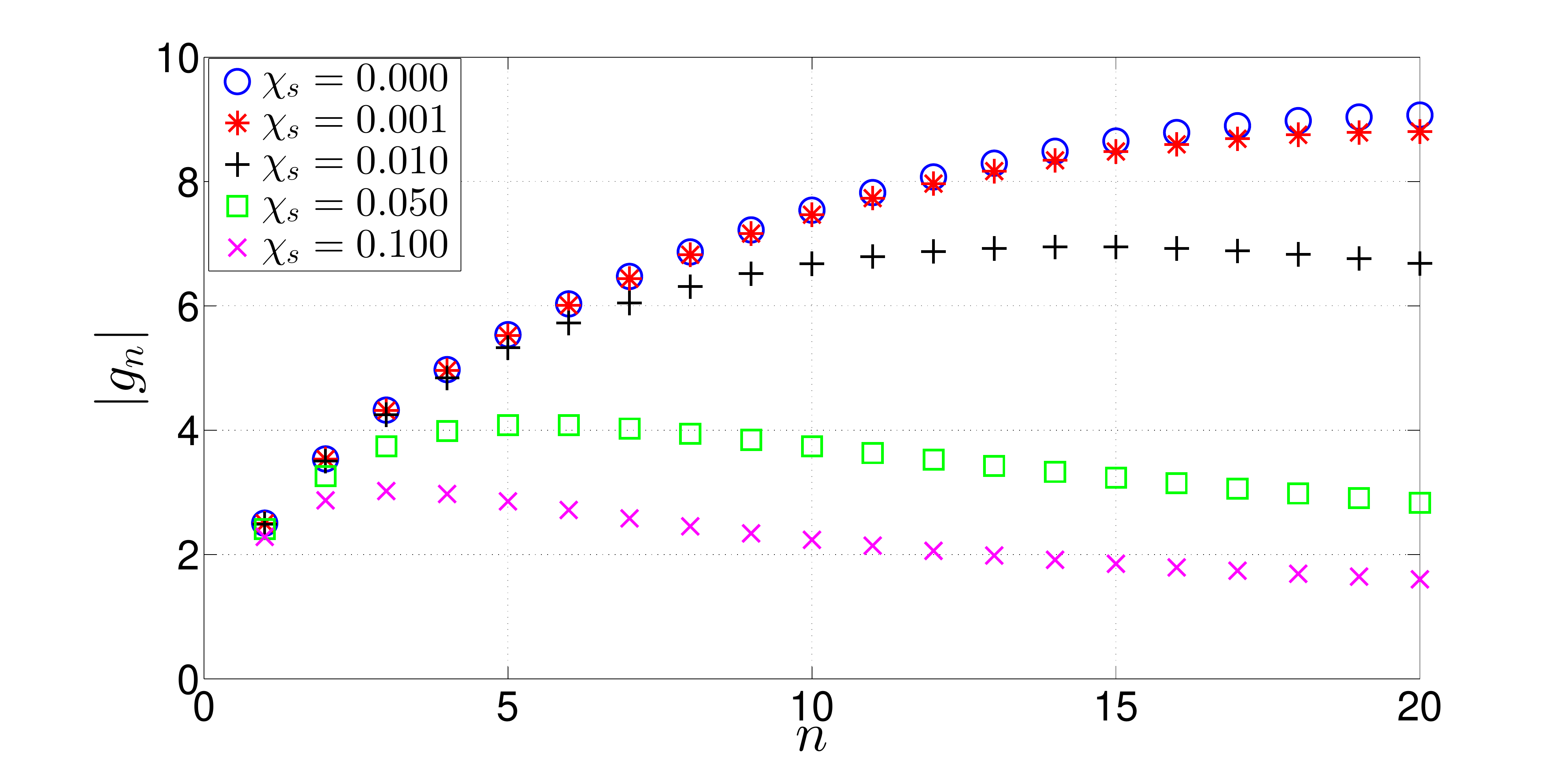}%
}}\hfill
\subfloat[\label{subfig:couplingsx001Largescale}]{%
\centerline{\includegraphics[scale=0.24]{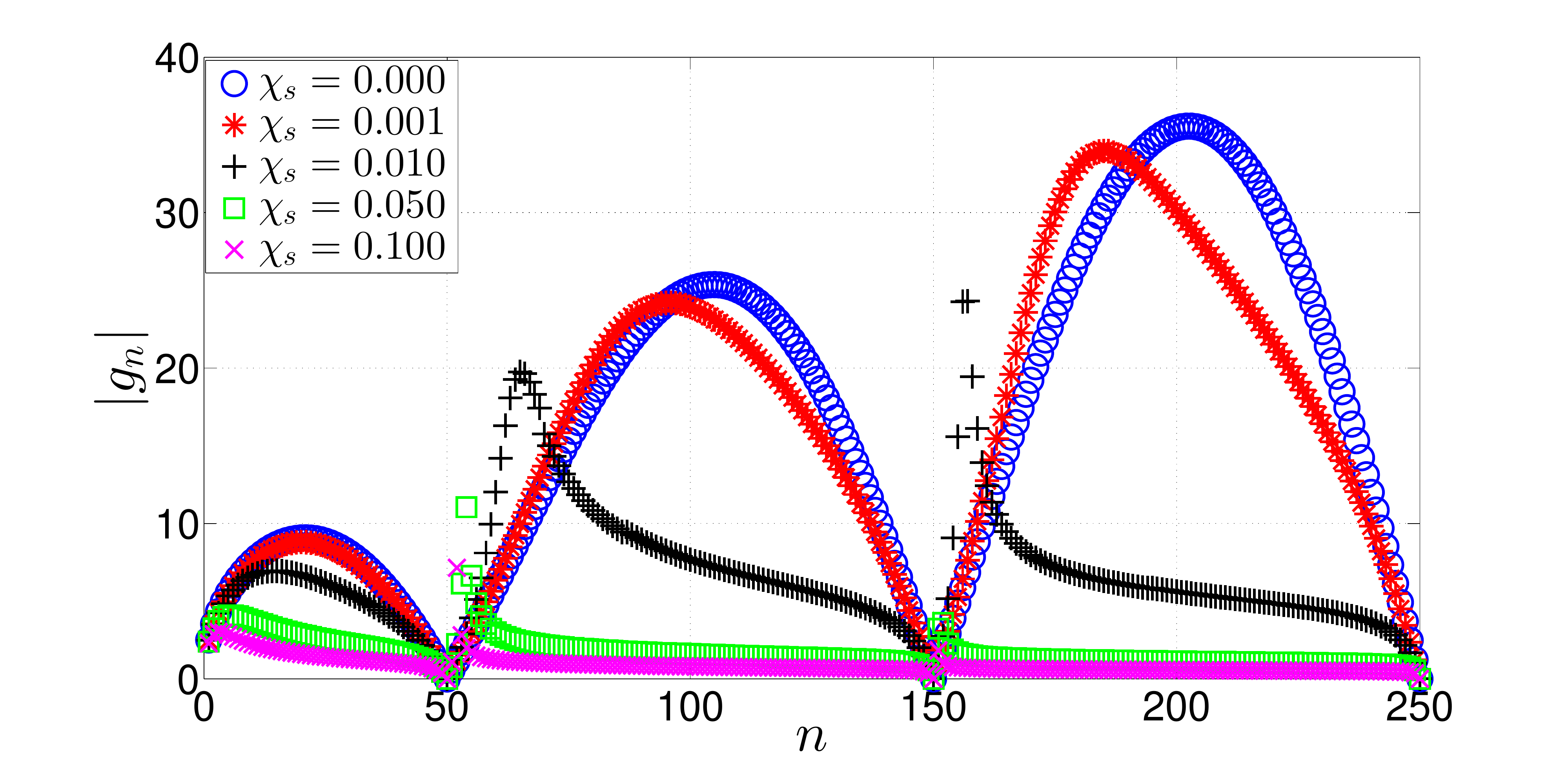}%
}}\hfill
\caption{Normalized coupling strength $g_n$ for $x_0=0.01L$ a) First 20 modes b) Large scale behavior for 250 modes. In both graphs, coupling strength is normalized such that only the normalized photonic dependence is kept i.e. $g_n=(k_n L)^{\frac{1}{2}}\tilde{\Phi}_n(x_0)$}
\label{fig:couplingsx001}
\end{figure}

In order to understand how much $g_n$ can deviate in practice from its former widely used expression in terms of the unmodified ($\chi_s = 0$) modes, in Fig. \ref{fig:couplingsx001}, we have compared the results for  various values of $\chi_s$. We note that in recent experiments on an ultra-long ($\sim70$ cm) transmission line cavity \cite{sundaresan_beyond_2015}, $\chi_s$ was found to be  around $10^{-3}$. For shorter, more standard transmission line resonators we should expect $\chi_s \sim 0.1$ because $\chi_s \propto \frac{1}{L}$. 

As we observe in these figures, CC couplings $g_n$ are very sensitive to a change in $\chi_s$. For instance, in Fig. \ref{subfig:couplingsx001} which is for the common case of connecting the qubit to one end, $x_0=0.01 L$, it is observed that even for $\chi_s=0.001$ (red stars) the relative shift in the highest mode shown (mode 20) is about $3\%$. This relative change increases to $26\%$, $69\%$ and $80\%$ for $\chi_s=0.01$, $\chi_s=0.05$ and $\chi_s=0.1 $ respectively. 

These modifications, even for small $\chi_s$, are clearly observable in the multi-mode regime, i.e. when the qubit is resonant with a very high order mode. To study the large scale behavior of couplings, we have plotted the first 250 CC couplings $g_n$ in Fig. \ref{subfig:couplingsx001Largescale} for the same parameters. As we mentioned earlier, due to the fact that the qubit is placed at a symmetry point, we expect that with a period of 100 modes, the couplings fall down to zero. The first mode that has a local minimum at $x_0=0.01 L$ is mode 50 and it occurs again at modes 150, 250 and so on.  As a general rule, higher CC modes experience a bigger shift in their coupling strength.   Another important observation is the suppression of coupling strength as $\chi_s$ increases such that the highest coupling strengths occur at the beginning of each cycle (black, blue and purple) rather than in the middle (red and blue). 

It remains to be seen whether the non-zero dispersion in frequencies and the modifications of coupling strengths to higher order modes is observable in practice, because in considering such large frequency intervals, the frequency dependent response function of superconductors would have to be taken into account \cite{catelani_effect_2010, reagor_reaching_2013}. 

The dependence of $g_n$ on $x_0$ has been studied in Appendix C for two other cases $x_0=\frac{L}{2}$ and $x_0=\frac{L}{4}$ along with their corresponding CC eigenfrequencies and eigenmodes.

\section{GENERALIZATION TO  AN OPEN-CAVITY: OPEN-BOUNDARY CC Basis}

We now discuss the quantization in an open geometry, where the resonator is coupled to two long microwave transmission lines, of length $L_L$ and $L_R$, at each side through nonzero capacitors $C_L$ and $C_R$ (Fig.~\ref{Circuit-QED-Closed}). In Appendix E, we discuss how these nonzero capacitances alter the boundary conditions at each end, and hence the mode structure as a result as well. The resulting real eigenfrequencies of the resonator can be found from the transcendental equation
\begin{align}
\begin{split}
&+\left(1-\chi_R\chi_L (k_nL)^2\right)\sin{(k_n L)}\\
&+\left(\chi_R+\chi_L\right)k_nL\cos{(k_n L)}\\
&+\chi_s k_n L\cos{(k_n x_0)}\cos{(k_n (L-x_0))}\\
&-\chi_R\chi_s (k_n L)^2 \cos{(k_n x_0)}\sin{(k_n (L-x_0))}\\
&-\chi_L\chi_s (k_n L)^2 \sin{(k_n x_0)}\cos{(k_n (L-x_0))}\\
&+\chi_R\chi_L\chi_s (k_n L)^3 \sin{(k_n x_0)}\sin{(k_n (L-x_0))}=0
\end{split}
\end{align}
where $\chi_{R,L}\equiv\frac{C_{R,L}}{cL}$ are normalized coupling constants to the left and right transmission line. Considering only the first two terms in the expression above, by setting $\chi_s=0 $, would lead to the well-known equation in the literature \cite{koch_time-reversal-symmetry_2010, schmidt_circuit_2013, nunnenkamp_synthetic_2011}
\begin{align}
\tan{(k_n L)}=\frac{(\chi_R+\chi_L)k_n L}{\chi_R\chi_L (k_n L)^2-1}
\end{align}
which only describes eigenfrequencies of an isolated resonator and does not contain appropriate current conservation at the qubit location. The third term is the same modification we have found in the closed case and has a significant influence as $\chi_s$ increases, while the others represent higher order corrections and are almost negligible except for very high order modes. The real-space representation of these eigenmodes are found as
\begin{align}
\tilde{\Phi}_n(x)\propto
\begin{cases}
\tilde{\Phi}_n^{<}(x) \quad 0<x<x_0\\
\tilde{\Phi}_n^{>}(x) \quad x_0<x<L
\end{cases}
\end{align}
where $\tilde{\Phi}_n^{<}(x)$ and $\tilde{\Phi}_n^{>}(x)$ are given by
\begin{align}
\begin{split}
\tilde{\Phi}_n^{<}(x)&=\left[\cos{(k_n(L-x_0))}-\chi_Rk_nL\sin{(k_n(L-x_0))}\right]\\
&\times\left[\cos{(k_n x)}-\chi_L k_nL\sin{(k_n x)}\right]
\end{split}\\
\begin{split}
\tilde{\Phi}_n^{>}(x)&=\left[\cos{(k_n x_0)}-\chi_L k_nL\sin{(k_n x_0)}\right]\\
&\times\left[\cos{(k_n(L-x))}-\chi_Rk_nL\sin{(k_n(L-x))}\right]
\end{split}
\end{align}
The {\it open-boundary CC basis} can be shown to satisfy the modified orthogonality relations 
\begin{align}
&\int_{0}^{L} dx \frac{c_{op}(x,x_0)}{c}\tilde{\Phi}_m(x)\tilde{\Phi}_n(x)=L\delta_{mn}
\end{align}
where the capacitance per unit length $c_{op}(x,x_0)$, due to the leaky boundary is given by 
\begin{align}
\begin{split}
c_{op}(x,x_0)&=c+C_s \delta(x-x_0)\\
&+ C_R\delta(x-L^-)+C_L \delta(x-0^+)
\end{split}
\end{align}
The remaining orthogonality relations for the current is also modified, 
\begin{align}
\begin{split}
&+\int _{0}^{L}dx \frac{\partial\tilde{\Phi}_m(x)}{\partial x}\frac{\partial\tilde{\Phi}_n(x)}{\partial x}\\
&-\frac{1}{2}\left(k_m^2+k_n^2\right)L\left[\chi_R\tilde{\Phi}_m(L^-)\tilde{\Phi}_n(L^-)+\chi_L\tilde{\Phi}_m(0^+)\tilde{\Phi}_n(0^+)\right]\\
&=k_m k_n L \delta_{mn}
\end{split}
\end{align}
The same argument holds for the CC modes of the left and right transmission lines, while the exact knowledge of these modes requires assigning appropriate boundary conditions at their outer boundaries. For instance, if the side resonators are assumed to be very long, an outgoing boundary condition is a very good approximation, since the time scale by which the escaped signal bounces back and reaches the original resonator is much larger than the round-trip time of the central resonator. On the other hand, if we have a lattice \cite{koch_time-reversal-symmetry_2010, nunnenkamp_synthetic_2011, schmidt_circuit_2013} of identical resonators each connected to a qubit and capacitively coupled to each other, then the same basis can be used for each of them. Assuming we also have the solution for the CC basis of right and left resonators as $\{\omega_{n,S},\tilde{\Phi}_{n,S}| n \in \mathbb{N}, S=\{R,L\}\}$, the quantum flux fields in each side resonator can be expanded in terms of these CC modes as
\begin{align}
\begin{split}
&\hat{\Phi}_S(x,t)=\sum \limits_{n,S} \left(\frac{\hbar}{2\omega_{n,S} cL_S}\right)^{\frac{1}{2}} \left(\hat{b}_{n,S}(t) + \hat{b}_{n,S}^{\dagger}(t) \right)\tilde{\Phi}_{n,S}(x)
\end{split}
\end{align}
where $\hat{b}_{n,S=\{R,L\}}$ are the annihilation and creation operators for the $n^{th}$ open CC mode in each side resonator. Following the quantization procedure discussed in Appendix E, we find the Hamiltonian in its 2nd quantized representation as
\begin{align}
\begin{split}
\hat{\mathcal{H}}&=\underbrace{\sum\limits_n \hbar\Omega_n \hat{P}_{nn}}_{\hat{\mathcal{H}}_A}+\underbrace{\sum\limits_n \hbar\omega_n \hat{a}_n^{\dagger} \hat{a}_n}_{\hat{\mathcal{H}}_C^{mod}} +\underbrace{\sum\limits_{n,S=\{L,R\}} \hbar\omega_{n,S} \hat{b}_{n,S}^{\dagger}\hat{b}_{n,S}}_{\hat{\mathcal{H}}_B}\\
&+\underbrace{\sum\limits_{m<n,l} \hbar g_{mnl} \left(\hat{P}_{mn}+\hat{P}_{nm}\right)\left(\hat{a}_l+\hat{a}^{\dagger}_l\right)}_{\hat{\mathcal{H}}_{int}}\\
&+\underbrace{\sum\limits_{m,n,S=\{L,R\}} \hbar\beta_{mn,S}\left(\hat{a}_m+\hat{a}_m^{\dagger}\right)\left(\hat{b}_{n,S}+ \hat{b}_{n,S}^{\dagger}\right)}_{\hat{\mathcal{H}}_{CB}}
\end{split}
\label{Eq:2nd Quantized Hamiltonian_Open Case_Body}
\end{align}

In this expression, $\beta_{mn,S}$ stand for coupling strength of $m^{th}$ open CC mode of the resonator to $n^{th}$ open CC mode of the side baths and is found as
\begin{align}
&\beta_{mn,S}=\frac{C_{S}}{2c\sqrt{L}\sqrt{L_{S}}}\omega_m^{\frac{1}{2}}\omega_{n,S}^{\frac{1}{2}}\tilde{\Phi}_m(L^-)\tilde{\Phi}_{n,S}(L^+)
\label{Eq:beta_{L,R}}
\end{align}
where $C_S$ here stands for side capacitors $C_{R,L}$ and shouldn't be confused with the series capacitance $C_s$ introduced earlier. Notice that light-matter coupling strength $g_{mnl}$ has the same form as before, but in terms of open CC eigenmodes and eigenfrequencies.

\section{Discussion}
\begin{figure}
\centering
\centerline{\includegraphics[scale=0.40]{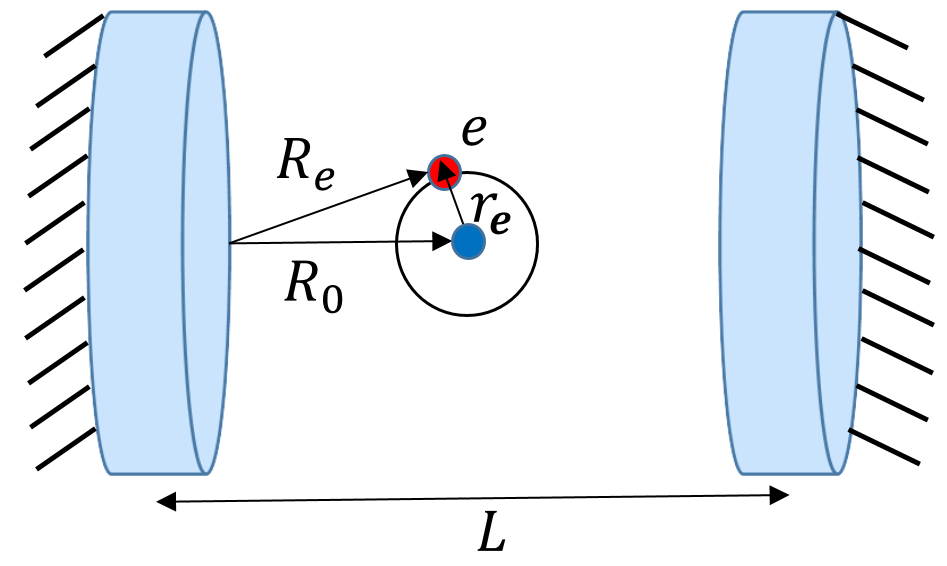}}
\caption{A single-electron atom interacting with the EM field inside a closed cavity of length $L$. } 
\label{fig:Cavity-QED-Closed}
\end{figure}

The corrections to the spectral structure of the resonator found in sections $\RN{2}$ and $\RN{3}$ are mathematically equivalent to the scattering corrections that result from the presence of an atom in atomic CQED systems. Electromagnetic field quantization has been studied in great detail for CQED systems including single-electron atoms~\cite{scully_quantum_1997, walls_quantum_2008}, multi-electron atoms \cite{knoll_action_1987}, and for atoms embedded in dispersive and absorptive dielectric media~\cite{knoll_action_1987, glauber_quantum_1991, huttner_quantization_1992, matloob_electromagnetic_1995, dung_three-dimensional_1998, knoll_qed_2000}. For completeness, in Appendix G we present a full derivation of the minimal coupling Hamiltonian(neglecting electron's spin) for this system starting from a Lagrangian formalism that yields the Maxwell's equations and the Lorentz force law \cite{knoll_action_1987}. The term $\mathcal{L}_{int}=\mathcal{H}_{int}=\frac{1}{2}C_g\left(\dot{\Phi}_J-\dot{\Phi}(x_0,t)\right)^2$ that appears in the canonical quantization of cQED systems is mathematically equivalent to the approximate(zero-order dipole approximated) minimal coupling term $\mathcal{T}_e \approx \frac{1}{2m_e}({\vec{p}_e-e\vec{A(\vec{R}_0,t)}})^2$ appearing in CQED Hamiltonian, thus their impact on the cavity modal structure is similar. It could be argued that the freedom in the choice of the point of reference for the generalized fluxes i.e. the choice of ground, is analogous to the gauge freedom. However, the fact that the cavity modes are modified due to  the existence of the qubit is a property that is gauge-independent. In Appendix G, we show that in a similar manner to the discussion here, the existence of the $A^2$ term in the Coulomb gauge gives rise to modified spectral properties of the resonator. However, in atomic CQED these corrections are tiny because of the smallness of typical atomic transition dipoles and the fine structure constant. In Appendix B.3, we have also studied the reverse question and proved it is feasible to retrieve an $A^2$-like term if one naively performs the quantization by the cosine basis. This completes the similarity between cQED and CQED Hamiltonians in the lowest order(zeroth order dipole approximation) where the dimension of transmon (atom) is completely neglected compared to the cavity's wavelength.

\section{ACKNOWLEDGEMENTS}
We acknowledge helpful discussions with Jonathan Keeling, Steven Girvin, Howard A. Stone, Gianluigi Catelani, and Rob Schoelkopf. This work was supported by the NSF grant DMR-1151810.
\appendix

\section{cQED NOTATION}
In order to describe the dynamics of any cQED system we follow a common quantization procedure \cite{bishop_circuit_2010, devoret_quantum_2014}. The first step is to write the Lagrangian in terms of a generalized coordinate $\mathcal{L}[q_n]$. Then, a Legendre transformation to find the Hamiltonian in terms of the coordinate and its conjugate momentum $p_n\equiv \frac{\partial \mathcal{L}}{\partial \dot{q}_n}$ as $\mathcal{H}[q_n,p_n]=\sum\limits_n \dot{q}_n p_n -\mathcal{L}$ . Finally, we need to apply the canonical quantization by imposing a nonzero commutation relation between the conjugate pairs as  $[q_n,p_n]=i\hbar $. Here we go after the convention used in cQED by choosing the generalized coordinate as $\Phi_n(t)=\int_0^t V_n(t')\,dt'\ $ in which $V_n(t)$ is the voltage at node $n$ and is measured with respect to a ground node. This quantity has the units of magnetic flux and it can be shown that its conjugate variable has the units of charge and we denote it by $Q_n(t)$. 
There is an additional rule one has to keep in mind. In the case of external magnetic flux applied on a certain loop, the algebraic sum of flux variables over that loop should be equal to the external flux. Taking into account all these considerations, the Lagrangian for any cQED system is found as
\begin{align}
\mathcal{L}[\Phi_n ,\dot{\Phi}_n]=\mathcal{T}[\dot{\Phi}_n ]-\mathcal{U}[\Phi_n]
\end{align}
where $\mathcal{T}$ represents the kinetic energy corresponding to capacitors as $ \mathcal{T}_C[\dot{\Phi}]=\frac{1}{2}C\dot{\Phi}_C^2 $ and $\mathcal{U}$ stands for the potential energy corresponding to inductors as $ \mathcal{U}_L\{\Phi\}=\frac{1}{2L}\Phi_L^2 $ or any other nonlinear magnetic device such as Josephson junction $ \mathcal{U}_{J}[\Phi_J]=-E_J\cos\left(2\pi\frac{\Phi_J}{\phi_0}\right)$ where $\phi_0=\frac{h}{2e}$ is the flux quantum.

\section{CLASSICAL HAMILTONIAN AND MODIFIED EIGENMODES OF A CLOSED cQED SYSTEM}
Here, we follow the procedure discussed in Appendix A for the system shown in Fig. \ref{subfig:circuit-QED-Closed-EquivalentCircuit}. We first use a discretized lumped element LC-model\cite{yurke_quantum_1984} for the microwave resonator and then take the limit where these infinitessimal elements go to zero while leaving the capacitance and inductance per length of the resonator invariant. 
\subsection{Discrete Limit}
\subsubsection{Classical Lagrangian for the Discretized Circuit}
In terms of the generalized coordinates introduced in Appendix A, the Lagrangian for the discretized circuit can be written as the difference between kinetic capacitive energy and potential inductive energy and it reads
\begin{align}
\begin{split}
\mathcal{L}&=\frac{1}{2}C_J\dot{\Phi}_J^2+E_J\cos\left(2\pi\frac{\Phi_J}{\phi_0}\right)\\
&+\sum\limits_{n}\left[\frac{1}{2}c\Delta x\dot{\Phi}_n^2-\frac{1}{2l\Delta x}\left(\Phi_{n+1}-\Phi_n\right)^2\right]\\
&+\frac{1}{2}C_g(\dot{\Phi}_0-\dot{\Phi}_J)^2	
\end{split}
\end{align}

In the expression above, we have labeled the discrete nodes such that the qubit is connected to the zeroth node.
\subsubsection{Classical Hamiltonian for the Discretized Circuit}

The first step to find the Hamiltonian is to derive the conjugate variables associated with the generalized coordinate $\{\{\Phi_{n}\};\Phi_J\}$. These conjugate variables will have the dimension of charge and we represent them as $\{\{Q_{n}\};Q_J\}$. By definition, these conjugate variables read
\begin{align}
Q_J\equiv\frac{\delta \mathcal{L}}{\delta \dot{\Phi}_J}&=(C_J+C_g)\dot{\Phi}_J-\sum\limits_{n} C_g\delta_{n0}\dot{\Phi}_n \\
Q_n\equiv\frac{\delta \mathcal{L}}{\delta \dot{\Phi}_n}&=(c\Delta x+C_g\delta_{n0})\dot{\Phi}_n-C_g\delta_{n0}\dot{\Phi}_J
\end{align}

The next step is to calculate the discrete Hamiltonian by a Legendre transformation 
\begin{align}
\begin{split}
\mathcal{H}&=\sum\limits_n Q_n\dot{\Phi}_n+Q_J\dot{\Phi}_J-\mathcal{L}\\
&=\frac{1}{2}(C_J+C_g)\dot{\Phi}_J^2-E_J\cos\left(2\pi\frac{\Phi_J}{\phi_0}\right)\\
&+\sum\limits_{n}\left[\frac{1}{2}(c\Delta x+C_g\delta_{n0})\dot{\Phi}_n^2+\frac{1}{2l\Delta x}\left(\Phi_{n+1}-\Phi_n\right)^2\right]\\
&-C_g\dot{\Phi}_0\dot{\Phi}_J
\end{split}
\end{align}

Now, we need to solve for $\dot{\Phi}_J$ and $\dot{\Phi}_n$ in terms of $Q_J$ and $Q_n$ to represent the Hamiltonian only in terms of generalized coordinates and their conjugate variables.

Before proceeding further, let's define a few quantities in order to simplify the calculation
\begin{align}
&\gamma\equiv\frac{C_g}{C_g+C_J}
\label{gamma}\\
&C_s\equiv\frac{C_gC_J}{C_g+C_J}
\label{Cs}\\
&C_{s,n}\equiv c\Delta x+C_s\delta_{n0}\\
&C_{g,n}\equiv c\Delta x+C_g\delta_{n0}
\end{align}
where $C_{s}$ represents the series combination of the coupling capacitor $C_g$ and Transmon's capacitor $C_J$. In terms of these new quantities we can write
\begin{align}
&\dot{\Phi}_n=\frac{Q_n}{C_{s,n}}+\frac{\gamma\delta_{n0}Q_J}{C_{s,n}}\\
&\dot{\Phi}_J=\left(\frac{\gamma}{C_J}+\frac{\gamma^2}{C_{s,0}}\right)Q_J+\sum\limits_n \frac{\gamma\delta_{n0}}{C_{s,n}}Q_n\\
\begin{split}
&\mathcal{H}=\frac{1}{2}\frac{C_g}{\gamma}\dot{\Phi}_J^2 -E_J\cos\left(2\pi\frac{\Phi_J}{\Phi_0}\right)\\
&+\sum\limits_{n}\left[\frac{1}{2}C_{g,n}\dot{\Phi}_n^2+\frac{1}{2l\Delta x}\left(\Phi_{n+1}-\Phi_n\right)^2\right]\\
&-C_g\dot{\Phi}_0\dot{\Phi}_J
\end{split}
\end{align}

By inserting the expressions for $\dot{\Phi}_n$ and $\dot{\Phi}_J$ into the one for the Hamiltonian one finds that
\begin{align}
\begin{split}
\mathcal{H}&=\frac{1}{2}\left[\frac{\gamma}{C_g}+\frac{\gamma^2 (C_{g,0}-\gamma C_g)}{C_{s,0}^2}\right]Q_J^2-E_J\cos\left(2\pi\frac{\Phi_J}{\phi_0}\right)\\
&+\sum\limits_{n}\left[\frac{1}{2}\frac{C_{g,n}-\gamma C_g\delta_{n0}}{C_{s,n}^2}Q_n^2+\frac{1}{2l\Delta x}\left(\Phi_{n+1}-\Phi_n\right)^2\right]\\
&+\frac{\gamma}{C_{s,0}}Q_JQ_0
\end{split}
\end{align}

Notice that this expression can be further simplified since $C_{g,n}$ and $C_{s,n}$ are related as $C_{g,n} -\gamma C_g \delta_{n0}=C_{s,n} $. Therefore, the final result for the discretized Hamiltonian will be
\begin{align}
\begin{split}
\mathcal{H}&=\frac{1}{2}\left(\frac{\gamma}{C_g}+\frac{\gamma^2}{C_{s,0}}\right)Q_J^2-E_J\cos\left(2\pi\frac{\Phi_J}{\phi_0}\right)\\
&+\sum\limits_{n}\left[\frac{Q_n^2}{2C_{s,n}}+\frac{1}{2l\Delta x}\left(\Phi_{n+1}-\Phi_n\right)^2\right]\\
&+\frac{\gamma}{C_{s,0}}Q_JQ_0 
\end{split}
\end{align}
where the conjugate variables obey the classical Poisson-bracket relations
\begin{align}
\{\Phi_n,Q_m\}&=\delta_{mn} \\
\{\Phi_J,Q_J\}&=1 \\
\{Q_n,Q_m\}&=\{\Phi_n,\Phi_m\}=0 \\
\{Q_J,Q_J\}&=\{\Phi_J,\Phi_J\}=0
\end{align}
\subsection{Continuum Limit} 

Now that we have the expressions for Lagrangian and Hamiltonian in the discrete limit, we can obtain the analogous continuous ones by simply taking the limit $\Delta x \to 0$, while keeping the capacitance and inductance per length constant. In order to do so, let's find first how some of the terms change in this limit. Let's first investigate the Kronecker delta. It is quite natural to argue that
\begin{align}
\lim\limits_{\Delta x \to 0} \frac{\delta_{n0}}{\Delta x}&=\delta(x)
\end{align}
$\delta(x)$ here represents the Dirac delta function. One can simply verify that it has all the properties of a Dirac delta function
\begin{enumerate}
\item  $\delta(x)=0 , \qquad\  x\neq0$
\item  $\delta(x) \to +\infty , \quad x \to 0$
\item  $\lim\limits_{\Delta x \to 0}\sum\limits_n \frac{\delta_{n0}}{\Delta x}\Delta x=\int_{-\epsilon}^{+\epsilon} \delta(x)\,dx\ =1$ \\
\end{enumerate}

Based on this result, it is possible to find how $C_{s,n}$ transform in the continuous case as 
\begin{align}
c(x)\equiv\lim_{\Delta x \to 0}\frac{C_{s,n}}{\Delta x}=c+C_s\delta(x) 
\label{Eq:modified capacitance per length}
\end{align}

We call this quantity \textit{modified capacitance per length} of the resonator, since it has the information regarding the position of the qubit and the way it changes the capacitance at the point of connection. By going to continuum limit, the charge variable $Q_n$ goes to zero, since it represents the charge of infinitesimal capacitors. However, the charge density remains a finite quantity
\begin{align}
\rho(x,t)\equiv \lim_{\Delta x \to 0} \frac{Q_n(t)}{\Delta x}
\end{align}
Finally, by definition
\begin{align}
\lim_{x \to 0}\frac{\Phi_{n+1}(t)-\Phi_n(t)}{\Delta x}&=\frac{\partial \Phi(x,t)}{\partial x}
\end{align}
\subsubsection{Classical Lagrangian and Euler-Lagrange E.O.M. in the Continuum Limit}
Applying the limits introduced in the previous section, Lagrangian in the continuum limit reads
\begin{align}
\begin{split}
\mathcal{L}&=\frac{1}{2}(C_J+C_g)\dot{\Phi}_J^2-U_J(\Phi_J)\\
&+\int_{-L/2}^{L/2}\,dx\ \left[\frac{1}{2}(c+C_g\delta(x))(\frac{\partial \Phi}{\partial t})^2-\frac{1}{2l}(\frac{\partial \Phi}{\partial x})^2\right]\\
&-\int_{-L/2}^{L/2}\,dx\ C_g\delta(x)\dot{\Phi}_J\frac{\partial \Phi}{\partial t} 
\end{split}
\end{align}
where $U_J(\Phi_J)=-E_J\cos\left(2\pi\frac{\Phi_J}{\phi_0}\right)$. Euler-Lagrange equations of motion are derived from the variational principle $\delta \mathcal{L}=0$ as

\begin{align}
&(C_J+C_g)\ddot{\Phi}_J-C_g\int_{-L/2}^{L/2} dx \delta(x)\frac{\partial^2 \Phi}{\partial t^2}+\frac{\partial U_J(\Phi_J)}{\partial\Phi_J}=0
\label{Eq: unuseful E.O.M for Phi_J ddot} \\
&-\frac{\partial^2 \Phi}{\partial x^2}+lc\frac{\partial^2 \Phi}{\partial t^2}+lC_g\delta(x)\left(\frac{\partial^2 \Phi}{\partial t^2}-\ddot{\Phi}_J\right)=0
\label{Eq: unuseful E.O.M for Phi_x ddot}
\end{align}

It is helpful to rewrite these equations by first finding $\ddot\Phi_J $ from \ref{Eq: unuseful E.O.M for Phi_J ddot} and plugging into \ref{Eq: unuseful E.O.M for Phi_x ddot}
\begin{align}
\frac{\partial^2 \Phi}{\partial x^2}-lc(x)\frac{\partial^2 \Phi}{\partial t^2}&=l\gamma\delta(x)\frac{\partial U_J(\Phi_J)}{\partial\Phi_J}
\end{align}
which is a wave equation with modified capacitance per length and the transmon qubit as a source on the right hand side. Therefore, the simplified equations of motion read
\begin{align}
&\ddot{\Phi}_J+\frac{\gamma}{C_g}\frac{\partial U_J(\Phi_J)}{\partial\Phi_J}=\gamma\frac{\partial^2 \Phi(0,t)}{\partial t^2}
\label{Eq:Transmon's E.O.M} \\
&\frac{\partial^2 \Phi}{\partial x^2}-lc(x)\frac{\partial^2 \Phi}{\partial t^2}=l\gamma\delta(x)\frac{\partial U_J(\Phi_J)}{\partial\Phi_J} 
\label{Eq:Resonator's E.O.M}
\end{align}

The Dirac delta function in the wave equation \ref{Eq:Resonator's E.O.M} can be translated into discontinuity in the spatial derivative of $\Phi(x,t)$. Therefore, equation \ref{Eq:Resonator's E.O.M} can be understood as
\begin{align}
&\frac{\partial^2 \Phi}{\partial x^2}-lc\frac{\partial^2 \Phi}{\partial t^2}=0 , x \neq 0 \\
&\Phi(0^+,t)=\Phi(0^-,t)
\label{Eq:Continuity of Voltage} \\
&\left.\frac{\partial \Phi}{\partial x}\right|_{x=0^+}-\left.\frac{\partial \Phi}{\partial x}\right|_{x=0^-}=lC_s\left.\frac{\partial^2 \Phi}{\partial t^2}\right|_{x=0}+l\gamma \frac{\partial U_J(\Phi_J)}{\partial\Phi_J}
\label{Eq:Discontinuity of current}
\end{align}

 In terms of voltage and current,  equation \ref{Eq:Continuity of Voltage} means that voltage is spatially continuous while \ref{Eq:Discontinuity of current} means that current is not continuous at the position of the transmon, since some of the current has to go into the qubit. The two terms on the right hand side of \ref{Eq:Discontinuity of current} are proportional to the current that enters $C_J$ and the Josephson junction respectively.

\subsubsection{Classical Hamiltonian and Heisenberg E.O.M. in the Continuum Limit}
Starting from our discrete Hamiltonian, we try to find the continuous one again by taking the limit $\Delta x\longrightarrow 0 $. Let's investigate each term seperately.
\begin{align}
\lim_{\Delta x \to 0}\left(\frac{\gamma}{C_g}+\frac{\gamma^2}{C_{s,0}}\right)=\frac{1}{C_J}
\end{align} 

Therefore, the transmon's Hamiltonian will be
\begin{align}
\frac{Q_J^2}{2C_J}-E_J\cos\left(2\pi\frac{\Phi_J}{\phi_0}\right)
\end{align}

The resonator's Hamiltonian transforms as
\begin{align}
\begin{split}
&\lim_{\Delta x\to0} \sum\limits_{n}\left[\frac{Q_n^2}{2c_n}+\frac{1}{2l\Delta x}\left(\Phi_{n+1}-\Phi_n\right)^2\right]\\
&=\lim\limits_{\Delta x\to0} \sum\limits_{n}\Delta x\left[\frac{1}{2}\frac{(\frac{Q_n}{\Delta x})^2}{\frac{c_n}{\Delta x}}+\frac{1}{2l}\left(\frac{\Phi_{n+1}-\Phi_n}{\Delta x}\right)^2\right]\\
&=\int_{-L/2}^{L/2}dx \left[\frac{\rho^2(x,t)}{2c(x)}+\frac{1}{2l}\left(\frac{\partial \Phi(x,t)}{\partial x}\right)^2\right]
\end{split}
\end{align}
and finally the interaction term can be written as
\begin{align}
\begin{split}
\lim\limits_{\Delta x\to 0}\frac{\gamma}{C_{s,0}}Q_JQ_0=&\lim\limits_{\Delta x\to 0}\gamma Q_J \sum\limits_{n} \frac{Q_n}{C_{s,n}}\delta_{n0}\\
=&\lim\limits_{\Delta x\to 0}\gamma Q_J \sum\limits_{n} \frac{\frac{Q_n}{\Delta x}}{\frac{C_{s,n}}{\Delta x}} \frac{\delta_{n0}}{\Delta x}\Delta x\\
=&\gamma Q_J \int_{-L/2}^{L/2}dx \frac{\rho(x,t)}{c(x)} \delta(x)
\end{split}
\end{align}

Putting all the terms together, the final expression for the Hamiltonian will be
\begin{align}
\begin{split}
\mathcal{H}&=\underbrace{\frac{Q_J^2}{2C_J}-E_J\cos\left(2\pi\frac{\Phi_J}{\phi_0}\right)}_{\mathcal{H}_{A}}\\
&+\underbrace{\int_{-L/2}^{L/2}dx\left[\frac{\rho^2(x,t)}{2c(x)}+\frac{1}{2l}\left(\frac{\partial \Phi(x,t)}{\partial x}\right)^2\right]}_{\mathcal{H}^{mod}_C} \\
&+\underbrace{\gamma Q_J \int_{-L/2}^{L/2}dx \frac{\rho(x,t)}{c(x)} \delta(x)}_{\mathcal{H}_{int}}
\end{split}
\label{Closed-Circuit-QED Hamiltonian}
\end{align}
where the Poisson bracket relations now change to
\begin{align}
&\{\Phi_J,Q_J\}=1 \\
&\{\Phi(x,t),\rho(x',t)\}=\delta(x-x') 
\end{align}

Notice that in our expression for Hamiltonian we have a Dirac delta function hidden in $ c(x) $ in the denominator of both resonator's capacitive energy and the interaction term. At the first sight, it might seem unconventional to have a Dirac delta function in the denominator. However, we will show that the charge density $\rho(x,t)$  is also proportional to $c(x)$ which makes these integrals have finite values.

we know that the time dependence of an operator $O\left(\{\Phi_n\},\{Q_n\};\Phi_J,Q_J;t\right)$ is determined by
\begin{align}
\frac{dO}{dt}=\{O,H\}+\frac{\partial O}{\partial t} 
\end{align}

Using the Poisson-bracket relations introduced above one can find the Hamiltonian E.O.M as follows
\begin{align}
&\frac{\partial \Phi(x,t)}{\partial t}=\frac{\rho(x,t)}{c(x)}+\frac{\gamma \delta(x)}{c(x)} Q_J \\
&\frac{\partial \rho(x,t)}{\partial t}=\frac{1}{l}\frac{\partial^2 \Phi(x,t)}{\partial x^2} \\
&\frac{\partial \Phi_J}{\partial t}=\frac{Q_J}{C_J}+\int_{-L/2}^{L/2} dx \frac{\gamma \delta(x)}{c(x)}\rho(x,t)
\label{Eq:Transmon's Flux variable time derivative}
\\
&\frac{\partial Q_J}{\partial t}=-\frac{\partial U_J(\Phi_J)}{\partial \Phi_J}=-\frac{2\pi}{\phi_0}E_J \sin\left(2\pi \frac{\Phi_J}{\phi_0}\right) 
\label{Eq:Transmon's Charge Variable time derivative}
\end{align}

The results here, can be generalized to a case where the transmon is connected to some arbitrary point $x_0$, where the modified capacitance per length now changes to $c(x,x_0)=c+C_s \delta(x-x_0)$. 
\subsection{Modified Resonator Eigenmodes and Eigenfrequencies}
Consider the second term $\mathcal{H}_{C}^{mod}$ in \ref{Closed-Circuit-QED Hamiltonian} which is the modified resonator Hamiltonian. The goal here is to find out how this modification in capacitance per length  influences the closed Hermitian eigenmodes and eigenfrequencies of the resonator. Assuming that the transmon is connected to some arbitrary point $x_0$ the Hamiltonian is given as
\begin{align}
\mathcal{H}_C^{mod}=\int_{0}^{L} dx \left[\frac{\rho^2(x,t)}{2c(x,x_0)}+\frac{1}{2l}\left(\frac{\partial \Phi(x,t)}{\partial x}\right)^2\right]
\end{align}

Applying the Poisson-braket relations discussed in the previous section, the Hamiltonian E.O.M for the conjugate fields read
\begin{align}
\frac{\partial \Phi(x,t)}{\partial t}&=\frac{\rho(x,t)}{c(x,x_0)} 
\label{Eq: E.O.M for Phi from H_c mod}\\
\frac{\partial \rho(x,t)}{\partial t}&=\frac{1}{l}\frac{\partial^2 \Phi(x,t)}{\partial x^2}
\label{Eq: E.O.M for rho from H_c mod}
\end{align} 

By combining the above equations and rewriting them in Fourier representation in terms of $\tilde{\Phi}(x,\omega)=\int_{-\infty}^{+\infty} dt \Phi(x,t)e^{i\omega t}$ we obtain
\begin{align}
\frac{\partial^2 \tilde{\Phi}(x,\omega)}{\partial x^2}+lc(x,x_0)\omega^2\tilde{\Phi}(x,\omega)&=0
\label{Eq:Modified Wave Equation-Closed Case}
\end{align}

Notice that there is a Dirac delta function hidden in $c(x,x_0)$. As we mentioned earlier, this can be translated into discontinuity in $\partial_x \tilde{\Phi}(x)$ which is proportional to the current $\tilde{I}(x)=-\frac{1}{l}\frac{\partial \tilde{\Phi}(x)}{\partial x}$ that enters and exits the point of connection to the transmon
\begin{align}
-\frac{1}{l}\left.\frac{\partial \tilde{\Phi}(x,\omega)}{\partial x}\right|_{x_0^+}+\frac{1}{l}\left.\frac{\partial \tilde{\Phi}(x,\omega)}{\partial x}\right|_{x_0^-} =C_s\omega^2 \tilde{\Phi}(x_0,\omega)
\end{align}

We are after a complete set of modes $\tilde{\Phi}_n(x)\equiv\tilde{\Phi}(x,\omega_n)$ where any solutions to the previous wave equation can be linearly decomposed on them. In order to find these modes, we have to solve
\begin{align}
\frac{\partial^2 \tilde{\Phi}_n(x)}{\partial x^2}+lc\omega_n^2\tilde{\Phi}_n(x)=0 ,\qquad x\neq x_0 
\end{align}
with boundary conditions
\begin{align}
&\left.\frac{\partial \tilde{\Phi}_n(x)}{\partial x}\right|_{x=0}=\left.\frac{\partial \tilde{\Phi}_n(x)}{\partial x}\right|_{x=L}=0 \\
&\left.\frac{\partial \tilde{\Phi}_n(x)}{\partial x}\right|_{x_0^+}-\left.\frac{\partial \tilde{\Phi}_n(x)}{\partial x}\right|_{x_0^-} + lC_s\omega_n^2 \tilde{\Phi}_n(x_0)=0 \\
&\tilde{\Phi}_n(x_0^+)=\tilde{\Phi}_n(x_0^-) 
\end{align}

Applying the boundary conditions we find a transcendental equation whose roots will give the Hermitian eigenfrequencies of this closed system as 
\begin{align}
\sin(k_nL)+\chi_s k_nL\cos(k_n x_0)\cos(k_n(L-x_0))=0 
\label{Eq: Closed CC Eigenfrequencies}
\end{align}

In the expression above, $ k_n $ represents the wavevector defined as $k_n^2 \equiv lc\omega_n^2$ and the quantity $\chi_{s}\equiv \frac{C_s}{cL}$ is a unitless measure for the discontinuity of current introduced by the transmon.\\

Eventually, the eigenfunctions are found as
\begin{align}
\tilde{\Phi}_n(x)\propto
\begin{cases}
\cos{\left(k_n(L-x_0)\right)}\cos{(k_n x)}&0<x<x_0\\
\cos{(k_n x_0)}\cos{(k_n(L-x))} &x_0<x<L
\end{cases}
\end{align}
where the proportionality constant is set by the orthogonality relation
\begin{align}
\int_0^L dx \frac{c(x,x_0)}{c} \tilde{\Phi}_n(x) \tilde{\Phi}_m(x) = L\delta_{mn}
\label{Eq: Orthogonality of Phi} 
\end{align}

Another important orthogonality condition can be derived in terms between $\{\partial_x \tilde{\Phi}_n \}$ as
\begin{align}
\int _{0}^{L}dx \frac{\partial\tilde{\Phi}_m(x)}{\partial x}\frac{\partial\tilde{\Phi}_n(x)}{\partial x}=k_m k_n L \delta_{mn}
\label{Eq: Orthogonality of d/dx Phi}
\end{align}

Finally, it is instructive to show explicitly the origin of an $A^2$-like term when instead of the CC-basis the conventional cosine modes are chosen. Replacing $\rho(x,t)$ from \ref{Eq: E.O.M for Phi from H_c mod} in $\mathcal{H}_C^{mod}$ gives
\begin{align}
\begin{split}
\mathcal{H}_C^{mod}&=\int_{0}^{L} dx \left[\frac{c(x,x_0)}{2}\left(\frac{\partial \Phi(x,t)}{\partial t}\right)^2+\frac{1}{2l}\left(\frac{\partial \Phi(x,t)}{\partial x}\right)^2\right]
\end{split}
\end{align}

Substituting $c(x,x_0)=c+C_s\delta(x-x_0)$ leads to
\begin{align}
\begin{split}
\mathcal{H}_C^{mod}&=\underbrace{\int_{0}^{L} dx \left[\frac{c}{2}\left(\frac{\partial \Phi(x,t)}{\partial t}\right)^2+\frac{1}{2l}\left(\frac{\partial \Phi(x,t)}{\partial x}\right)^2\right]}_{\mathcal{H}_C}\\
&+\underbrace{\frac{1}{2}C_s \left(\frac{\partial \Phi(x_0,t)}{\partial t}\right)^2}_{\mathcal{H}^{mod}}
\end{split}
\end{align}  

As discussed in \cite{devoret_quantum_2014} $\mathcal{H}_C$ has a diagonal representation in terms of cosine basis. However, by choosing this basis $\mathcal{H}^{mod}$ remains as an $A^2$-like term giving rise to intermode  interaction.
\section{CANONICAL QUANTIZATION OF A CLOSED cQED SYSTEM}
We have all the tools to extend the classical variables into quantum operators by introducing a set of creation and annihilation operators as
\begin{align}
\hat{O}(x,t)=\sum\limits_n C_{\hat{O}}(\omega_n)\left(\hat{a}_n \tilde{O}_n(x)+ \hat{a}_n^{\dagger} \tilde{O}^*_n(x)\right) 
\end{align}
where $O(x,t)$ is any arbitrary classical field that we already know its set of classical eigenmodes $\{\tilde{O}_n(x)\}$ and eigenfrequencies $\{\omega_n\}$ and $\hat{O}(x,t)$ represent its quantum analog. $C_{\hat{O}}(\omega_n)$ represents the appropriate normalization constant for each mode. The creation and annihilation operators obey the usual bosonic commutation relations
\begin{align}
[\hat{a}_n,\hat{a}^{\dagger}_m]&=i\hbar \delta_{nm} \\
[\hat{a}_n,\hat{a}_m]&=0 \\
[\hat{a}^{\dagger}_n,\hat{a}^{\dagger}_m]&=0
\end{align}

Remembering that $\{\tilde{\Phi}_n(x) \}$ represent Hermitian modes and thus real functions, we find the quantum operators $\hat{\Phi}(x,t)$ and $\hat{\rho}(x,t)$ to be
\begin{align}
&\hat{\Phi}(x,t)=\sum\limits_n \left(\frac{\hbar}{2\omega_n cL}\right)^{\frac{1}{2}}\left(\hat{a}_n + \hat{a}_n^{\dagger}\right)\tilde{\Phi}_n(x)\\
&\hat{\rho}(x,t)=-i\sum\limits_n \left(\frac{\hbar \omega_n}{2cL}\right)^{\frac{1}{2}}\left(\hat{a}_n - \hat{a}_n^{\dagger}\right)c(x,x_0)\tilde{\Phi}_n(x) 
\end{align}

Substituting these expressions into $\hat{\mathcal{H}}_C^{mod}$ and using the orthogonality relations \ref{Eq: Orthogonality of Phi} and \ref{Eq: Orthogonality of d/dx Phi}  will result in
\begin{align}
\hat{\mathcal{H}}_C^{mod}=\sum\limits_n \frac{\hbar \omega_n}{2}\left(a_n^{\dagger}a_n +a_n a_n^{\dagger}\right)=\sum\limits_n \hbar \omega_n a_n^{\dagger}a_n+ const.
\end{align}
which is a sum over energy of each independent mode as we expected. Having found the the resonator's Hamiltonian in second quantized form, we have to calculate now the spectrum of transmon whose Hamiltonian is given as

\begin{align}
\hat{\mathcal{H}}_{A}=\frac{\hat{Q}_J^2}{2C_J}-E_J\cos{\left(2\pi\frac{\hat{\Phi}_J}{\phi_0}\right)}
\end{align}

choosing to solve for the spectrum in the flux basis $\{\ket{\Phi_J}\}$ where $\hat{Q}_J\equiv\frac{h}{i}\frac{\partial}{\partial\Phi_J}$, we find

\begin{align}
\left[-\frac{\hbar^2}{2C_J}\frac{d^2}{d \Phi_J^2}-E_J\cos{\left(2\pi\frac{\Phi_J}{\phi_0}\right)}\right]\Psi_n(\Phi_J)=\hbar\Omega_n\Psi_n(\Phi_J)
\end{align}

The solution to the above equation is a set of real eigenenergies and eigenmodes $\{\hbar\Omega_n,\Psi_n(\phi_J)| n\in\mathbb{N}^{0}\}$ where any operator in the transmon's space has a spectral representation over them as

\begin{align}
\begin{split}
\hat{O}_T(t)=&\hat{U}(t)\hat{O}_T(0)\hat{U}^{\dagger}(t)\\
=&\hat{U}(t)\left(\sum_{m,n}\bra{m}\hat{O}_T(0)\ket{n}\hat{P}_{mn}\right)\hat{U}^{\dagger}(t)\\
=&\sum_{m,n}\bra{m}\hat{O}_T(0)\ket{n}\hat{P}_{mn}(t)
\end{split}
\end{align}
where $\{\hat{P}_{mn}=\ket{m}\bra{n}\}$ is a set of projection operators between states m and n , and

\begin{align}
\bra{m}\hat{O}_T(0)\ket{n}\equiv \int d\phi_J \Psi_m(\phi_J)\hat{O}_T\left[\phi_J,\frac{h}{i}\frac{\partial}{\partial\phi_J}\right]\Psi_n(\phi_J) 
\end{align}

We are now able to express $\hat{\Phi}_J$ and $\hat{Q}_J$ in their spectral representation. Notice that since the potential is an even function of $\phi_J$, the eigenmodes are either even or odd functions of $\phi_J$, so the diagonal matrix elements are zero, since

\begin{align} 
&\bra{n}\hat{\Phi}_J\ket{n}=\int d\Phi_J \underbrace{\Phi_J\Psi_n(\Phi_J)\Psi_n(\Phi_J)}_{Odd}=0 \\
&\bra{n}\hat{Q}_J\ket{n}=\frac{\hbar}{i}\int d\Phi_J \underbrace{\Psi_n(\Phi_J)\frac{\partial}{\partial \Phi_J} \Psi_n (\Phi_J)}_{Odd}=0
\end{align}

Therefore, we can express $\hat{\Phi}_J$ and $\hat{Q}_J$ as
\begin{align}
\begin{split}
\hat{\Phi}_J(t)&=\sum\limits_{m\neq n}\bra{m}\hat{\Phi}_J(0)\ket{n}\hat{P}_{mn}(t)\\
&=\sum\limits_{m<n}\bra{m}\hat{\Phi}_J(0)\ket{n}\left(\hat{P}_{mn}(t)+\hat{P}_{nm}(t)\right)
\end{split}\\
\begin{split}
\hat{Q}_J(t)&=\sum\limits_{m\neq n}\bra{m}\hat{Q}_J(0)\ket{n}\hat{P}_{mn}(t)\\
&=\sum\limits_{m<n}\bra{m}\hat{Q}_J(0)\ket{n}\left(\hat{P}_{mn}(t)-\hat{P}_{nm}(t)\right)
\end{split}
\end{align}
where the second lines are written based on the observation that by working in a flux basis, matrix elements of $\hat{Q}_J$ and $\hat{\Phi}_J$ are purely real and imaginary respectively. Now that we know the spectrum of both the resonator and the qubit, we can easily write the interaction term as
\begin{align}
\begin{split}
&\gamma \hat{Q}_J \int_{0}^{L} dx \frac{\hat{\rho}(x,t)}{c(x,x_0)} \delta(x-x_0) =\\
&-i\gamma \sum\limits_{m<n,l}Q_{J,mn} (\hat{P}_{mn} -\hat{P}_{nm})\left(\frac{\hbar \omega_l}{2cL}\right)^{\frac{1}{2}}\left(\hat{a}_l - \hat{a}_l^{\dagger}\right)\tilde{\Phi}_l(x_0)
\end{split}
\end{align}

Defining the coupling intensity $g_{mnl}$ as
\begin{align}
\hbar g_{mnl} \equiv \gamma \left(\frac{\hbar \omega_l}{2cL}\right)^{\frac{1}{2}}(iQ_{J,mn})\tilde{\Phi}_l(x_0) 
\label{Eq:Expression for g- closed case}
\end{align}
the interaction takes the form
\begin{align}
-\sum\limits_{m<n,l} \hbar g_{mnl}(\hat{P}_{mn}-\hat{P}_{nm}) (\hat{a}_l-\hat{a}_l^{\dagger})
\end{align} 

Finally, up to a unitary transformation $\hat{a}_l\rightarrow i\hat{a}_l$ and $\hat{P}_{mn}\rightarrow i\hat{P}_{mn}$ for $m<n$,  the Hamiltonian reads

\begin{align}
\begin{split}
\hat{\mathcal{H}}&=\underbrace{\sum\limits_n \hbar\Omega_n \hat{P}_{nn}}_{\hat{\mathcal{H}}_A}+\underbrace{\sum\limits_n \hbar \omega_n \hat{a}_n^{\dagger} \hat{a}_n}_{\hat{\mathcal{H}}_C^{mod}}\\
&+\underbrace{\sum\limits_{m<n,l} \hbar g_{mnl} \left(\hat{P}_{mn}+\hat{P}_{nm}\right)\left(\hat{a}_l+\hat{a}^{\dagger}_l\right)}_{\hat{\mathcal{H}}_{int}}
\end{split}
\end{align}

Notice that by truncating transmon's space into its first two levels, we are able to recover a multimode Rabi Hamiltonian
\begin{align}
\begin{split}
\hat{\mathcal{H}}&=\underbrace{\frac{1}{2} \hbar \omega_{01} \hat{\sigma}^z}_{\hat{\mathcal{H}}_A^{tru}}+\underbrace{\sum\limits_n \hbar \omega_n \hat{a}_n^{\dagger} \hat{a}_n}_{\hat{\mathcal{H}}_C^{mod}}\\
&+\underbrace{\sum\limits_n \hbar g_n (\hat{\sigma}^- + \hat{\sigma}^+)(\hat{a}_n+\hat{a}^{\dagger}_n)}_{\hat{\mathcal{H}}_{int}}
\end{split}
\end{align}
where we have used the shorthand notation $\hat{\sigma}^{-}=\hat{P}_{01}$, $\hat{\sigma}^{+}=\hat{P}_{10}$ and $\omega_{01}=\Omega_1-\Omega_0$. $g_{mnl}$ is also reduced to $g_{n}\equiv g_{01n}$ given as
\begin{align}
\hbar g_{n} \equiv \gamma \left(\frac{\hbar \omega_n}{2cL}\right)^{\frac{1}{2}}(iQ_{J,01})\tilde{\Phi}_n(x_0) 
\end{align}
\begin{figure}[h!]
\centering
\subfloat[\label{subfig:Eigfreqsx050}]{%
\centerline{\includegraphics[scale=0.24]{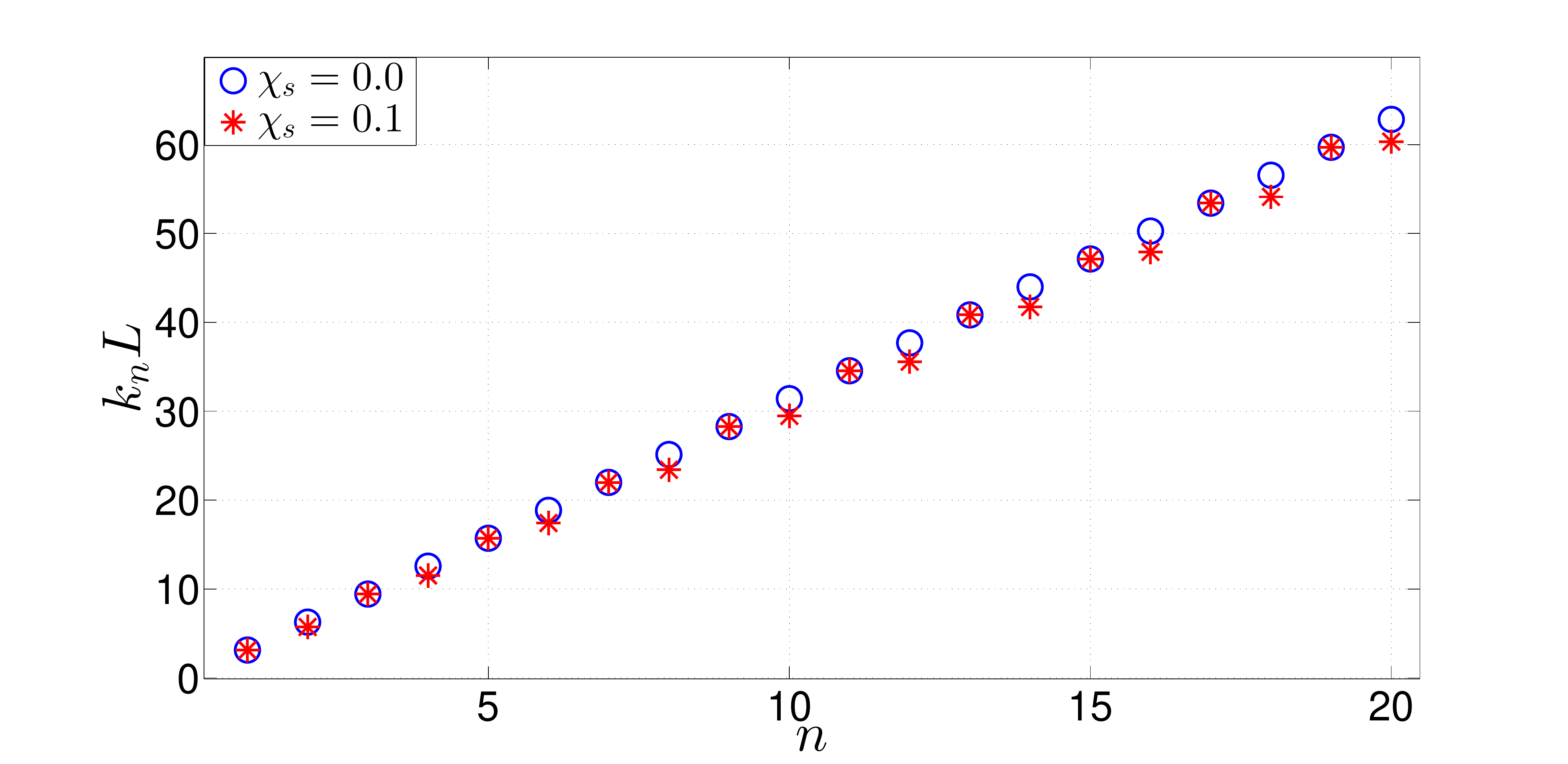}%
}}\hfill
\subfloat[\label{subfig:mode1x050}]{%
\includegraphics[scale=0.29]{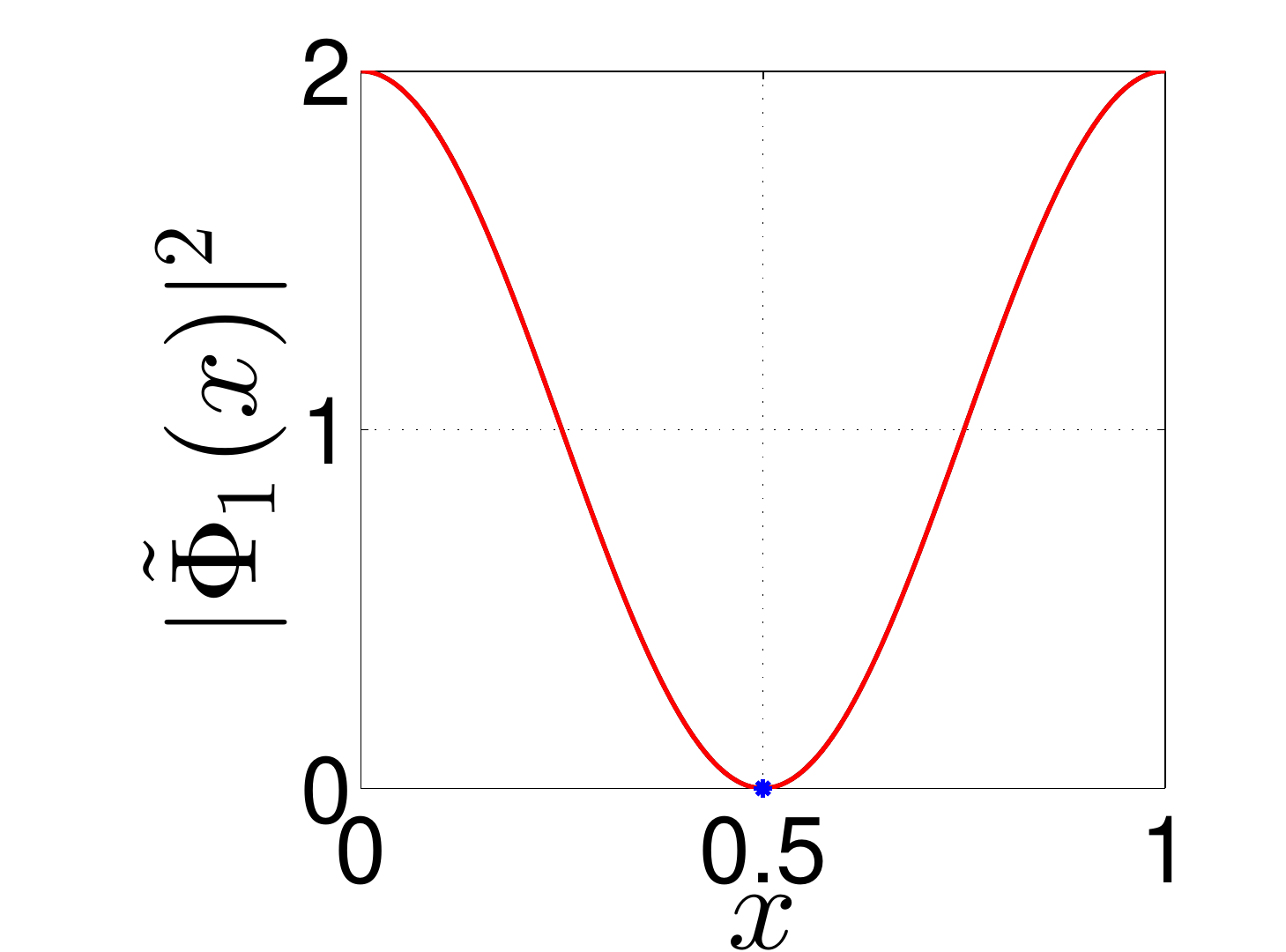}%
}
\subfloat[\label{subfig:mode2x050}]{%
\includegraphics[scale=0.29]{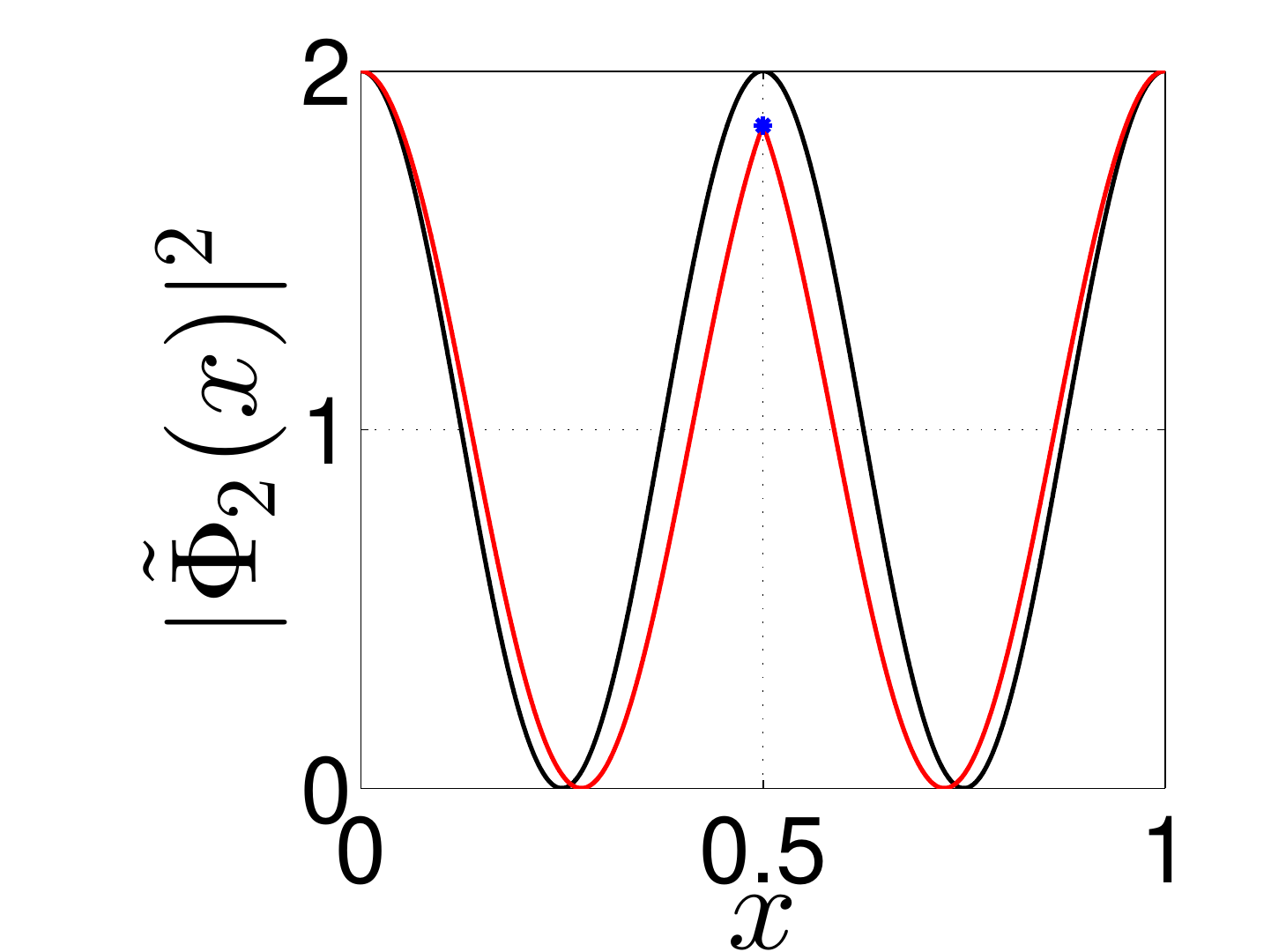}%
}\hfill
\subfloat[\label{subfig:mode3x050}]{%
\includegraphics[scale=0.29]{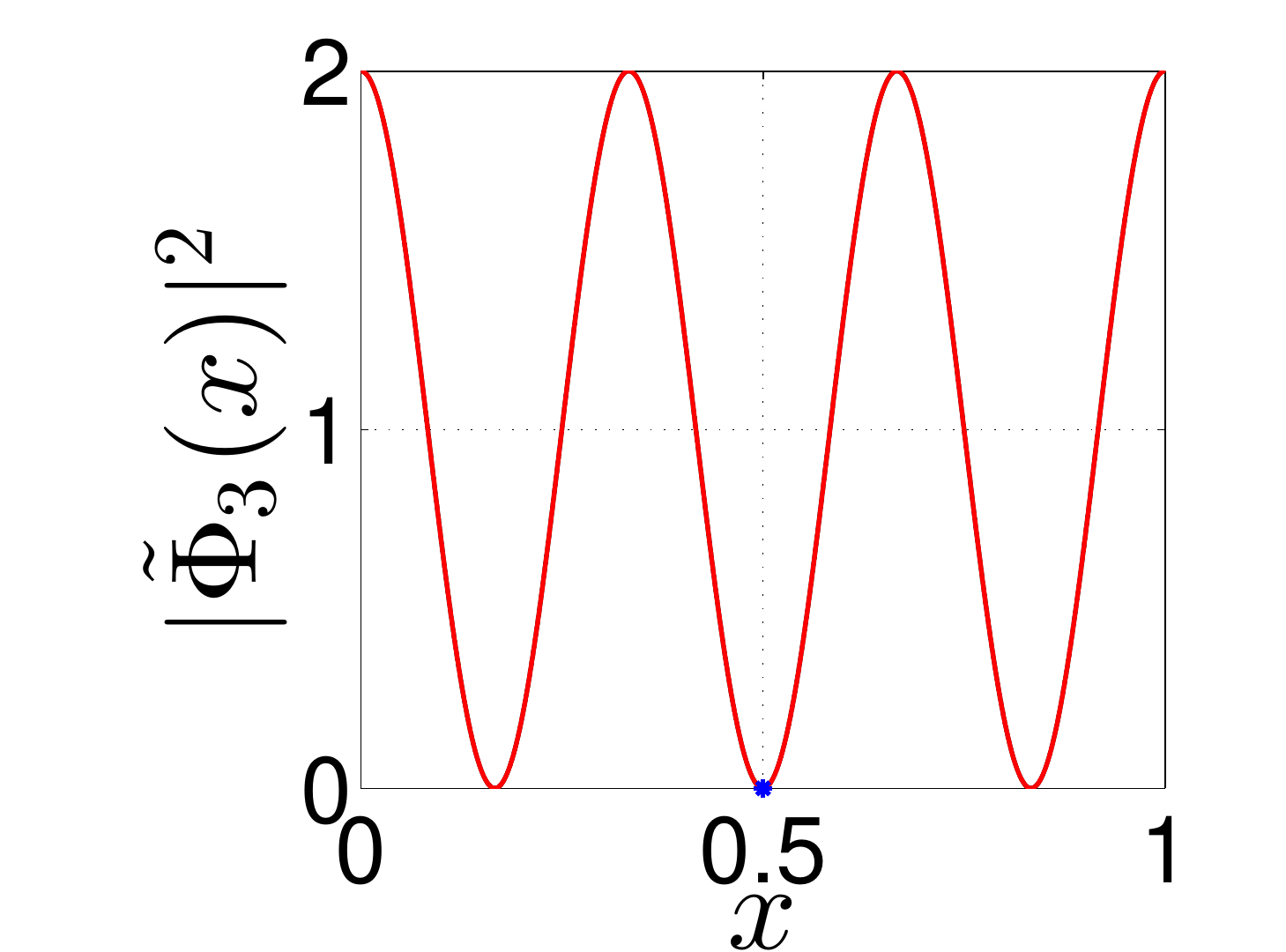}%
}
\subfloat[\label{subfig:mode4x050}]{%
\includegraphics[scale=0.29]{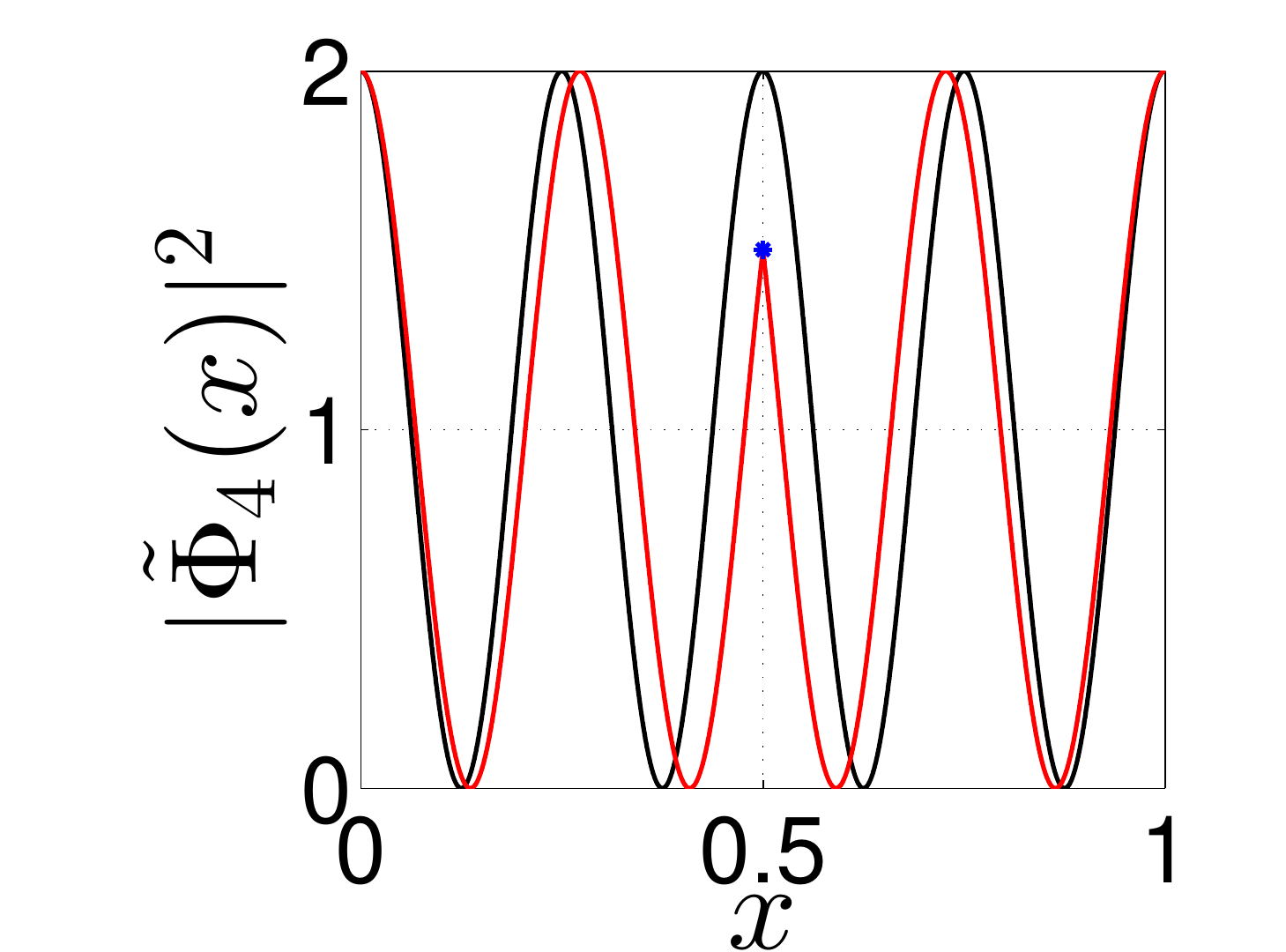}%
}\hfill
\subfloat[\label{subfig:Eigfreqsx050}]{%
\centerline{\includegraphics[scale=0.24]{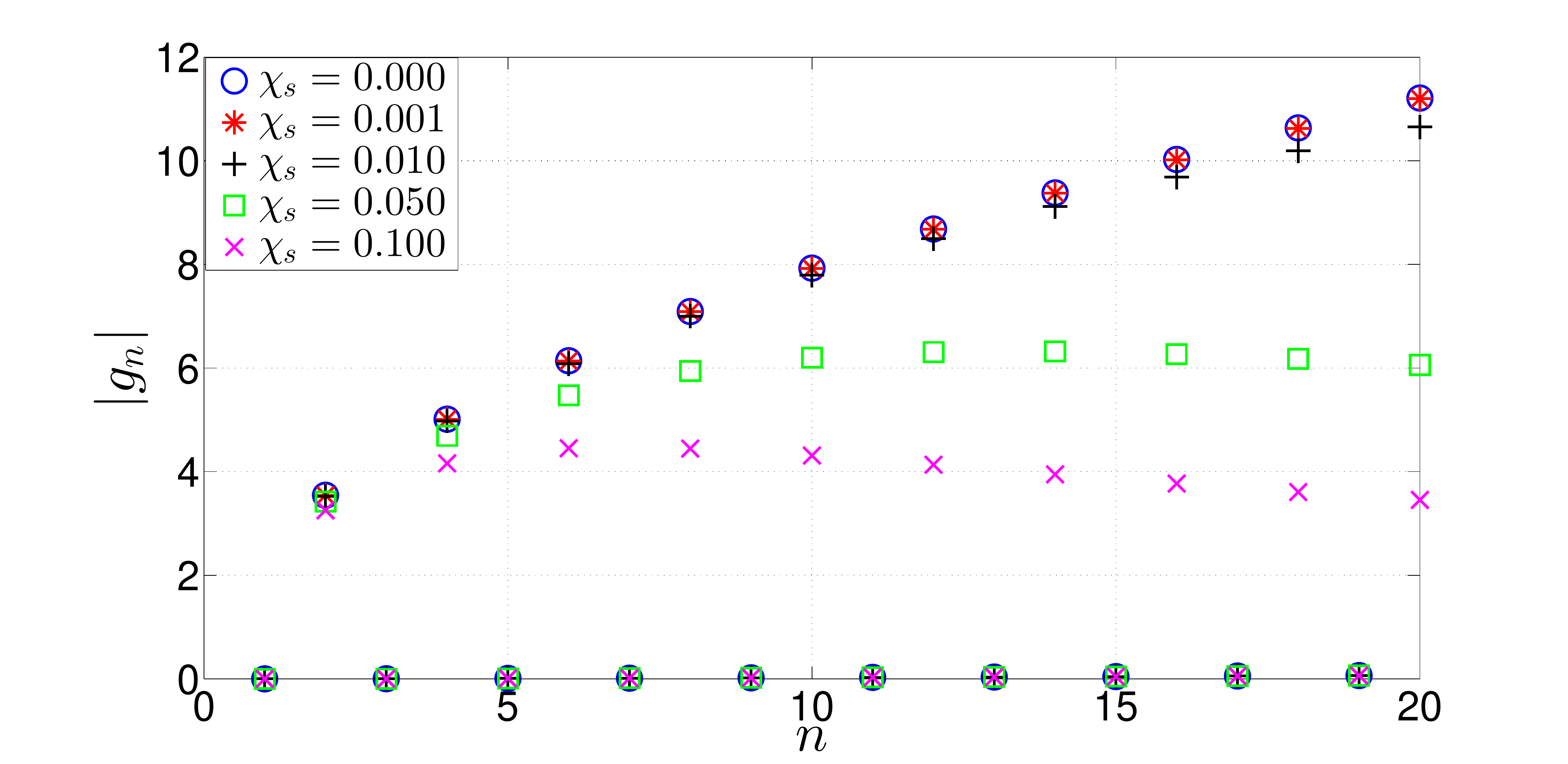}%
}}
\caption{Closed-boundary CC modes for $x_0=0.5L$ (a)- First 20 Eigenfrequencies for $\chi_s=0.1$ b-e) Normalized energy density of the first 4 modes for $\chi_s=0.1$. The black curve shows cosine modes while the red ones represent CC modes. The blue star shows where the qubit is connected. f) First 20 coupling strengths $g_n$ for various values of $\chi_s$. }
\label{fig:Eigx050}
\end{figure}
\begin{figure}[h!]
\centering
\subfloat[\label{subfig:Eigfreqsx025}]{%
\centerline{\includegraphics[scale=0.24]{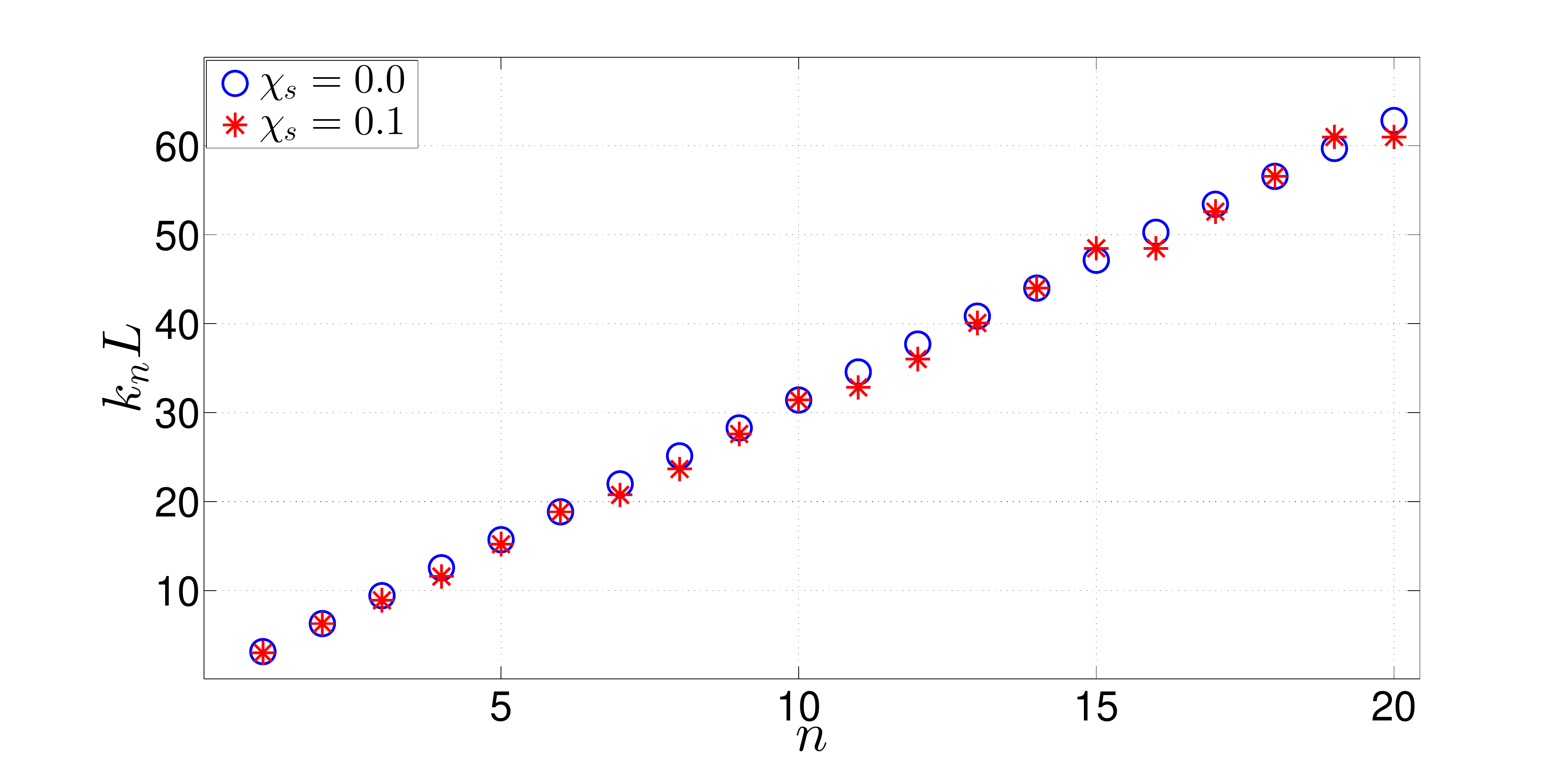}%
}}\hfill
\subfloat[\label{subfig:mode1x025}]{%
\includegraphics[scale=0.29]{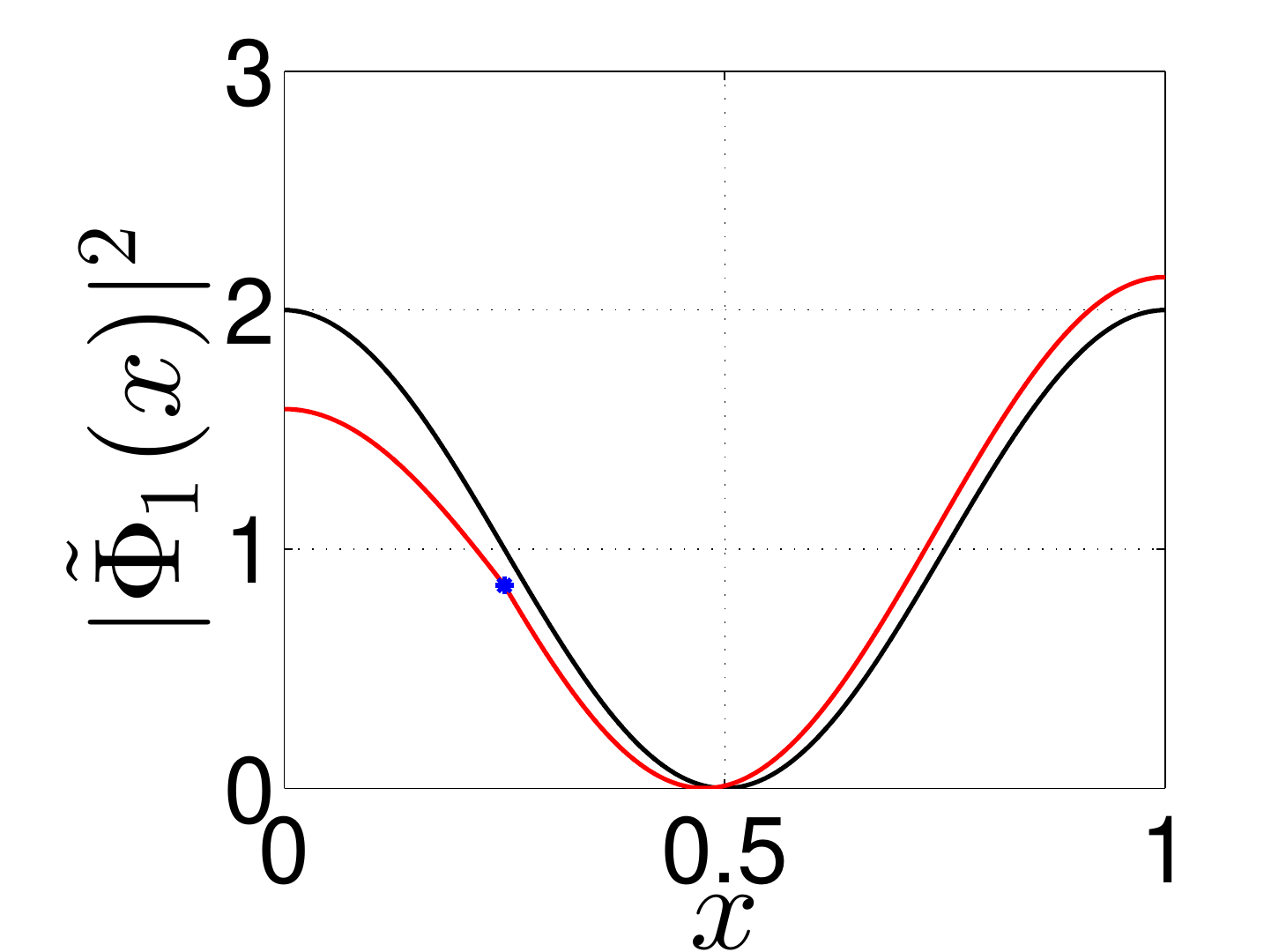}%
}
\subfloat[\label{subfig:mode2x025}]{%
\includegraphics[scale=0.29]{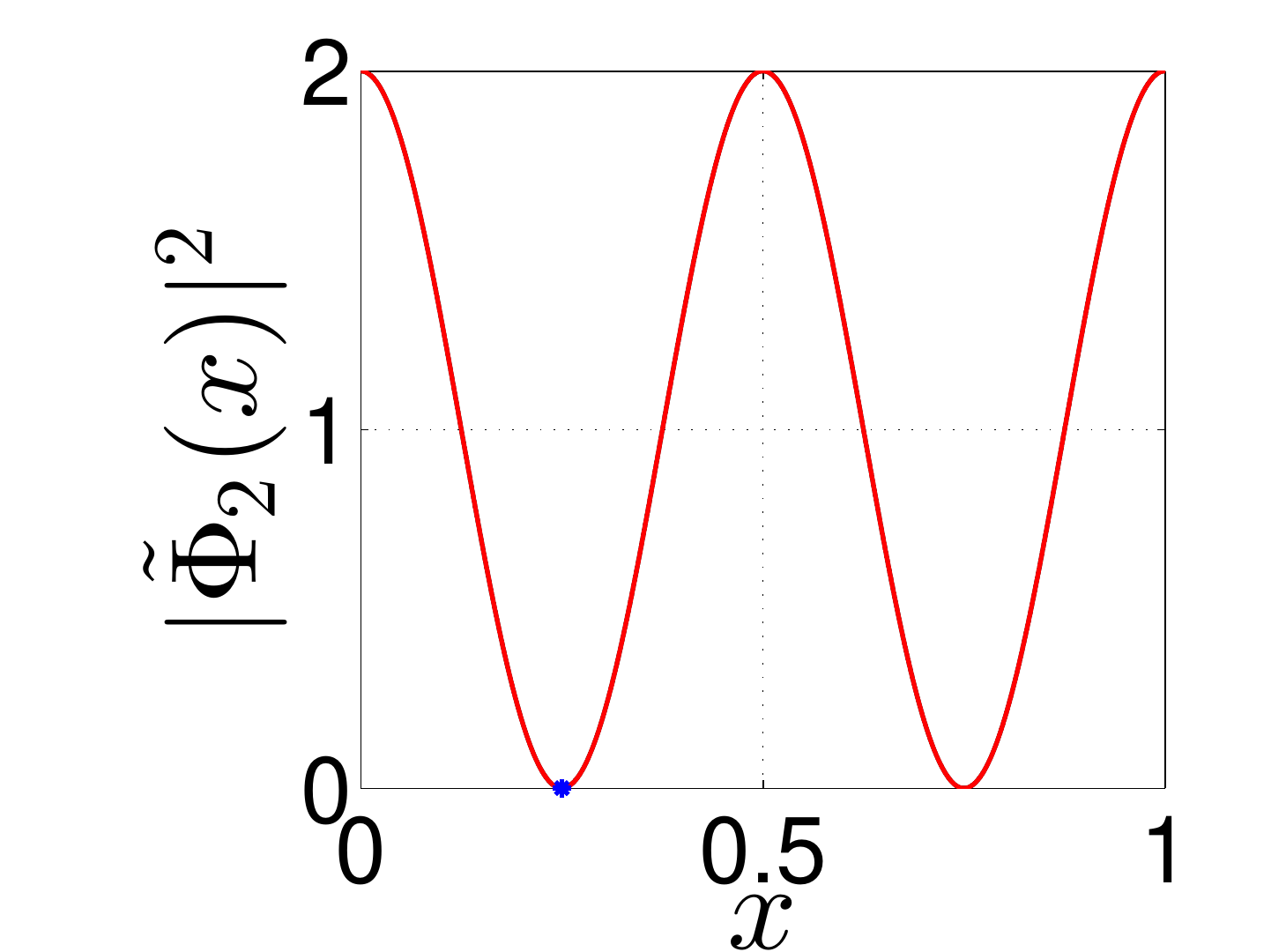}%
}\hfill
\subfloat[\label{subfig:mode3x025}]{%
\includegraphics[scale=0.29]{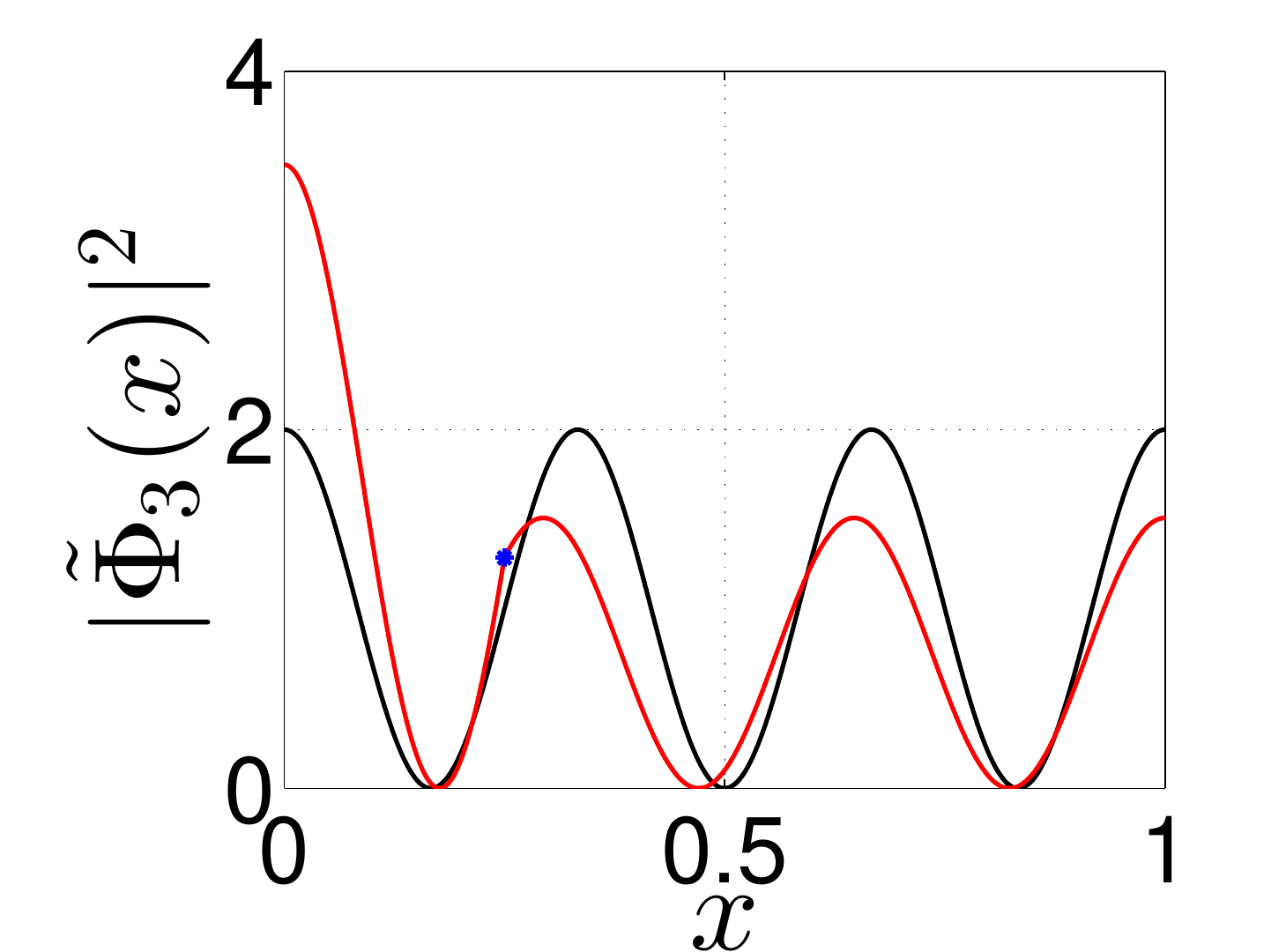}%
}
\subfloat[\label{subfig:mode4x025}]{%
\includegraphics[scale=0.29]{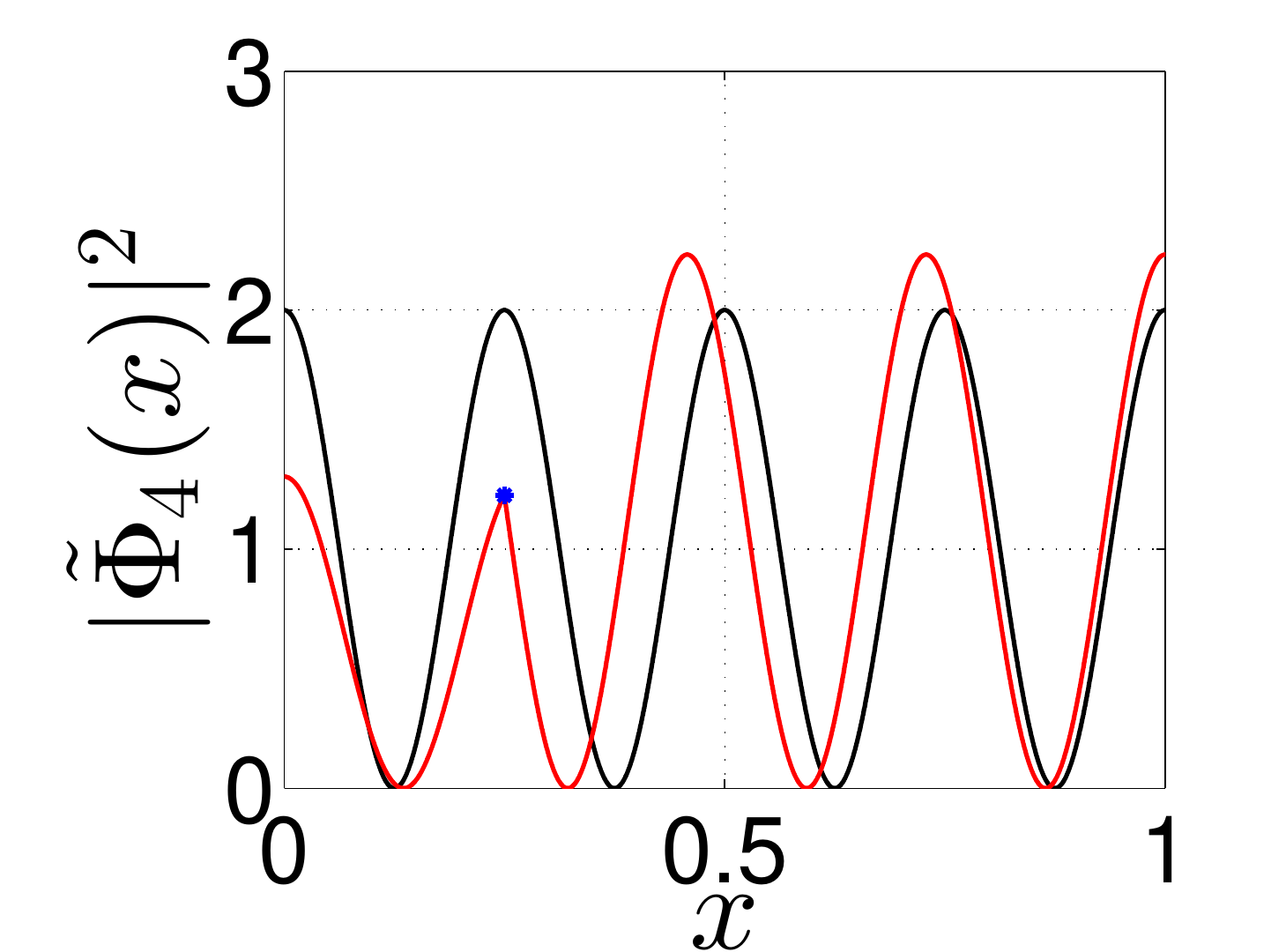}%
}\hfill
\subfloat[\label{subfig:Eigfreqsx025}]{%
\centerline{\includegraphics[scale=0.24]{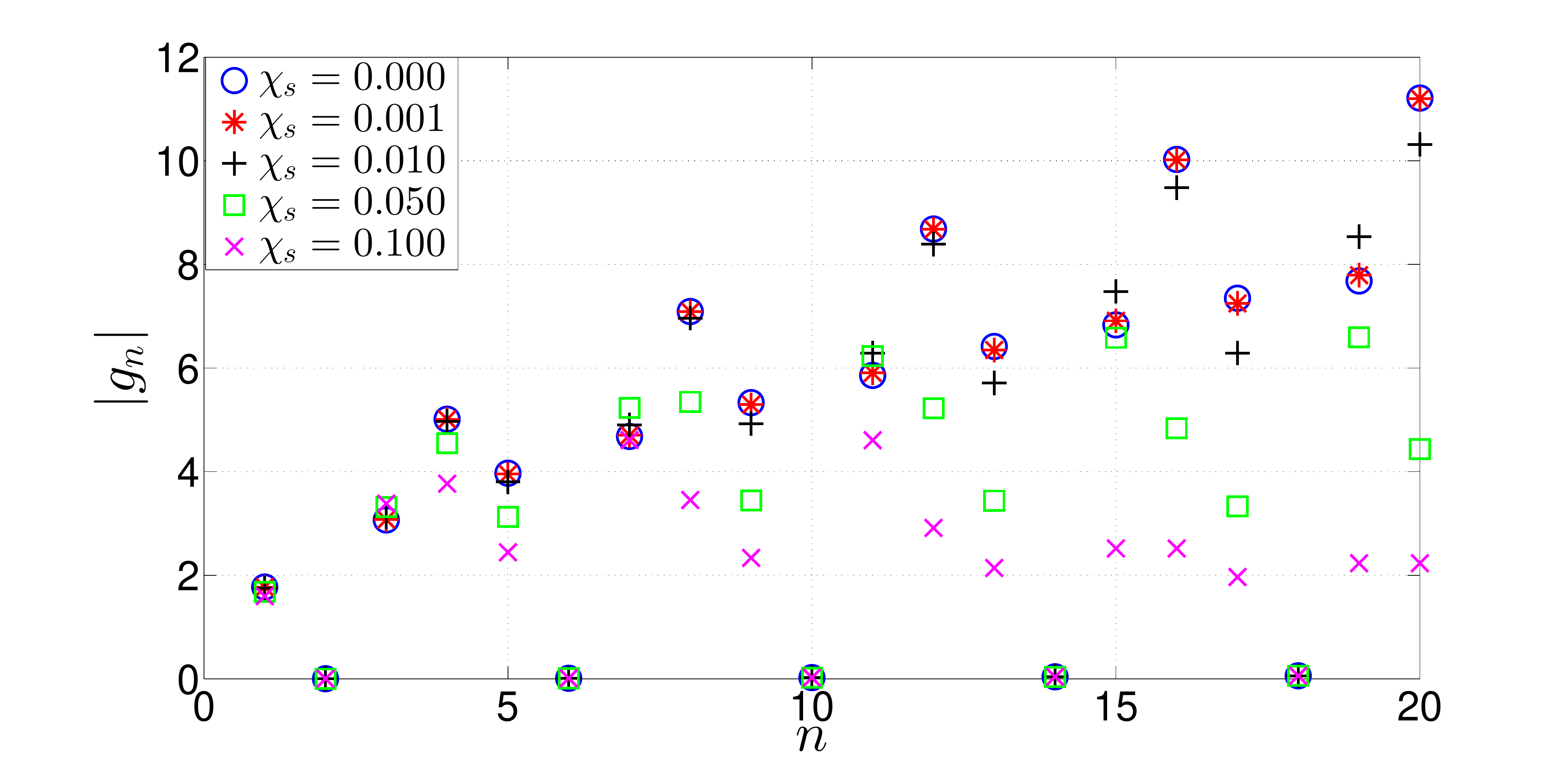}%
}}
\caption{Closed-boundary CC modes for $x_0=0.25L$ (a)- First 20 Eigenfrequencies b-e) Normalized energy density of the first 4 modes for $\chi_s=0.1$. The black curve shows cosine modes while the red ones represent CC modes. The blue star shows where the qubit is connected. f)  First 20 coupling strengths $g_n$ for various values of $\chi_s$.}
\label{fig:Eigx025}
\end{figure}

In the main body of this paper we have only included the results for the case $x_0=0.01L$. In Figs. \ref{fig:Eigx050} and \ref{fig:Eigx025}, we have considered the cases of $x_0=0.50L$ and $x_0=0.25L$ respectively. For $x_0=0.50L$ we observe that all even numbered CC modes are unperturbed, while for odd numbered CC modes, both eigenmodes and eigenfrequencies are found to be less than cosine ones. The reason for invariance of even numbered modes is that originally qubit sits on a local minimum of the photonic energy density and therefore does not interact with these modes. In addition, regardless of the value for $\chi_s$, all odd numbered coupling strengths are zero due to high symmetry of this point.  For $\chi_s=0 $, the even modes follow an envelope that goes like $\sqrt{\omega_n}$ and as $\chi_s$ grows they are suppressed and fall below this envelope.

This behavior is not specific to $x_0=0.50L$. Generally, if the qubit is placed at $x_0=\frac{L}{n},n\in \mathbb{N}$ then there is a periodicity in the mode structure such that every $n$ modes remain unperturbed. This is for example observed in Fig. \ref{fig:Eigx025} where $ x_0=0.25L $ and thus modes indexed as $4n-2,n\in\mathbb{N}$ are unchanged. In this case, depending on the value of $\chi_s$ the CC coupling strengths can be either below or above the solutions for $\chi_s=0$.
\section{TRK SUM RULES FOR A TRANSMON QUBIT}
In this Appendix, first we are after finding a general sum rule in quantum mechanics and then we apply the results to calculate upper bounds for matrix elements of charge and flux operator i.e. $\bra{m}\hat{Q}_J\ket{n}$ and $\bra{m}\hat{\Phi}_J\ket{n}$ for the case of a transmon qubit. Assume a Hamiltonian $\hat{\mathcal{H}}$ where its eigenmodes and eigenenergies are known as $\left\{\ket{n},E_n | n\in \mathbb{N}^{0}\right\}$. Consider an arbitrary Hermitian operator $\hat{\mathcal{A}}=\hat{\mathcal{A}}^{\dagger}$ where we define successive commutation of $\hat{\mathcal{H}}$ and $\hat{\mathcal{A}}$ as
\begin{align}
\hat{\mathcal{C}}_{\hat{\mathcal{A}}}^{(k)}\equiv\left[
\hat{\mathcal{H}},\hat{\mathcal{C}}_{\hat{\mathcal{A}}}^{(k-1)}\right],\quad 
\hat{\mathcal{C}}_{\hat{\mathcal{A}}}^{(0)}\equiv \hat{\mathcal{A}}
\end{align}

By this definition we find that for any two arbitrary eigenstates $\ket{m}$ and $\ket{n}$  
\begin{align}
\begin{split}
\bra{m}\hat{\mathcal{C}}_{\hat{\mathcal{A}}}^{(k)}\ket{n}=&(E_m-E_n)\bra{m}\hat{\mathcal{C}}_{\hat{\mathcal{A}}}^{(k-1)}\ket{n}\\
&\vdots\\
=&\left(E_m-E_n\right)^k \bra{m}\hat{\mathcal{A}}\ket{n}
\end{split}
\end{align}

Using the above identity we find that 
\begin{align}
\begin{split}
\bra{m}\left[\hat{\mathcal{A}},\hat{\mathcal{C}}_{\hat{\mathcal{A}}}^{(k)}\right]\ket{m}&=\bra{m}\hat{\mathcal{A}}\underbrace{\hat{\mathbf{1}}}_{\sum\limits_n \ket{n}\bra{n}}\hat{\mathcal{C}}_{\hat{\mathcal{A}}}^{(k)}\ket{m}\\
&-\bra{m}\hat{\mathcal{C}}_{\hat{\mathcal{A}}}^{(k)}\underbrace{\hat{\mathbf{1}}}_{\sum\limits_n \ket{n}\bra{n}}\hat{\mathcal{A}}\ket{m}\\
=\sum\limits_n \left(E_n-E_m \right)&\left[1-(-1)^k\right]\left|\bra{m}\hat{\mathcal{A}}\ket{n}\right|^2
\end{split}
\label{Eq:General Sum Rule in QM}
\end{align}
Now, consider the Hamiltonian for a transmon qubit 
\begin{align}
\hat{\mathcal{H}}=\frac{\hat{Q}_J^2}{2C_J}+U(\hat{\Phi}_J)
\end{align} 
where $U(\hat{\Phi}_J)=-E_J\cos{\left(\frac{2\pi}{\phi_0}\hat{\Phi}_J\right)}$. Applying the result found in \ref{Eq:General Sum Rule in QM} we can write
\begin{align}
\bra{0}[\hat{\Phi}_J,\underbrace{[\hat{\mathcal{H}},\hat{\Phi}_J]}_{\hat{\mathcal{C}}_{\hat{\Phi}}^{(1)}}]\ket{0}=\sum\limits_{n>0} 2\left(E_n-E_0 \right)\left|\bra{0}\hat{\Phi}_J\ket{n}\right|^2
\end{align}
where $\ket{0}$ represents the ground state. The L.H.S can be calculated explicitly as $\frac{\hbar^2}{C_J}$ which leads to the sum rule for $\hat{\Phi}_J$ as
\begin{align}
\sum\limits_{n>0} 2\left(E_n-E_0 \right)\left|\bra{0}\hat{\Phi}_J\ket{n}\right|^2=\frac{\hbar^2}{C_J}
\label{Eq:Sum Rule for Phi_J}
\end{align}
Noticing that all terms on the L.H.S are positive, we can find an upper bound for $\Phi_{J,01}$ as
\begin{align}
|\Phi_{J,01}|^2<\frac{\hbar^2}{2C_J(E_1-E_0)}\approx \frac{E_C}{\sqrt{8E_JE_C}-E_C}\left(\frac{\hbar}{e}\right)^2 
\end{align}
where we have defined the charging energy $E_C\equiv\frac{e^2}{2C_J}$. In a similar manner, it is possible to obtain a sum rule for $\hat{Q}_J$ as
\begin{align}
\bra{0}[\hat{Q}_J,\underbrace{[\hat{\mathcal{H}},\hat{Q}_J]}_{\hat{\mathcal{C}}_{\hat{Q}}^{(1)}}]\ket{0}=\sum\limits_{n>0} 2\left(E_n-E_0 \right)\left|\bra{0}\hat{Q}_J\ket{n}\right|^2
\end{align}
The L.H.S can be explicitly found as
\begin{align}
\begin{split}
[\hat{Q}_J,[\hat{\mathcal{H}},\hat{Q}_J]]=\hbar^2\frac{\partial^2U(\hat{\Phi}_J)}{\partial\hat{\Phi}_J^2}=\left(\frac{2\pi\hbar}{\phi_0}\right)^2 E_J\cos{\left(\frac{2\pi}{\phi_0}\hat{\Phi}_J\right)}
\end{split}
\end{align}
Noticing that $\frac{2\pi\hbar}{\phi_0}=2e$ brings us the sum rule for $\hat{Q}_J$ as
\begin{align}
\begin{split}
\sum\limits_{n>0} 2\left(E_n-E_0 \right)\left|\bra{0}\hat{Q}_J\ket{n}\right|^2&=(2e)^2E_J \bra{0}\cos{\left(\frac{2\pi}{\phi_0}\hat{\Phi}_J\right)}\ket{0}\\
&<(2e)^2E_J
\end{split}
\label{Eq:Sum Rule for Q_J _app}
\end{align}
Again, due to positivity of terms on the L.H.S we find
\begin{align}
|Q_{J,01}|^2<\frac{2e^2 E_J}{E_1-E_0}\approx \frac{2E_J}{\sqrt{8E_JE_C}-E_C}e^2 
\end{align}
\section{GENERALIZATION TO AN OPEN cQED SYSTEM}
\subsection{Lagrangian and Modified Eigenmodes}
The results from the previous section make it very easy to find the Lagrangian and hence the dynamics for the open case where now the end capacitors $C_R$ and $C_L$ have finite values as shown in Fig.\ref{subfig:circuit-QED-Closed-EquivalentCircuit}. Here, we have a finite length resonator which is capacitively coupled to two other microwave resonators at each end. Assuming that the transmon qubit is connected to the resonator at some arbitrary point $x=x_0$, the Lagrangian for this system can be written as
\begin{align}
\begin{split}
\mathcal{L}&=\frac{1}{2}C_J\dot{\Phi}_{J}(t)^2-U(\Phi_J(t)) \\
&+\int_{0^+}^{L^-}\,dx\left[\frac{1}{2} c(\frac{\partial \Phi}{\partial t})^2-\frac{1}{2l}(\frac{\partial \Phi}{\partial x})^2\right]\\
&+\int_{L^+}^{L+L_R}\,dx\left[\frac{1}{2} c(\frac{\partial \Phi_R}{\partial t})^2-\frac{1}{2l}(\frac{\partial \Phi_R}{\partial x})^2\right]\\
&+\int_{-L-L_L}^{0^-}\,dx\left[\frac{1}{2} c(\frac{\partial \Phi_L}{\partial t})^2-\frac{1}{2l}(\frac{\partial \Phi_L}{\partial x})^2\right]\\
&+\frac{1}{2}C_L\left(\dot{\Phi}_L(0^-,t)-\dot{\Phi}(0^+,t)\right)^2\\&
+\frac{1}{2}C_R\left(\dot{\Phi}_R(L^+,t)-\dot{\Phi}(L^-,t)\right)^2\\
&+\frac{1}{2}C_g\left(\dot{\Phi}_J(t)-\dot{\Phi}(x_0,t)\right)^2
\end{split}
\end{align}

We have already learned how the coupling intensity depends on the Hermition eigenmodes and eigenfrequencies of the resonator as well as the dipole moment of the transmon. Here we have the same situation except that due to the opening introduced by the finite end capacitors $C_R$ and $C_L$, we need to find the modified Hermitian modes of the open system. Therefore, let's for the moment forget about Lagrangian of the transmon and its coupling to the resonator and focus on the modification introduced by one of the end capacitors, for instance $C_L$. The trick is that we can write this contribution as sum of three separate terms

\begin{align}
\begin{split}
\frac{1}{2}C_L \left(\dot{\Phi}_L(0^-,t)-\dot{\Phi}(0^+,t)\right)^2&=\frac{1}{2}C_L \dot{\Phi}_L(0^-,t)^2\\
&+\frac{1}{2}C_L \dot{\Phi}(0^+,t)^2\\
&-C_L \dot{\Phi}(0^+,t)\dot{\Phi}_L(0^-,t) 
\end{split}
\end{align}

Notice that only the term $-C_L \dot{\Phi}(0^+,t)\dot{\Phi}_L(0^-,t)$ is responsible for coupling of the resonator to the bath and the other two can be considered as a modification on top of the closed case. Applying the same method for the right capacitor, we can define the modified Lagrangian for the left and right baths

\begin{align}
\begin{split}
\mathcal{L}_{R}^{op}&=\int_{L^+}^{\infty}\,dx\left[\frac{1}{2} c(\frac{\partial \Phi_R}{\partial t})^2-\frac{1}{2l}(\frac{\partial \Phi_R}{\partial x})^2\right]\\
&+\frac{1}{2}C_R \dot{\Phi}_R(L^+,t)^2
\end{split}\\
\begin{split}
\mathcal{L}_{L}^{op}&=\int_{-\infty}^{0^-}\,dx\left[\frac{1}{2} c(\frac{\partial \Phi_L}{\partial t})^2-\frac{1}{2l}(\frac{\partial \Phi_L}{\partial x})^2\right]\\
&+\frac{1}{2}C_L \dot{\Phi}_R(0^-,t)^2
\end{split}
\end{align}

In addition, the interaction Lagrangian is found as
\begin{align}
\mathcal{L}_{C,LR}=-C_L \dot{\Phi}(0^+,t)\dot{\Phi}_L(0^-,t)-C_R \dot{\Phi}(L^-,t)\dot{\Phi}_R(L^+,t)
\end{align}
which is also equal to the interaction Hamiltonian since by going from Lagrangian to Hamiltonian capacitive contributions(kinetic contributions) don't change sign. The idea is to find the Hermition modes governed only by each of these uncoupled modified contributions and finally write the interaction in terms of Hermitian modes of each subsystem.

Up to here, we haven't considered the effect of transmon on capacitance per length as we found in \ref{Eq:modified capacitance per length}. It is not necessary to go over the derivation again, since these two effects, modification due to opening and due to transmon, are completely independent. By considering the inhomogeneity introduced by the transmon we have
\begin{align}
\mathcal{L}_{C}^{op}=\int_{0^+}^{L^-}\,dx\left[\frac{1}{2} c_{op}(x)(\frac{\partial \Phi}{\partial t})^2-\frac{1}{2l}(\frac{\partial \Phi}{\partial x})^2\right]
\end{align}
where $c_{op}(x,x_0)$ is given as
\begin{align}
\begin{split}
c_{op}(x,x_0)&=c+C_s \delta(x-x_0)\\
&+ C_R\delta(x-L^-)+C_L \delta(x-0^+)
\end{split}
\end{align}

The new delta functions in $c_{op}(x)$ are only important at the boundaries, which can be found by integrating the equation along an infinitesimal interval that includes the delta functions. In order to find the modes, we need to solve

\begin{align}
\frac{\partial^2 \tilde{\Phi}_n(x)}{\partial x^2}+lc\omega_n^2\tilde{\Phi}_n(x)=0 ,\qquad x\neq x_0 
\end{align}
with boundary conditions given as
\begin{align}
&\left.\frac{\partial}{\partial x}\tilde{\Phi}_n(x)\right|_{x=L^-}=lC_R \omega_n^2\tilde{\Phi}_n(L^-)\\
&\left.\frac{\partial}{\partial x}\tilde{\Phi}_n(x)\right|_{x=0^+}=-lC_L\omega_n^2\tilde{\Phi}_n(0^+)\\
&\left.\frac{\partial \tilde{\Phi}_n(x)}{\partial x}\right|_{x_0^+}-\left.\frac{\partial \tilde{\Phi}_n(x)}{\partial x}\right|_{x_0^-} + lC_s\omega_n^2 \tilde{\Phi}_n(x_0)=0 \\
&\tilde{\Phi}_n(x_0^+)=\tilde{\Phi}_n(x_0^-) 
\end{align}

Defining unitless parameters $\chi_{R,L}\equiv \frac{C_{R,L}}{cL}$ as we did for $\chi_s$, we find eigenfrequencies satisfy a transcendental equation as
\begin{align}
\begin{split}
&+\left(1-\chi_R\chi_L (k_nL)^2\right)\sin{(k_n L)}\\
&+\left(\chi_R+\chi_L\right)k_nL\cos{(k_n L)}\\
&+\chi_s k_n L\cos{(k_n x_0)}\cos{(k_n (L-x_0))}\\
&-\chi_R\chi_s (k_n L)^2 \cos{(k_n x_0)}\sin{(k_n (L-x_0))}\\
&-\chi_L\chi_s (k_n L)^2 \sin{(k_n x_0)}\cos{(k_n (L-x_0))}\\
&+\chi_R\chi_L\chi_s (k_n L)^3 \sin{(k_n x_0)}\sin{(k_n (L-x_0))}=0
\label{Eq: Open CC Eigenfrequencies}
\end{split}
\end{align}
and the real-space representation of these modes read
\begin{align}
\tilde{\Phi}_n(x)\propto
\begin{cases}
\tilde{\Phi}_n^{<}(x) \quad 0<x<x_0\\
\tilde{\Phi}_n^{>}(x) \quad x_0<x<L
\end{cases}
\end{align}
where $\tilde{\Phi}_n^{<}(x)$ and $\tilde{\Phi}_n^{>}(x)$ are found as
\begin{align}
\begin{split}
\tilde{\Phi}_n^{<}(x)&=\left[\cos{(k_n(L-x_0))}-\chi_Rk_nL\sin{(k_n(L-x_0))}\right]\\
&\times\left[\cos{(k_n x)}-\chi_L k_nL\sin{(k_n x)}\right]
\end{split}\\
\begin{split}
\tilde{\Phi}_n^{>}(x)&=\left[\cos{(k_n x_0)}-\chi_L k_nL\sin{(k_n x_0)}\right]\\
&\times\left[\cos{(k_n(L-x))}-\chi_Rk_nL\sin{(k_n(L-x))}\right]
\end{split}
\end{align}
The normalization constant will be set by the orthogonality conditions
\begin{align}
\int_{0}^{L} dx \frac{c_{op}(x,x_0)}{c}\tilde{\Phi}_m(x)\tilde{\Phi}_n(x)=L\delta_{mn}
\end{align}
\begin{align}
\begin{split}
&+\int _{0}^{L}dx \frac{\partial\tilde{\Phi}_m(x)}{\partial x}\frac{\partial\tilde{\Phi}_n(x)}{\partial x}\\
&-\frac{1}{2}\left(k_m^2+k_n^2\right)L\left[\chi_R\tilde{\Phi}_m(L^-)\tilde{\Phi}_n(L^-)+\chi_L\tilde{\Phi}_m(0^+)\tilde{\Phi}_n(0^+)\right]\\
&=k_m k_n L \delta_{mn}
\end{split}
\end{align}
\subsection{Canonical Quantization}
Following the same quantization procedure as for the closed case we can write the field operators in terms of the eigenmodes and eigenfrequencies of each part of the circuit as 
\begin{align}
\hat{\Phi}(x,t)=\sum\limits_n \left(\frac{\hbar}{2\omega_n cL}\right)^{\frac{1}{2}}\left(\hat{a}_n(t) + \hat{a}_n^{\dagger}(t)\right)\tilde{\Phi}_n(x)
\end{align}
\begin{align}
\begin{split}
&\hat{\Phi}_R(x,t)=\sum \limits_n \left(\frac{\hbar}{2\omega_{n,R} cL_R}\right)^{\frac{1}{2}} \left(\hat{b}_{n,R}(t) + \hat{b}_{n,R}^{\dagger}(t)\right)\tilde{\Phi}_{n,R}(x)
\end{split}\\
\begin{split}
&\hat{\Phi}_L(x,t)=\sum \limits_n \left(\frac{\hbar}{2\omega_{n,L} cL_L}\right)^{\frac{1}{2}} \left(\hat{b}_{n,L}(t) + \hat{b}_{n,L}^{\dagger}(t)\right)\tilde{\Phi}_{n,L}(x)
\end{split}
\end{align}

where we have also considered some finite length for the left and right resonators as well to keep the normalization constants meaningful. Now, we can easily derive an expression for resonator-bath coupling in terms of modes of each part. Consider coupling to the left bath for the moment
\begin{align}
&\hat{\mathcal{H}}_{CL}=-C_L \dot{\hat{\Phi}}(0^+,t)\dot{\hat{\Phi}}_L(0^-,t)\\
&\hat{\mathcal{H}}_{CR}=-C_R \dot{\hat{\Phi}}(L^+,t)\dot{\hat{\Phi}}_R(L^-,t)
\end{align}

Considering that the time-dynamics of annihiliation and creation operators up to here are only governed by the free Lagrangian of each part, we have
\begin{align}
\begin{split}
\dot{\hat{\Phi}}(x,t)&=\sum\limits_n \left(\frac{\hbar}{2\omega_n cL}\right)^{\frac{1}{2}}\left(-i\omega_n \hat{a}_n(t) +i\omega_n \hat{a}_n^{\dagger}(t)\right)\tilde{\Phi}_n(x)\\
&=-i\sum\limits_n \left(\frac{\hbar \omega_n}{2cL}\right)^{\frac{1}{2}}\left(\hat{a}_n(t) - \hat{a}_n^{\dagger}(t)\right)\tilde{\Phi}_n(x)
\end{split}
\end{align} 
And we have the same type of expression for the end resonators as well. The interaction Hamiltonian then reads
\begin{align}
\begin{split}
\hat{\mathcal{H}}_{C,LR}=&-\sum_{m,n}\hbar \beta_{mn,R}\left(\hat{a}_m-\hat{a}_m^{\dagger}\right)\left(\hat{b}_{n,R}-\hat{b}_{n,R}^{\dagger}\right)\\
&-\sum_{m,n}\hbar \beta_{mn,L}\left(\hat{a}_m-\hat{a}_m^{\dagger}\right)\left(\hat{b}_{n,L}-\hat{b}_{n,L}^{\dagger}\right)
\end{split}
\end{align}
where we find $\beta_{mn,R}$ and $\beta_{mn,L}$ as

\begin{align}
&\beta_{mn,R}=\frac{C_{R}}{2c\sqrt{L}\sqrt{L_{R}}}\omega_m^{\frac{1}{2}}\omega_{n,R}^{\frac{1}{2}}\tilde{\Phi}_m(L^-)\tilde{\Phi}_{n,R}(L^+)\\
&\beta_{mn,L}=\frac{C_{L}}{2c\sqrt{L}\sqrt{L_{L}}}\omega_m^{\frac{1}{2}}\omega_{n,L}^{\frac{1}{2}}\tilde{\Phi}_m(0^+)\tilde{\Phi}_{n,L}(0^-)
\label{Eq:beta{L,R}}
\end{align}

The expression for $g_n$ is the same as in \ref{Eq:Expression for g- closed case} but with the new set of Hermitian modes satisfying the open-boundary conditions discussed before. The interaction Hamiltonian then is found as
\begin{align}
-\sum\limits_{m<n,l} \hbar g_{mnl}(\hat{P}_{mn}-\hat{P}_{nm}) (\hat{a}_l-\hat{a}_l^{\dagger})
\end{align}  

Gathering all different contributions together and moving to a new frame where $\hat{P}_{mn}\to i\hat{P}_{mn}$ for $m<n$, $ a_n \to i a_n $ and $b_{n,L/R} \to i b_{n,L/R} $, the Hamiltonian in its $2^{nd}$ quantized form reads 
\begin{align}
\begin{split}
\hat{\mathcal{H}}&=\underbrace{\sum\limits_n \hbar\Omega_n \hat{P}_{nn}}_{\hat{\mathcal{H}}_A}+\underbrace{\sum\limits_n \hbar\omega_n \hat{a}_n^{\dagger} \hat{a}_n}_{\hat{\mathcal{H}}_C} +\underbrace{\sum\limits_{n,S=\{L,R\}} \hbar\omega_{n,S} \hat{b}_{n,S}^{\dagger}\hat{b}_{n,S}}_{\hat{\mathcal{H}}_B}\\
&+\underbrace{\sum\limits_{m<n,l} \hbar g_{mnl} \left(\hat{P}_{mn}+\hat{P}_{nm}\right)\left(\hat{a}_l+\hat{a}^{\dagger}_l\right)}_{\hat{\mathcal{H}}_{int}}\\
&+\underbrace{\sum\limits_{m,n,S=\{L,R\}} \hbar\beta_{mn,S}\left(\hat{a}_m+\hat{a}_m^{\dagger}\right)\left(\hat{b}_{n,S}+ \hat{b}_{n,S}^{\dagger}\right)}_{\hat{\mathcal{H}}_{CB}}
\end{split}
\label{Eq:2nd Quantized Hamiltonian_Open Case_APP}
\end{align}
\section{LARGE TRANSMON LIMIT}
It is very insightful to see how the previous results change if we assume transmon's length $d$ is comparable with resonator's wavelength $L$. In such a case we assume that the coupling is not local and spreads over whole length of transmon with mutual capacitance per length $c_g$. Following the same discrete to continuous approach one finds the Lagrangian as
\begin{align}
\begin{split}
\mathcal{L}&=\frac{1}{2}(C_J+c_g d)\dot{\Phi}_J^2-U_J(\Phi_J)\\
&+\int_{0}^{L}\,dx\ \left[\frac{1}{2}(c+c_g\pi_d(x,x_0))(\frac{\partial \Phi}{\partial t})^2-\frac{1}{2l}(\frac{\partial \Phi}{\partial x})^2\right]\\
&-\int_{0}^{L}\,dx\ c_g\pi_d(x,x_0)\dot{\Phi}_J\frac{\partial \Phi}{\partial t} 
\end{split}
\end{align}
where $\pi_d(x,x_0)$ is a unit pulse of width $d$ that is defined in terms of Heaviside function $\theta(x)$ as $\pi_d(x,x_0)\equiv \theta(x-x_0+d/2)-\theta(x-x_0-d/2)$. The Euler-Lagrange E.O.M then read
\begin{align}
&(C_J+c_gd)\ddot{\Phi}_J+\frac{\partial U_J(\Phi_J)}{\partial\Phi_J}=\int_{0}^{L} dx c_g \pi_d(x,x_0)\frac{\partial^2 \Phi}{\partial t^2}\\
&\frac{\partial^2 \Phi}{\partial x^2}-l\left(c+c_g\pi_d(x,x_0)\right)\frac{\partial^2 \Phi}{\partial t^2}=-lc_g\pi_d(x,x_0)\ddot{\Phi}_J
\end{align}

Comparing this to the Euler-Lagrange E.O.M drived earlier in appendix $B.2$ It is clear that the structure of the equations has remained the same while $C_g\delta(x-x_0)$ has been replaced with $c_g\pi(x,x_0)$ . The Hamiltonian can be found through the usual Legendre transformation as
\begin{align}
\begin{split}
\mathcal{H}&=\underbrace{\frac{Q_J^2}{2C_J^{mod}}-E_J\cos\left(2\pi\frac{\Phi_J}{\phi_0}\right)}_{\mathcal{H}_A^{mod}}\\
&+\underbrace{\int_{0}^{L}dx \left[\frac{\rho^2(x,t)}{2c_d(x,x_0)}+\frac{1}{2l}\left(\frac{\partial \Phi(x,t)}{\partial x}\right)^2\right]}_{\mathcal{H}_C^{mod}} \\
&+\underbrace{\frac{Q_J}{C_J^{mod}}\int_{0}^{L}dx c_g\pi_d(x,x_0)\frac{\rho(x,t)}{c_d(x,x_0)}}_{\mathcal{H}_{int}}
\end{split}
\end{align}
where $c_d(x,x_0)$ is the modified capacitance per length and reads
\begin{align}
c_d(x,x_0)=c+c_g\pi_d(x,x_0)\circledS \frac{C_J}{d}
\end{align}
The $\circledS$-notation represents series combination of two capacitors. Surprisingly, we observe that when the dimension of transmon is taken into account, the modification is mutual and transmon's spectrum is also influenced by the coupling such that $C_J^{mod}$ reads
\begin{align}
\begin{split}
C_J^{mod}&=C_J+\int_{0}^{L} dx \underbrace{\frac{cc_g\pi_d(x,x_0)}{c+c_g\pi_d(x,x_0)}}_{c_g\pi_d(x,x_0)\circledS c}\\
&=C_J+(c_g \circledS c)d
\end{split}
\end{align}
The results above are general such that one can replace the rectangular pulse $c_g\pi_d(x,x0)$ in capacitance per length by any smooth capacitance per length $c_g(x,x_0)$ and the form of $\mathcal{H}$, $c_d(x,x_0)$ and $C_J^{mod}$ remain the same. 
\section{HAMILTONIAN AND MODIFIED EIGENMODES OF A CLOSED CAVITY-QED SYSTEM}
In this Appendix, first we derive the classical Hamiltonian for a general system containing finite number of point charges interacting with the EM field inside a closed cavity. This is achieved by expressing the Maxwell's and Newton's equations of motion in a Lagrangian formalism and then a Legendre transformation to find the Hamiltonian. This model is then reduced to describe a one-dimensional cavity shown in Fig. \ref{fig:Cavity-QED-Closed}. In order to emphasize on the resemblance to the cQED results we found earlier, it is assumed that there is only a single electron at  $\vec{R}_e(t)$, while all other electronic or nuclear degrees of freedom are frozen at $\vec{R}_0$.
\subsection{Classical Lagrangian}
Following the usual canonical quantization scheme, first we have to find the classical Lagranian. We already know the equations of motion for the EM fields to be the Maxwell's equations as
\begin{align}
&\nabla.\vec{E}(\vec{r},t)=\frac{\rho(\vec{r},t)}{\epsilon_0}
\label{Eq:Maxwell Equations-Divergence of E}\\
&\nabla.\vec{B}(\vec{r},t)=0
\label{Eq:Maxwell Equations-Divergence of B}\\
&\nabla\times\vec{E}(\vec{r},t)=-\frac{\partial\vec{B}(\vec{r},t)}{\partial t}
\label{Eq:Maxwell Equations-Curl of E}\\
&\nabla\times\vec{B}(\vec{r},t)=\mu_0 \vec{J}(\vec{r},t)+\mu_0\epsilon_0\frac{\partial\vec{E}(\vec{r},t)}{\partial t}
\label{Eq:Maxwell Equations-Curl of B}
\end{align}
where $\rho(\vec{r},t)$ and $\vec{J}(\vec{r},t)$ are scalar charge density and vector current density  and are given as 
\begin{align}
&\rho(\vec{r},t)=\sum\limits_n q_n\delta^{(3)}\left(\vec{r}-\vec{r}_n(t)\right)
\label{Eq:Maxwell Eqautions-Charge density}\\
&\vec{J}(\vec{r},t)=\sum\limits_n q_n\dot{\vec{r}}_n(t)\delta^{(3)}\left(\vec{r}-\vec{r}_n(t)\right)
\label{Eq:Maxwell Eqautions-current density}
\end{align}

The remaining equation of motion is a Newton equation regarding the mechanical motion of the electron which reads

\begin{align}
m_n\ddot{\vec{r}}_n(t)&=\underbrace{q_n\left[\vec{E}(\vec{r}_n(t),t)+\dot{\vec{r}}_n(t)\times \vec{B}(\vec{r}_n(t),t)\right]}_{\rm Lorentz\quad Force}
\label{Eq:Newton's Law for Lorentz Force}
\end{align}

Based on \ref{Eq:Maxwell Equations-Divergence of B} and \ref{Eq:Maxwell Equations-Curl of E} we are able to express the physical fields $\vec{E}(\vec{r},t)$ and $\vec{B}(\vec{r},t)$ in terms of scalar potential $V(\vec{r},t)$ and vector potential $\vec{A}(\vec{r},t)$ up to a gauge degree of freedom as
\begin{align}
&\vec{E}(\vec{r},t)=-\frac{\partial \vec{A}(\vec{r},t)}{\partial t}-\nabla V(\vec{r},t) ) \\
&\vec{B}(\vec{r},t)=\nabla\times\vec{A}(\vec{r},t)
\end{align}

It is possible to write a Lagrangian that produces all previous equations of motion \ref{Eq:Maxwell Equations-Divergence of E}-\ref{Eq:Maxwell Equations-Curl of B} and \ref{Eq:Newton's Law for Lorentz Force} as a result of the variational principle $\delta \mathcal{L}=0$. This Lagrangian reads 
\begin{align}
\begin{split}
\mathcal{L}&=\sum\limits_n\frac{1}{2}m_n\dot{\vec{r}}_n^2\\
&+\int d^3r\left[\frac{1}{2}\epsilon_0\left(\partial_t\vec{A}+\nabla V\right)^2-\frac{1}{2\mu_0}\left(\nabla \times \vec{A}\right)^2\right]\\
&+\int d^3r \left[\vec{J}.\vec{A}-\rho V\right]
\end{split}
\end{align}

In order to proceed further, we need to fix the gauge. Choosing to work in Coulomb gauge defined as $\nabla .\vec{A}=0$ and using \ref{Eq:Maxwell Equations-Divergence of E} we find that the scalar potential $V(\vec{r},t)$ satisfies a Poisson equation as
\begin{align}
\nabla^2 V(\vec{r},t)=-\frac{\rho(\vec{r},t)}{\epsilon_0}
\end{align}

Having the charge density as \ref{Eq:Maxwell Eqautions-Charge density} we can easily solve this equation to obtain
\begin{align}
V(\vec{r},t)=\sum\limits_n \frac{q_n}{4\pi\epsilon_0 |\vec{r}-\vec{r}_n(t)|}
\label{Eq:solution for V(r,t)}
\end{align}

Furthermore, this choice of gauge helps to simplify the Lagrangian since due to the divergence theorem
\begin{align}
\begin{split}
\int d^3r \partial_t \vec{A}.\nabla V&=\int d^3r \nabla.(V\partial_t \vec{A})-\int d^3r V\partial_t \underbrace{(\nabla. \vec{A})}_{0}\\ 
&=\oint d\vec{S}. \left(V\partial_t\vec{A}\right)
\end{split}
\end{align}
which means this term only contributes at the boundaries. Boundary terms do not affect equations of motion inside the cavity, however their existence are necessary to ensure the correct boundary conditions, i.e. continuity of parallel electric field and perpendicular magnetic field at the interface of cavity with the outside environment. As far as we fix these conditions properly, we can remove all surface terms form the Lagrangian. In a similar manner 
\begin{align}
\begin{split}
\int d^3 r \frac{1}{2}\epsilon_0(\nabla V)^2&=\frac{1}{2}\epsilon_0\oint d\vec{S}.\left(V\nabla V\right)-\int d^3 r\frac{1}{2}\epsilon_0 V\nabla^2 V\\
&=\frac{1}{2}\epsilon_0\oint d\vec{S}.\left(V\nabla V\right)+\int d^3r \frac{1}{2}\rho V
\end{split}
\end{align}

Finally, by putting everything together and neglecting surface terms, we find the simplified Lagrangian as
\begin{align}
\begin{split}
\mathcal{L}&=\sum\limits_n\frac{1}{2}m_n\dot{\vec{r}}_n^2-\int d^3r \, \frac{1}{2}\rho V\\
&+\int d^3r\left[\frac{1}{2}\epsilon_0\left(\partial_t\vec{A}\right)^2-\frac{1}{2\mu_0}\left(\nabla \times \vec{A}\right)^2\right]\\
&+\int d^3r \, \vec{J}.\vec{A}\\
\end{split}
\end{align}
\subsection{Classical Hamiltonian}

The first step is find the conjugate momenta as
\begin{align}
&\vec{p}_n \equiv \frac{\partial \mathcal{L}}{\partial{\dot{\vec{r}}_n}}=m_n\dot{\vec{r}}_n+q_n\vec{A}(\vec{r}_n(t),t)
\label{Eq: conjugate momentum of r(t)}\\
&\vec{\Pi}(\vec{r},t)\equiv \frac{\partial \mathcal{L}}{\partial{\dot{\vec{A}}}}= \epsilon_0 \partial_t \vec{A}(\vec{r},t)
\label{Eq: conjugate momentum of A(r,t)}
\end{align}

Then, the Hamiltonian is calculated via a Legendre transformation of the Lagrangian as
\begin{align}
\mathcal{H}=\sum\limits_n \vec{p}_n.\dot{\vec{r}}_n+\int d^3r \vec{\Pi}(\vec{r},t).\partial_t \vec{A}(\vec{r},t)-\mathcal{L}
\end{align}

substituting \ref{Eq: conjugate momentum of r(t)} and \ref{Eq: conjugate momentum of A(r,t)} into the expression for Hamiltonian we find
\begin{align}
\begin{split}
\mathcal{H}&=\sum\limits_n \frac{1}{2}m\dot{\vec{r}}_n^2+\sum\limits_n \frac{1}{2}q_n V(\vec{r}_n)\\
&+\int d^3 r\left[\frac{1}{2}\epsilon_0(\partial_t \vec{A}(r,t))^2+\frac{1}{2 \mu_0}\left(\vec{\nabla}\times\vec{A}(\vec{r},t)\right)^2\right]
\end{split}
\end{align}

By replacing $\partial_t \vec{A}(\vec{r},t)$ and $\dot{\vec{r}}_n(t)$ in terms of conjugate momenta $\vec{\Pi}(\vec{r},t)$ and $\vec{p}_n(t)$ respectively, the Hamiltonian can be rewritten as
\begin{align}
\begin{split}
\mathcal{H}&= \sum\limits_n\frac{\left[\vec{p}_n-q_n\vec{A}(\vec{r}_n,t)\right]^2}{2m_n}+\sum\limits_n \frac{1}{2}q_n V(\vec{r}_n)\\
&+\int d^3 r \left[\frac{\vec{\Pi}^2(\vec{r},t)}{2\epsilon_0}+\frac{\left(\nabla \times \vec{A}(\vec{r},t)\right)^2}{2\mu_0}\right]
\end{split}
\end{align}

which can be written in a more instructive way as
\begin{align}
\begin{split}
\mathcal{H}&=\underbrace{\sum\limits_n \frac{\vec{p}_n^2}{2m_n}+\sum\limits_n\frac{1}{2}q_n V(\vec{r}_n)}_{\mathcal{H}_A}\\
&+\underbrace{\int d^3 r \left[\frac{\vec{\Pi}^2(\vec{r},t)}{2\epsilon_0}+\frac{\left(\nabla \times \vec{A}(\vec{r},t)\right)^2}{2\mu_0}\right]}_{\mathcal{H}_C}\\
&+\underbrace{\int d^3r \sum\limits_n\frac{q_n^2}{2m_n}\vec{A}^2(\vec{r},t)\delta^{(3)}(\vec{r}-\vec{r}_n)}_{\mathcal{H}_C^{mod}}\\
&\underbrace{-\sum\limits_n \frac{q_n}{m_n}\vec{p}_n.\vec{A}(\vec{r}_n,t)}_{\mathcal{H}_{int}}
\end{split}
\label{Eq:closed cavity-QED Hamiltonian}
\end{align}

This is the most general form of the classical Hamiltonian of a finite number of charges interacting with the EM field inside a closed cavity. In what follows, we make a few assumptions to reduce this model for the system shown in Fig. \ref{fig:Cavity-QED-Closed}. First of all, we assume that the wavelength of EM field is much larger than atomic scale $\vec{r}_e$ such that we can apply $0^{th}$-order dipole approximation $\vec{A}(\vec{R}_e(t),t)\approx \vec{A}(\vec{R}_0,t)$ in both $\mathcal{H}_C^{mod}$ and $\mathcal{H}_{int}$. Finally, by considering that there is only a single charge located at $\vec{R}_e(t)$ and the nucleus is fixed at $\vec{R}_0$ the Hamiltonian is reduced to
\begin{align}
\begin{split}
\mathcal{H}&=\underbrace{\frac{\vec{p}_e^2}{2m_e}-eV(\vec{r}_e)}_{\mathcal{H}_A}\\
&+\underbrace{\int d^3 r \left[\frac{\vec{\Pi}^2(\vec{r},t)}{2\epsilon_0}+\frac{\left(\nabla \times \vec{A}(\vec{r},t)\right)^2}{2\mu_0}\right]}_{\mathcal{H}_C}\\
&+\underbrace{\int d^3r \frac{e^2}{2m_e}\vec{A}^2(\vec{r},t)\delta^{(3)}(\vec{r}-\vec{R}_0)}_{\mathcal{H}_C^{mod}}\\
&\underbrace{-\frac{e}{m_e}\vec{p}_e.\vec{A}(\vec{R}_0,t)}_{\mathcal{H}_{int}}
\end{split}
\label{Eq:closed cavity-QED Hamiltonian}
\end{align}

\subsection{Modified Cavity Eigenmodes and Eigenfrequencies}
Having found the modification introduced by the $\vec{A}^2$ term in the previous section, we can now calculatae the effect it has on the structure of the modes. Specifically, we are after eigenmodes of the modified Hamiltonian for the cavity given as $\mathcal{H}_C^{mod}\equiv\mathcal{H}_C+\mathcal{H}^{mod}$. This can be done by finding the Hamiltonian E.O.M for the conjugate fields as
\begin{align}
\frac{\partial}{\partial t} \vec{A}(\vec{r},t)&=\frac{1}{\epsilon_0}\vec{\Pi}(\vec{r},t)
\label{Eq: time-evolution of A(x,t)}\\
\begin{split}
\frac{\partial}{\partial t} \vec{\Pi}(\vec{r},t)&=-\frac{1}{\mu_0}\nabla \times (\nabla \times \vec{A}(\vec{r},t))\\
&-\frac{e^2}{m_e}\vec{A}(\vec{r},t)\delta^{(3)}(\vec{r}-\vec{R}_0)
\end{split}
\label{Eq: time-evolution of Pi(x,t)}
\end{align}

By combining these two equations and applying the gauge condition $\nabla .\vec{A}=0$ we find
\begin{align}
\left(\nabla^2 -\mu_0\epsilon_0 \frac{\partial^2}{\partial t^2}\right)\vec{A}(\vec{r},t)=\frac{\mu_0 e^2}{m_e}\vec{A}(\vec{r},t)\delta^{(3)}(\vec{r}-\vec{R}_0)
\end{align}
which is a wave equation with an extra term on the R.H.S due to $\vec{A}^2$ modification. The coefficient $\frac{\mu_0 e^2}{m_e}$ can be expressed in terms of fine structure constant $\alpha$ and Bohr's radius $a_0$ as $4\pi\alpha^2 a_0$. By doing a Fourier transform 
\begin{align}
\tilde{\vec{A}}(\vec{r},t)=\frac{1}{2\pi}\int_{-\infty}^{+\infty} d\omega \tilde{\vec{A}}(\vec{r},\omega)e^{-i\omega t}
\end{align}
we can easily separate the time and spacial dependences to obtain
\begin{align}
&\left(\nabla^2+\left(\frac{\omega}{c}\right)^2-4\pi\alpha^2a_0\delta^{(3)}(\vec{r}-\vec{R}_0)\right)\tilde{\vec{A}}(\vec{r},\omega)=0
\label{Eq:seperation of variables A(r,omega)}
\end{align}

Assuming a closed cavity case and remembering that $\vec{E}_{\parallel}$ and $\vec{B}_{\bot}$ are continuous across the cavity, the boundary conditions read
\begin{align}
&\vec{n}_{\parallel}\times \left.\left(- i\omega\tilde{\vec{A}}(\vec{r},\omega)+\nabla \tilde{V}\right)\right|_{B}=0
\label{Eq:continuity of parallel Electric field}\\
&\vec{n}_{\bot}. \left.\left(\nabla \times \tilde{\vec{A}}(\vec{r},\omega)\right)\right|_{B}=0
\label{Eq:continuity of perpendicular Electric field}
\end{align}
where $\vec{n}_{\bot}$ and $\vec{n}_{\parallel}$ represent perpendicular and parallel unit vectors on the boundaries of the cavity and $\tilde{V}$ is the time-Fourier transform of the scalar potential. Equation \ref{Eq:seperation of variables A(r,omega)} with the boundary conditions above provide a discrete set of modes due to finite volume of the cavity. For notation simplicity, we label the eigenfrequencies as $\omega_{\lambda}$ and the modes as $\tilde{\vec{A}}_{\lambda}(\vec{r})\equiv\tilde{\vec{A}}(\vec{r},\omega_{\lambda})$ , while in reality $\lambda $ denotes multiple sets of discrete numbers each for a separate dimension of the cavity. These modes satisfy the general orthogonality relation
\begin{align}
&\int d^3 r \tilde{\vec{A}}_{\lambda}(\vec{r}).\tilde{\vec{A}}_{\lambda'}(\vec{r})=\mathcal{V}\delta_{\lambda\lambda'}
\end{align}
where we have set the normalization such that the modes are dimensionless. Another orthogonality relation can be found in terms of $\nabla\tilde{\vec{A}}_{\lambda}(\vec{r})$ as
\begin{align}
\begin{split}
&+\int d^3 r \nabla\tilde{\vec{A}}_{\lambda}(\vec{r}).\nabla\tilde{\vec{A}}_{\lambda'}(\vec{r})\\
-&\frac{1}{2}\oint d\vec{S}.\left[\tilde{\vec{A}}_{\lambda}(\vec{r}).\nabla\tilde{\vec{A}}_{\lambda'}(\vec{r})+\tilde{\vec{A}}_{\lambda'}(\vec{r}).\nabla\tilde{\vec{A}}_{\lambda}(\vec{r})\right]\\
&+4\pi\alpha^2 a_0\tilde{\vec{A}}_{\lambda}(\vec{R}_0).\tilde{\vec{A}}_{\lambda'}(\vec{R}_0)=k_{\lambda}k_{\lambda'}\mathcal{V}\delta_{\lambda\lambda'}
\end{split}
\end{align}

Up to this point, we have considered the mode structure for a general cavity with any arbitrary geometry. In order to demonstrate the connection to the results for a one dimensional cQED system, we have to make a few assumptions about the geometry of the cavity.  We assume that the cavity's length is much larger than the diameter of its cross section, i.e. $L\gg\sqrt{S}$. By considering variation of the eigenmodes only along this dimension we can write $\vec{A}(\vec{r},t)=\vec{u}_z A(x,t)$ and thus $\vec{B}(\vec{r},t)=-\vec{u}_y \partial_x A(x,t)$. The Hamiltonian is then reduced to
\begin{align}
\begin{split}
\mathcal{H}&=\underbrace{\frac{\vec{p}_e^2}{2m_e}-eV(\vec{r}_e)}_{\mathcal{H}_A}\\
&+\underbrace{\int d^2s \int_{0}^{L} dx \left[\frac{\Pi^2(x,t)}{2\epsilon_0}+\frac{\left(\partial_x A(x,t)\right)^2}{2\mu_0}\right]}_{\mathcal{H}_C}\\
&+\underbrace{\int d^2s \int_{0}^{L} dx \frac{e^2}{2m_e}A^2(x_0,t)\delta^2(\vec{s}-\vec{s}_0)\delta(x-x_0)}_{\mathcal{H}^{mod}}\\
&\underbrace{-\frac{e}{m_e}p_e^z A(x_0,t)}_{\mathcal{H}_{int}}
\end{split}
\end{align}

Following the same procedure, we can find a modified wave equation as a result of $\hat{\mathcal{H}}_C^{mod}$  
\begin{align}
\left(\frac{d^2}{dx^2}+\left(\frac{\omega}{c}\right)^2-\frac{4\pi\alpha^2a_0}{S}\delta(x-x_0)\right)\tilde{A}(x,\omega)=0
\label{Eq:seperation of variables A(x,omega)}
\end{align}

Assuming that the atom is fixed at point $x_0$, $\nabla \tilde{V}$ only affects the bouundary condtion for the zero frequency mode which we are not interested in. Therefore, by applying the boundary conditions
\begin{align}
\left. \tilde{A}(x,\omega)\right|_{x=0,L}=0
\end{align}
we find the normalized eigenfrequencies to satisfy a transcendental equation as
\begin{align}
\sin{(k_nL)}+\chi_c \frac{\sin{(k_nx_0)}\sin{(k_n(L-x_0))}}{k_nL}=0
\label{Eq: Closed Cavity Eigenfrequencies}
\end{align}
where we have defined the unitless parameter $\chi_c \equiv 4\pi\alpha^2 a_0 \frac{L}{S}$. 
The real-space representation of the eigenmodes read
\begin{align}
\tilde{A}_n(x)\propto
\begin{cases}
\sin{\left(k_n(L-x_0)\right)}\sin{(k_n x)}&0<x<x_0\\
\sin{(k_n x_0)}\sin{(k_n(L-x))} &x_0<x<L
\end{cases}
\label{Eq: Closed Cavity Eigenmodes}
\end{align}

Eventually, one can show that these eigenmodes satisfy the orthogonality relations:
\begin{align}
&\int_{0}^{L}dx \tilde{A}_m(x)\tilde{A}_n(x)=L\delta_{mn}
\label{Eq:Orthogonality condition for A _ app}\\
&\int_{0}^{L} dx \frac{\partial \tilde{A}_m}{\partial x}\frac{\partial \tilde{A}_n}{\partial x}+\frac{\chi_c}{L}\tilde{A}_m(x_0)\tilde{A}_n(x_0)=k_mk_nL\delta_{mn}
\label{Eq:Orthogonality condition for d/dx A _app}
\end{align}

Note that only the ratio $\frac{L}{S}$ is determined by the geometry of the cavity, while the pre-factor $4\pi\alpha^2 a_0 \approx 3.54\times 10^{-14} m$ is a universal length scale. This implies that the modification is only visible when $\frac{S}{L}$ is around the same order as $4\pi\alpha^2 a_0$. 
\subsection{Canonical Quantization}
Having found the proper set of eigenmodes and eigenfrequencies that diagonalizes the classical Hamiltonian for a one-dimensional closed cavity-QED system, we can move forward and extend the classical variables into quantum operators by introducing the necessary commutation relation between conjugate pairs. Let's consider the conjugate fields for the cavity first. We can expand these fields in terms of the proper eigenmodes as 
\begin{align}
\hat{A}(x,t)=\sum\limits_n \left(\frac{\hbar}{2\omega_n \epsilon_0 SL}\right)^{\frac{1}{2}}\left(\hat{a}_n+\hat{a}_n^{\dagger}\right)\tilde{A}_n(x)\\
\hat{\Pi}(x,t)=-i\sum\limits_n \left(\frac{\hbar \epsilon_0 \omega_n}{2SL}\right)^{\frac{1}{2}}\left(\hat{a}_n-\hat{a}_n^{\dagger}\right)\tilde{A}_n(x)
\end{align}

where $\hat{a}_n$  and $\hat{a}_n^{\dagger}$ are annihilation and creation operators for each mode. By inserting the above equations and using the orthogonality conditions \ref{Eq:Orthogonality condition for A _ app} and \ref{Eq:Orthogonality condition for d/dx A _app}, $\hat{\mathcal{H}}_C^{mod}=\hat{\mathcal{H}}_C+\hat{\mathcal{H}}^{mod}$ become diagonal as
\begin{align}
\hat{\mathcal{H}}_C^{mod}=\sum\limits_n \frac{\hbar \omega_n}{2}\left(a_n^{\dagger}a_n +a_n a_n^{\dagger}\right)=\sum\limits_n \hbar \omega_n a_n^{\dagger}a_n+ const.
\end{align}

The next step is to obtain the spectrum of $\hat{\mathcal{H}}_A$ by solving a Schrodinger equation in real-space basis as
\begin{align}
\left(-\frac{\hbar^2}{2m_e}\nabla_e^2-e V(\vec{r}_e)\right)\Psi_n(\vec{r}_e)=\hbar\Omega_n \Psi_n(\vec{r}_e)
\end{align}
where we have denoted the eigenmodes and eigenenergies by $\{\Psi_n(\vec{r}_e),E_n=\hbar \Omega_n | n\in \mathbb{N}^{0}\}$. $\hat{\mathcal{H}}_A$ can be decomposed as
\begin{align}
\hat{\mathcal{H}}_A=\sum\limits_n \hbar\Omega_n \hat{P}_{nn} 
\end{align} 

$\vec{p}_e$ also has a spectral decomposition over this basis. Since the Coulomb potential $V(\vec{r}_e)$ is an even function of $\vec{r}_e$, as we explained in the case of charge qubit only diagonal matrix elements of $\vec{p}_e$ are nonzero and we can write
\begin{align}
\begin{split}
\hat{\vec{p}}_e =\sum\limits_{m\neq n} \bra{m}\vec{p}_{e}\ket{n}\hat{P}_{mn}\\
\end{split}
\end{align}
where matrix elements $\vec{p}_{e,mn}$ can be calculated as
\begin{align}
\bra{m}\vec{p}_e\ket{n}=\int d^3 \vec{r}_e \Psi_m(\vec{r}_e)\frac{\hbar}{i} \vec{\nabla} \Psi_n(\vec{r}_e)
\end{align}
By working in a basis where $\Psi_{n}(\vec{r}_e)$ are real functions, the dipole matrix elements are purely imaginary which allows us to write 
\begin{align}
\begin{split}
\hat{\vec{p}}_e =\sum\limits_{m>n} \bra{m}\vec{p}_{e}\ket{n}(\hat{P}_{mn}-\hat{P}_{nm})\\
\end{split}
\end{align}

Finally, for sake of resemblance to the cQED results we move to a new frame $\hat{P}_{mn} \to i \hat{P}_{mn}$ for $m<n$ to obtain the Hamiltonian as
\begin{align}
\begin{split}
\hat{\mathcal{H}}&=\underbrace{\sum\limits_n \hbar\Omega_n \hat{P}_{nn}}_{\hat{\mathcal{H}}_A}+\underbrace{\sum\limits_n \hbar \omega_n \hat{a}_n^{\dagger} \hat{a}_n}_{\hat{\mathcal{H}}_C^{mod}}\\
&+\underbrace{\sum\limits_{m>n,l} \hbar g_{mnl} \left(\hat{P}_{mn}+\hat{P}_{nm}\right)\left(\hat{a}_l+\hat{a}^{\dagger}_l\right)}_{\hat{\mathcal{H}}_{int}}
\end{split}
\end{align}
where the coupling strength $g_{mnl}$ reads
\begin{align}
\hbar g_{mnl}=\frac{e}{m_e}(i\vec{p}_{e,mn}.\vec{u}_z)\left(\frac{\hbar}{2\epsilon_0\omega_l SL}\right)^{\frac{1}{2}}\tilde{A}_l(x_0)
\end{align}
\bibliography{cQEDquantization}
\end{document}